\documentclass[acmtog,nonacm]{acmart}
\acmSubmissionID{1841}

\usepackage{booktabs} 

\usepackage{xspace} 

\usepackage[ruled]{algorithm2e} 

\SetAlFnt{\small}
\SetAlCapFnt{\small}
\SetAlCapNameFnt{\small}
\SetAlCapHSkip{0pt}
\usepackage[export]{adjustbox}
\usepackage{xspace}
\usepackage{array}
\newcolumntype{P}[1]{>{\centering\arraybackslash}p{#1}}
\usepackage{tikz}
\usepackage{amsmath}
\usepackage{newunicodechar}
\newunicodechar{↔}{\ensuremath{\leftrightarrow}}
\usetikzlibrary{arrows.meta, positioning, fit, shapes.geometric}
\usepackage{multirow}
\usepackage{graphicx}
\usepackage{float}
\usepackage{standalone}
\newcommand{\SA}[1]{{#1}}
\newcommand{\FI}[1]{{#1}}
\usepackage{framed}
\newcommand{\placeholderFailed}{
    \begin{tikzpicture}[baseline=(current bounding box.center)]
        \draw[gray, thick] (0,0) rectangle (2,2); 
        \draw[gray, thick] (0,0) -- (2,2); 
        \draw[gray, thick] (0,2) -- (2,0); 
        \node at (1,1) {Failed}; 
    \end{tikzpicture}
}
\newcommand{\placeholderUnav}{
    \begin{tikzpicture}[baseline=(current bounding box.center)]
        \draw[gray, thick] (0,0) rectangle (2,2); 
        \node at (1,1) {Unavailable}; 
    \end{tikzpicture}
}

\acmJournal{TOG}

\acmDOI{10.1145/3763336}

\setcopyright{none}
\renewcommand\footnotetextcopyrightpermission[1]{}
\begin{document}
\title{\SA{Detail Enhanced Gaussian Splatting for Large-Scale Volumetric Capture}}

\author[Julien Philip]{Julien Philip}
\orcid{0000-0003-3125-1614}
\authornotemark[1]
\affiliation{
  \institution{Eyeline Labs}
  \city{London}
  \country{United Kingdom}
}
\email{julien.philip@scanlinevfx.com}

\author[Li Ma]{Li Ma}
\orcid{0000-0002-6992-0089}
\authornotemark[1]
\affiliation{
  \institution{Eyeline Labs}
  \city{Los Angeles}
  \country{United States of America}
}
\email{li.ma@scanlinevfx.com}

\author[Pascal Clausen]{Pascal Clausen}
\orcid{0009-0000-2534-9266}
\authornotemark[1]
\affiliation{
  \institution{Eyeline Labs}
  \city{Geneve}
  \country{Switzerland}
}
\email{pascal.clausen@scanlinevfx.com}

\author[Wenqi Xian]{Wenqi Xian}
\orcid{0009-0007-4678-6458}
\authornote{These authors made equal technical contributions.}
\affiliation{
  \institution{Eyeline Labs}
  \city{Los Angeles}
  \country{United States of America}
}
\email{wenqi.xian@scanlinevfx.com}

\author[Ahmet Levent Ta\c{s}el]{Ahmet Levent Taşel}
\orcid{0009-0002-7150-0160}
\affiliation{
  \institution{Eyeline Labs}
  \city{Vancouver}
  \country{Canada}
}
\email{ahmet.tasel@scanlinevfx.com}

\author[Mingming He]{Mingming He}
\orcid{0000-0002-9982-7934}
\affiliation{
  \institution{Eyeline Labs}
  \city{Los Angeles}
  \country{United States of America}
}
\email{mingming.he@scanlinevfx.com}

\author[Xueming Yu]{Xueming Yu}
\orcid{0009-0009-8189-6024}
\affiliation{
  \institution{Eyeline Labs}
  \city{Los Angeles}
  \country{United States of America}
}
\email{xueming.yu@scanlinevfx.com}

\author[David M. George]{David M. George}
\orcid{0009-0001-1570-3708}
\affiliation{
  \institution{Eyeline Labs}
  \city{Los Angeles}
  \country{United States of America}
}
\email{david.george@scanlinevfx.com}

\author[Ning Yu]{Ning Yu}
\orcid{0009-0004-6865-1325}
\affiliation{
  \institution{Eyeline Labs}
  \city{Los Angeles}
  \country{United States of America}
}
\email{ning.yu@scanlinevfx.com}

\author[Oliver Pilarski]{Oliver Pilarski}
\orcid{0009-0005-7803-2314}
\affiliation{
  \institution{Eyeline Labs}
  \city{Munich}
  \country{Germany}
}
\email{oliver.pilarski@scanlinevfx.com}

\author[Paul Debevec]{Paul Debevec}
\orcid{0000-0001-7381-2323}
\affiliation{
  \institution{Eyeline Labs}
  \city{Los Angeles}
  \country{United States of America}
}
\email{debevec@scanlinevfx.com}

\newcommand{\TODO}[1]{{\bf\color{red}[TODO: #1]}}

\newcommand{\JULIENTEXT}[1]{{\bf\color{purple}#1}}
\newcommand{\JULIEN}[1]{{\bf\color{purple}[JULIEN: #1]}}
\newcommand{\LITEXT}[1]{{\color{cyan}#1}}
\newcommand{\li}[1]{{\bf\color{cyan}[LI: #1]}}
\newcommand{\lirm}[1]{{\color{gray}#1}}
\newcommand{\AHMETTEXT}[1]{{\color{brown}#1}}
\newcommand{\AHMET}[1]{{\bf\color{brown}[AHMET: #1]}}
\newcommand{\WENQITEXT}[1]{{\color{purple}#1}}
\newcommand{\WENQI}[1]{{\bf\color{purple}[WENQI: #1]}}
\newcommand{\MINGMINGTEXT}[1]{{\bf\color{orange}#1}}
\newcommand{\MINGMING}[1]{{\bf\color{orange}[MINGMING: #1]}}
\newcommand{\OLIVERTEXT}[1]{{\color{cyan}#1}}
\newcommand{\OLIVER}[1]{{\bf\color{cyan}[OLIVER: #1]}}
\newcommand{\PASCALTEXT}[1]{{\color{magenta}#1}}
\newcommand{\PASCAL}[1]{{\bf\color{magenta}[PASCAL: #1]}}
\newcommand{\PAULTEXT}[1]{{\color{teal}#1}}
\newcommand{\PAUL}[1]{{\bf\color{teal}[PAUL: #1]}}
\newcommand{\NINGTEXT}[1]{{\color{red}#1}}
\newcommand{\NING}[1]{{\bf\color{red}[NING: #1]}}

\newcommand{\scenerig}{\emph{Scene Rig}\xspace}
\newcommand{\facerig}{\emph{Face Rig}\xspace}

\begin{abstract}

We present a unique system for large-scale, multi-performer, high resolution 4D volumetric capture providing realistic free-viewpoint video up to and including 4K resolution facial closeups.
To achieve this, we employ a novel volumetric \SA{capture, reconstruction and rendering pipeline based on Dynamic Gaussian Splatting and Diffusion-based Detail Enhancement.
We design our pipeline} specifically to meet the demands of high-end media production.
We employ two capture rigs: the \scenerig, which captures multi-actor performances at a resolution which falls short of 4K production quality, and the \facerig, which records \SA{high-fidelity} single-actor facial detail to serve as a reference for detail enhancement.
We first reconstruct dynamic performances from the \scenerig using 4D Gaussian Splatting, incorporating new model designs and training strategies to improve reconstruction, \SA{dynamic range}, and rendering quality.
\SA{Then to render high-quality images for facial closeups, we introduce a diffusion-based detail enhancement model.
This model is} fine-tuned with high-fidelity data from the same actors recorded in the \facerig.
We train on paired data generated from low- and high-quality Gaussian Splatting (GS) models, using the low-quality input to match the quality of the \scenerig, with the high-quality GS as ground truth.
Our results demonstrate the effectiveness of this pipeline in bridging the gap between the scalable performance capture of a large-scale rig and the high-resolution standards required for film and media production.

\end{abstract}

%
%


%
%


\begin{teaserfigure}
\centering
    \begin{minipage}{0.24\linewidth}
    \centering
    \setlength{\fboxrule}{0.5pt} 
    \setlength{\fboxsep}{0pt}
    \fbox{\adjincludegraphics[trim= {0.32\width} {0.18\height} {0.335\width} {0.2\height}, clip, width=0.48\linewidth]{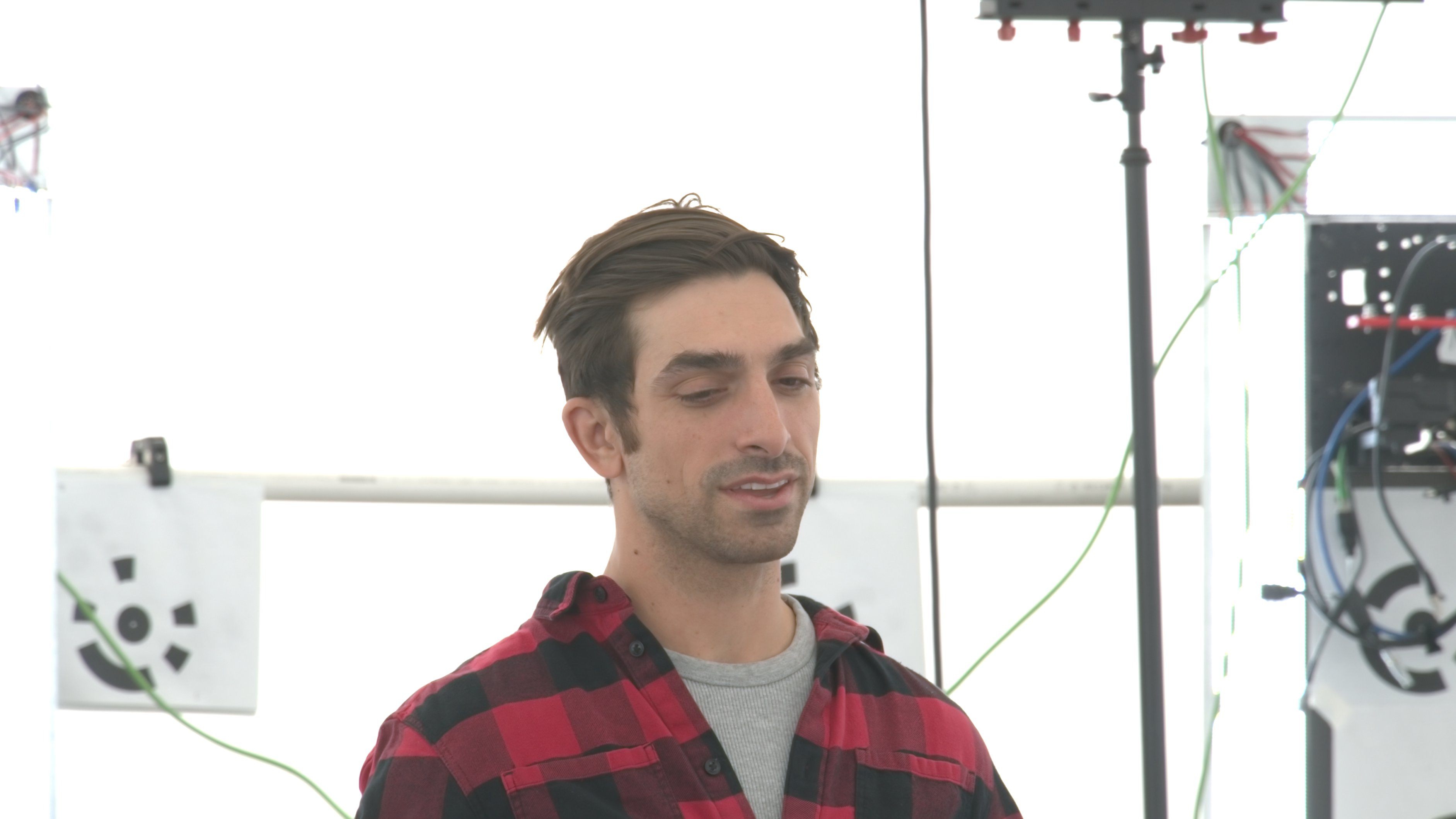}}
    \fbox{\adjincludegraphics[trim= {0.32\width} {0.18\height} {0.335\width} {0.2\height}, clip, width=0.48\linewidth]{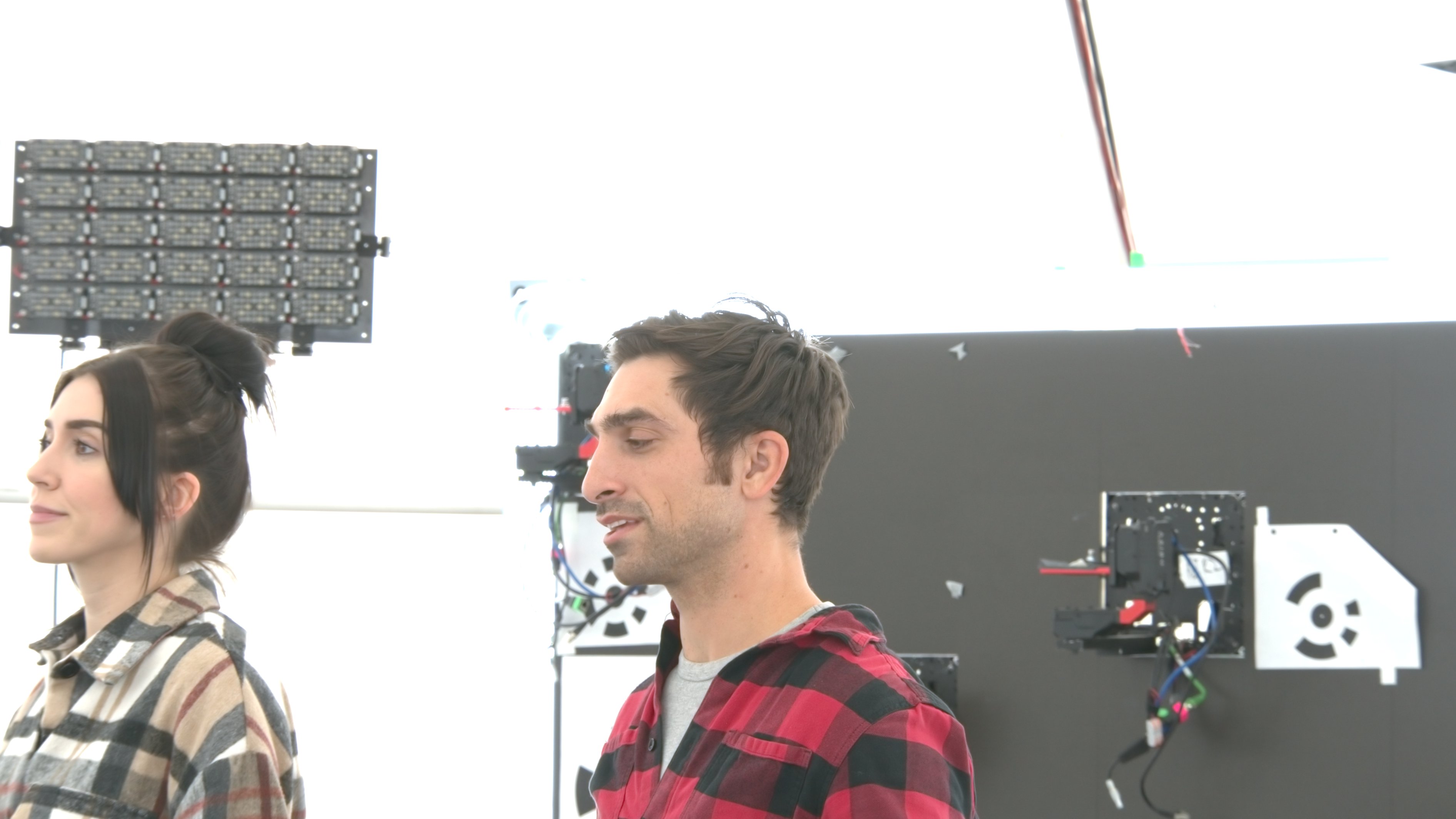}}\\
    \fbox{\adjincludegraphics[trim= {0.0\width} {0.435\height} {0.0\width} {0.0\height}, clip, width=0.48\linewidth]{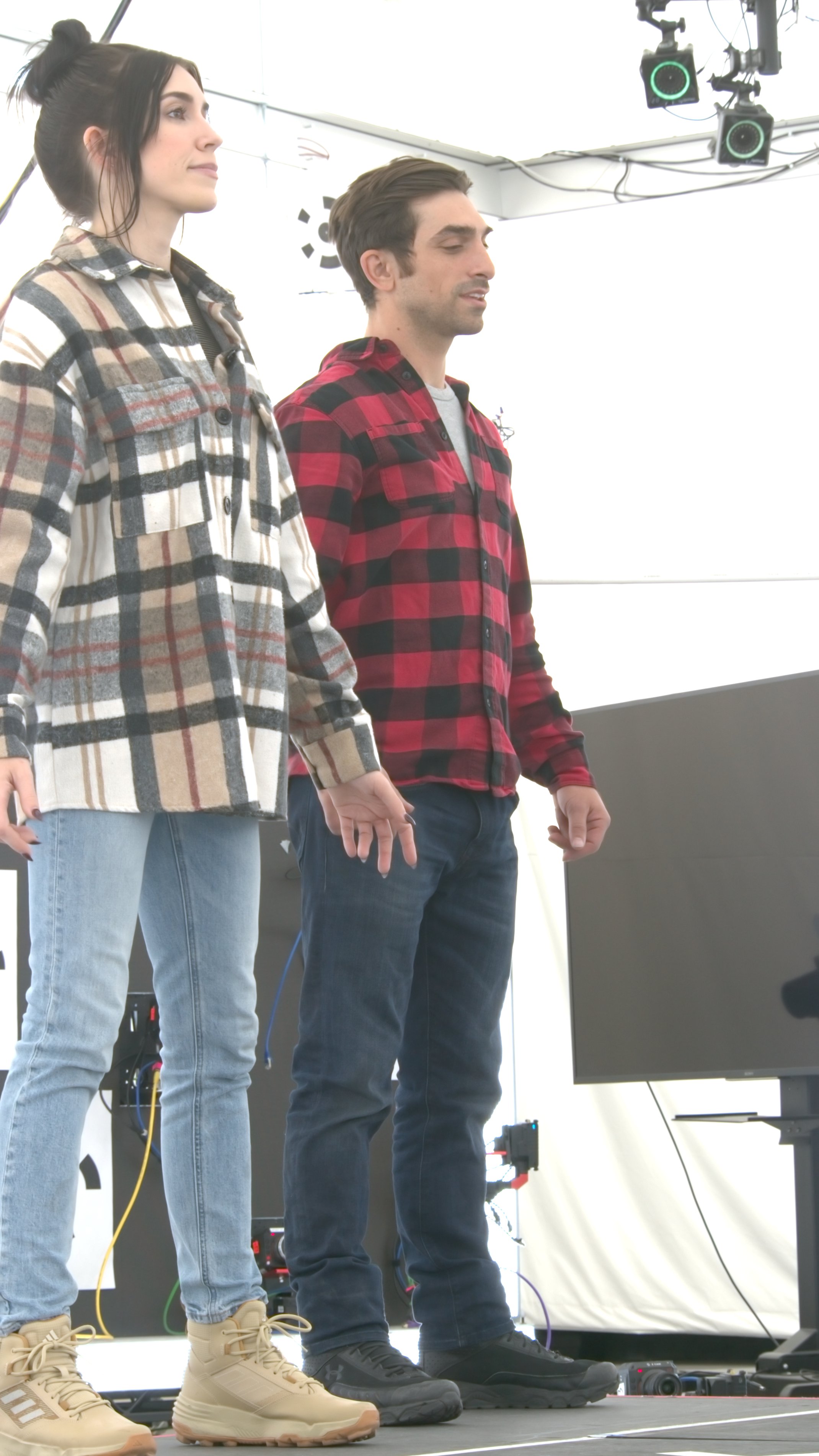}}
    \fbox{\adjincludegraphics[trim= {0.335\width} {0.18\height} {0.32\width} {0.2\height}, clip, width=0.48\linewidth]{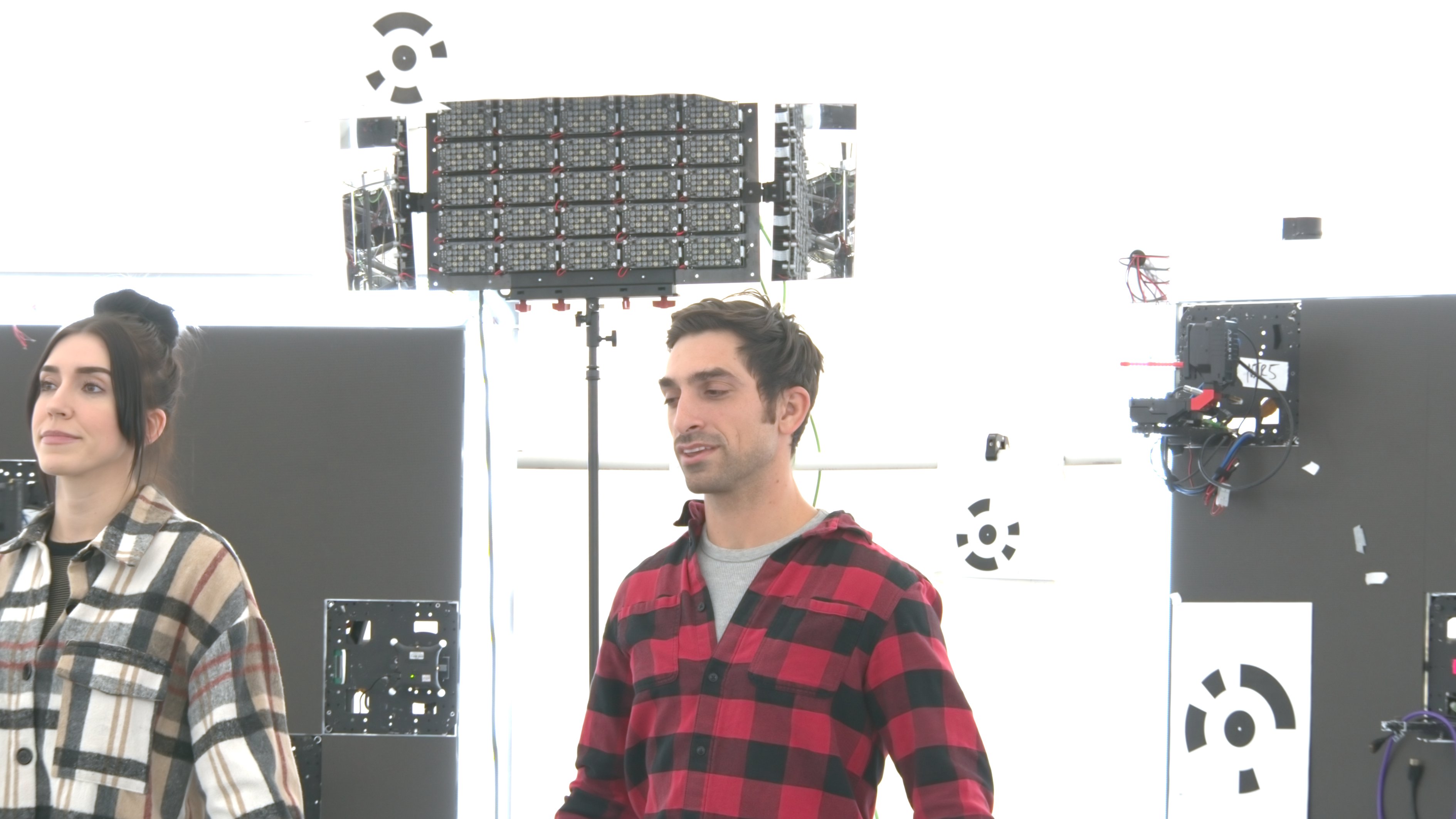}}
    \end{minipage}
    \hfill
    \begin{minipage}{0.159\linewidth}
    \adjincludegraphics[trim={0.32\width} {0.1\height} {0.38\width} {0.1\height}, clip, width=1.0\linewidth]{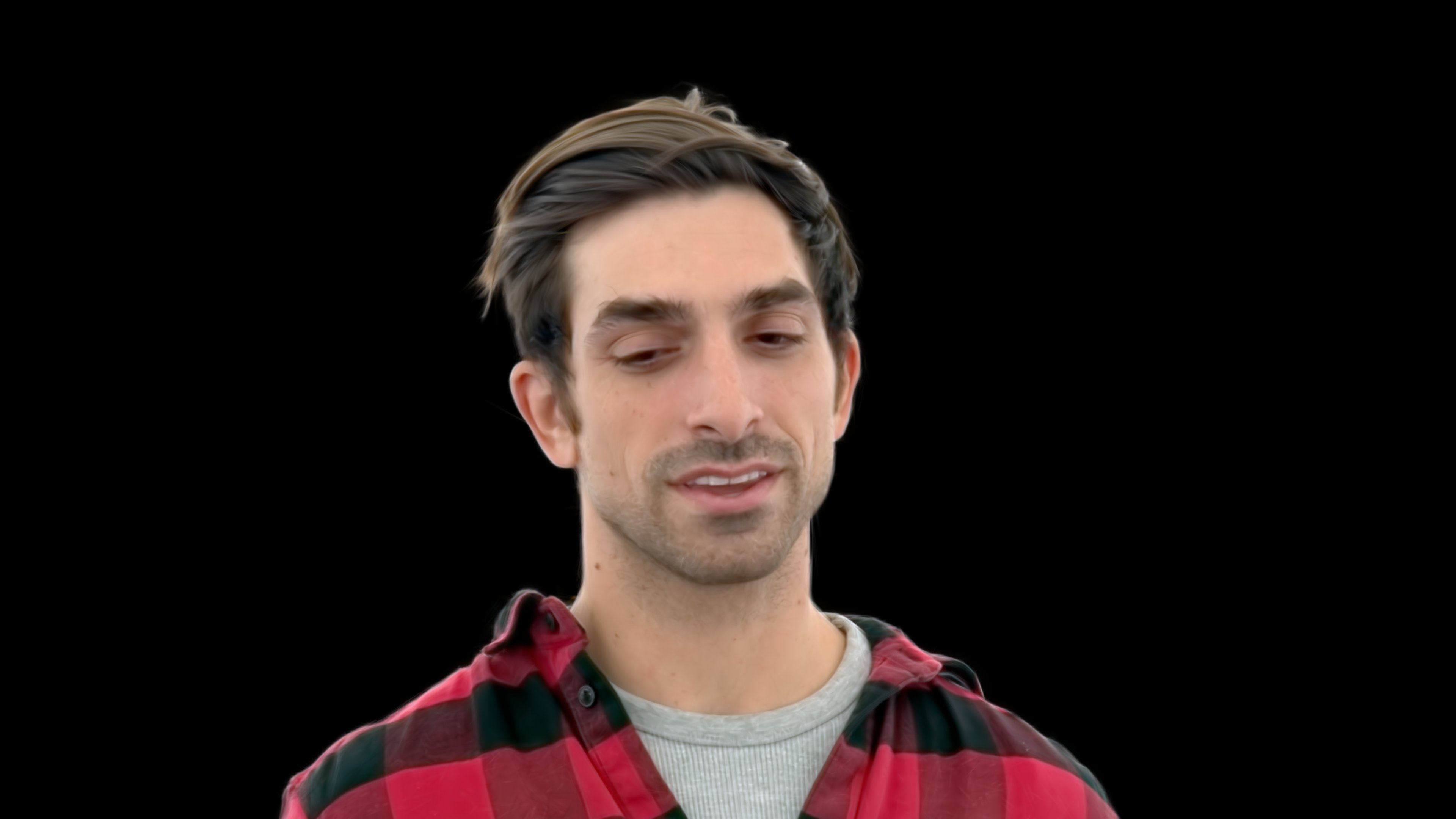}
    \end{minipage}
    \hspace{-1.3mm}
    \begin{minipage}{0.21\linewidth}
    \adjincludegraphics[trim= {0.38\width} {0.53\height} {0.52\width} {0.37\height}, clip, width=1.0\linewidth]{images/teaserv2/cam0000_1005.jpg}
    \adjincludegraphics[trim= {0.45\width} {0.37\height} {0.45\width} {0.53\height}, clip, width=1.0\linewidth]{images/teaserv2/cam0000_1005.jpg}
    \end{minipage}
    \hfill
    \begin{minipage}{0.159\linewidth}
    \centering
    \adjincludegraphics[trim={0.32\width} {0.1\height} {0.38\width} {0.1\height}, clip, width=1.0\linewidth]{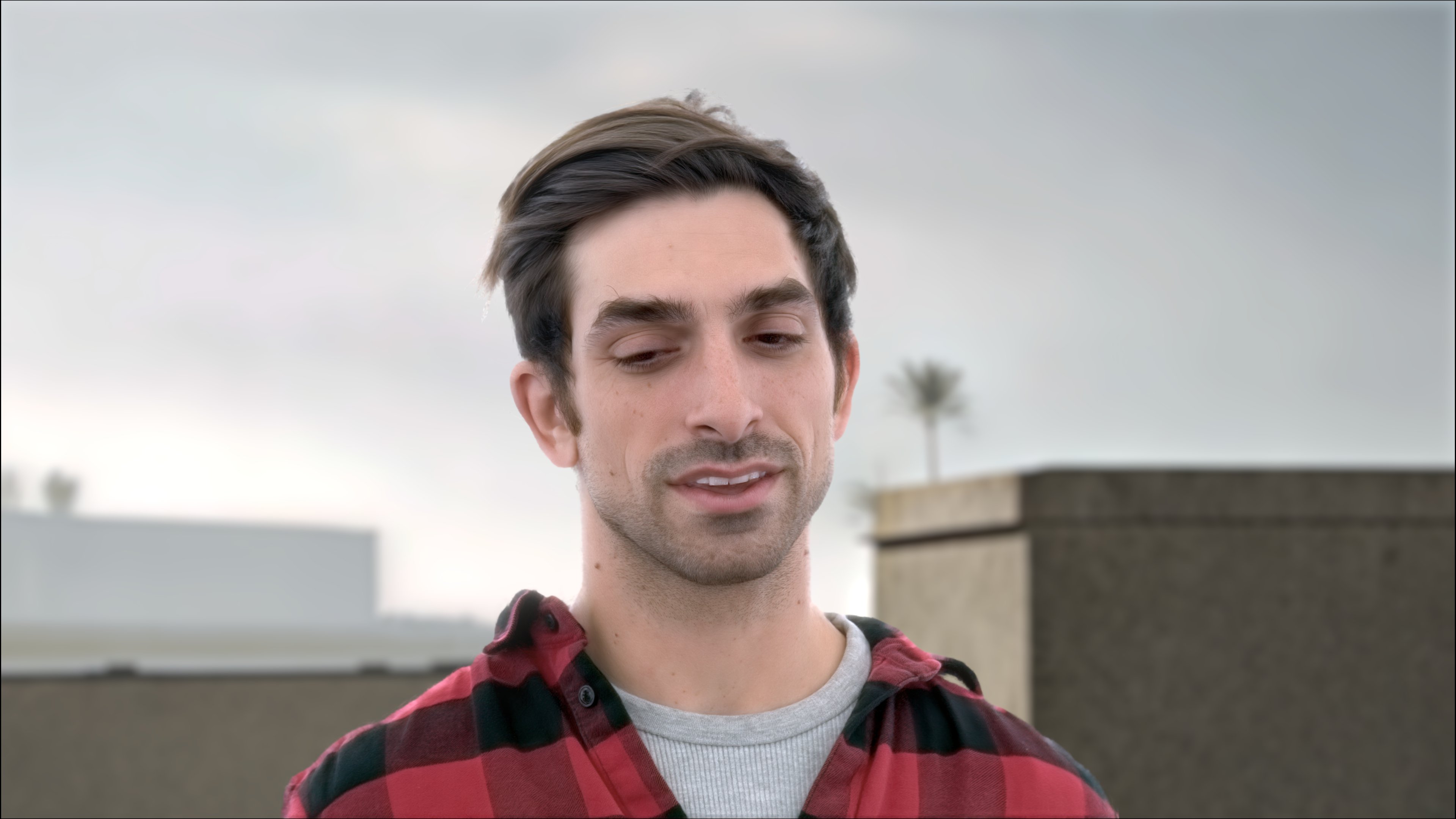}
    \end{minipage}
    \hspace{-1.3mm}
    \begin{minipage}{0.21\linewidth}
    \centering
    \adjincludegraphics[trim= {0.38\width} {0.53\height} {0.52\width} {0.37\height}, clip, width=1.0\linewidth]{images/teaserv2/papers_1841_teaser_AS_brighter.jpg}
    \adjincludegraphics[trim={0.45\width} {0.37\height} {0.45\width} {0.53\height}, clip, width=1.0\linewidth]{images/teaserv2/papers_1841_teaser_AS_brighter.jpg}
    \end{minipage}
\caption{From images taken in a dynamic multi-view capture stage (left), we reconstruct 4D Gaussian Splats tailored for production rendering (middle). The renderings are  refined with a detail enhancement diffusion model before compositing (right). \SA{Insets show the extent of our detail enhancement.} The enhancement model is trained on data of the actors acquired in a smaller facial capture stage.}
\end{teaserfigure}

\maketitle

\raggedbottom

\pagebreak
\section{Introduction}

4D volumetric performance capture systems are being leveraged with increasing frequency in media production, including for film and television where 4K resolution output is a requirement.
Film and TV applications also introduce the need to capture the interaction of multiple actors over an extended area, and to produce recordings which appear high-resolution in wide, medium, and closeup shots.  
Placing the cameras around a larger performance area -- ours is 6m $\times$ 9m -- increases their distance to the subjects, which makes it harder to capture high-resolution details of the dynamic performances.

In this paper, we present a novel volumetric recording, \SA{reconstruction, and detail enhancement} pipeline designed to address these challenges.
Our approach leverages two complementary physical capture rigs built at different scales.
The \scenerig is designed for multi-view, multi-actor performance capture, enabling high-quality reconstuctions, but not enough to render production-quality facial closeups.
The \facerig records the head of each actor with production-quality resolution for closeups, but cannot capture full-body performances.  


We first reconstruct performances of actors captured by the \scenerig using a novel Gaussian Splatting-based approach optimized for this capture setup.
This approach integrates a temporally stable camera calibration method and \SA{an HDR-aware} 4D Gaussian Splatting method, \SA{which accounts for practical issues.
Indeed,} our rendering pipeline incorporates carefully designed components and training strategies optimized for color, exposure, and black levels, ensuring enhanced color fidelity that meets production needs.

Next, to bridge the quality gap between the \scenerig captures and production resolution (especially for close-ups), we introduce a detail enhancement \SA{Diffusion Model}.
\SA{We modify a pre-trained image generation diffusion model, to support conditioning, to be temporally stable and to jointly generate RGB and Alpha channels.}
\SA{We fine-tune this model on high-fidelity \facerig data of the actors who performed in the \scenerig.}
\SA{Specifically, we use paired RGBA sequences of low-quality and high-quality renderings, obtained from pairs of low- and high-quality Dynamic Gaussian Splatting models}.
\SA{We limit the Gaussian count in the low-quality models to mimic the \scenerig's quality, with the high-quality renderings serving as ground truth}.
We demonstrate our method on several sequences of \SA{three sub-groups of} actors showing various novel camera paths, including facial close-ups which significantly exceed the quality of the original \scenerig capture.
We demonstrate the importance of our system components through a set of baseline and ablation comparisons.

To summarize, the main contributions of this work are as follows:  
\begin{itemize}
    \item A two-stage approach to performance capture, combining a scene-scale capture rig and a single-actor facial capture rig.
    \item A novel high-quality scene-scale volumetric performance capture rig, incorporating both static and dynamic cameras to track the performance of multiple actors.
    \item A reconstruction pipeline for dynamic performance capture, featuring stable calibration of moving cameras and \SA{4DGS with improved dynamic range and color fidelity.}
    \item A detail enhancement \SA{Diffusion Model, which supports 4K, RGB and Alpha and with improved temporal stability}.
\end{itemize}


\section{Related Work}
\subsection{Volumetric Data Capture and Reconstruction}
\SA{Rendering photorealistic, view-controllable human performances from volumetric capture remains an active research area. Pioneering works focus on reconstructing 3D meshes from multi-view setups, addressing facial performance~\cite{Guenter:MF:1998,DBLP:journals/cgf/FyffeHWMD11,DBLP:journals/tog/BeelerHBBBGSG11} and full-body geometry~\cite{Kanade:1997:VR,DBLP:conf/cvpr/AhmedTRTS08,DBLP:journals/tog/VlasicPBDPRM09,DBLP:conf/eccv/CagniartBI10,DBLP:journals/tog/AguiarSTAST08,DBLP:journals/tog/VlasicBMP08}.} Some methods rely on template priors such as shape-from-silhouettes~\cite{DBLP:conf/cvpr/AhmedTRTS08,DBLP:journals/tog/VlasicBMP08}, or track and estimate the performer’s deforming geometry \SA{using} canonical or reference geometry~\cite{DBLP:journals/tog/AguiarSTAST08,DBLP:journals/tog/VlasicPBDPRM09,DBLP:conf/eccv/CagniartBI10,DBLP:journals/tog/BeelerHBBBGSG11}. Others leverage various illumination patterns to capture both geometry and reflectance information~\cite{Einarsson:RHL:2006,DBLP:journals/tog/GuoLDBYWHOPDTTK19,DBLP:journals/cgf/FyffeHWMD11}. However, these approaches either rely on parameterized templates or fail to capture detailed geometry and appearance.

\SA{To reconstruct more details from multi-view videos, subsequent works \SA{propose using} IR video cameras~\cite{DBLP:journals/tog/ColletCSGECHKS15,DBLP:journals/tog/DouDFKKRTI17} or custom high-resolution depth sensors~\cite{DBLP:journals/tog/GuoLDBYWHOPDTTK19} to capture depth as geometry guidance, or incorporate custom color LED lights~\cite{DBLP:journals/tog/GuoLDBYWHOPDTTK19}}
Unfortunately, these mesh-based methods still struggle to reconstruct high-frequency details such as hair details, leading to a lack of photorealism.


With advances in neural rendering, learning-based approaches~\cite{DBLP:journals/tog/HedmanPPFDB18,DBLP:journals/tog/XuBSHSR19,DBLP:journals/tog/LombardiSSSLS19,KPLD21,ruckert2022adop} have been proposed for novel view synthesis \SA{using multi-view capture, but are limited to static scenes.}.
To address dynamic objects,\SA{~\citet{DBLP:journals/tog/LombardiSSSLS19} introduce} a real-time capture and rendering system with a deep architecture.
Recent methods ~\cite{DBLP:journals/tog/MekaPHOBDEZ0BLM20,DBLP:journals/tog/ZhaoJYZWDZZWXY22} combine traditional geometric pipelines with neural rendering, integrating volumetric and primitive-based rendering~\cite{DBLP:journals/tog/LombardiSSZSS21}, or hybrid scene representations~\cite{DBLP:journals/corr/abs-2310-08585,DBLP:conf/cvpr/PengYSBZ23,DBLP:conf/cvpr/Fridovich-KeilM23,DBLP:conf/cvpr/0008PLHSSBZ24}.


Among the new scene representations, neural radiance fields (NeRFs)~\cite{mildenhall2020nerf} have played a pivotal role in achieving high-quality novel view synthesis, \SA{with further improvements from InstantNGP\cite{mueller2022instant}, Mip-NeRF~\cite{barron2021mipnerf}, Point-NeRF~\cite{xu2022point}, Zip-NeRF~\cite{barron2023zipnerf}, Deblur-NeRF~\cite{ma2021deblur}, and RawNeRF~\cite{mildenhall2022rawnerf}}. \SA{NeRFs are also used for performance synthesis}~\cite{DBLP:conf/iccv/ParkSBBGSM21,DBLP:journals/tog/ParkSHBBGMS21,DBLP:conf/iccv/DuZYT021,DBLP:journals/tog/ZhaoJYZWDZZWXY22,DBLP:journals/tog/IsikRGKSAN23}.
Some approaches learn a canonical static template with deformable fields~\cite{DBLP:conf/iccv/ParkSBBGSM21,DBLP:journals/tog/ParkSHBBGMS21,TiNeuVox}, \SA{showing promise for short videos but struggling with long sequences and complex motion. Others use 4D NeRFs}~\cite{DBLP:conf/iccv/DuZYT021,DBLP:journals/tog/IsikRGKSAN23}, \SA{incorporating timestamps with spatial location and view direction to model spatio-temporal changes}.


\SA{Recently, 3D Gaussian Splatting (3DGS)~\cite{kerbl3Dgaussians} emerged achieving a significant breakthrough in both rendering speed and quality.}
\SA{Further innovations followed, including Cinematic Gaussians~\cite{wang2024cinematic} that enhances the framework by introducing HDR rendering capabilities and depth awareness, and 4D Gaussian Splatting (4DGS)~\cite{Wu_2024_CVPR,4drotor} which extends to dynamic scenes by incorporating temporal consistency.} 
\SA{Recently, 4DGS has been improved, enhancing rendering quality \cite{SpacetimeGS}, reducing the number of input views \cite{SplatFields}, enabling streaming \cite{3DGStream} and supporting longer sequences \cite{DBLP:conf/siggrapha/HeCTMPXRYBYD24,DBLP:journals/tog/XuXYPSBZ24,SWinGS}. These methods are constrained to synchronized static camera arrays and fail to meet key production requirements like linearity and 4K resolution.}
\SA{More aligned with our approach, \SA{\citet{Shen2024SuperGaussian}} uses diffusion-based video upsampling during training to enhance reconstruction quality, but is limited to $256 \times 256$ output.
In contrast, we customize 4DGS with a new calibration strategy to handle dynamic cameras, HDR and capture artifacts. Then we apply detail enhancement as a post-process to add 4K details.}

\begin{figure*}[h!tb]
    \centering
    \includegraphics[width=0.99\linewidth]{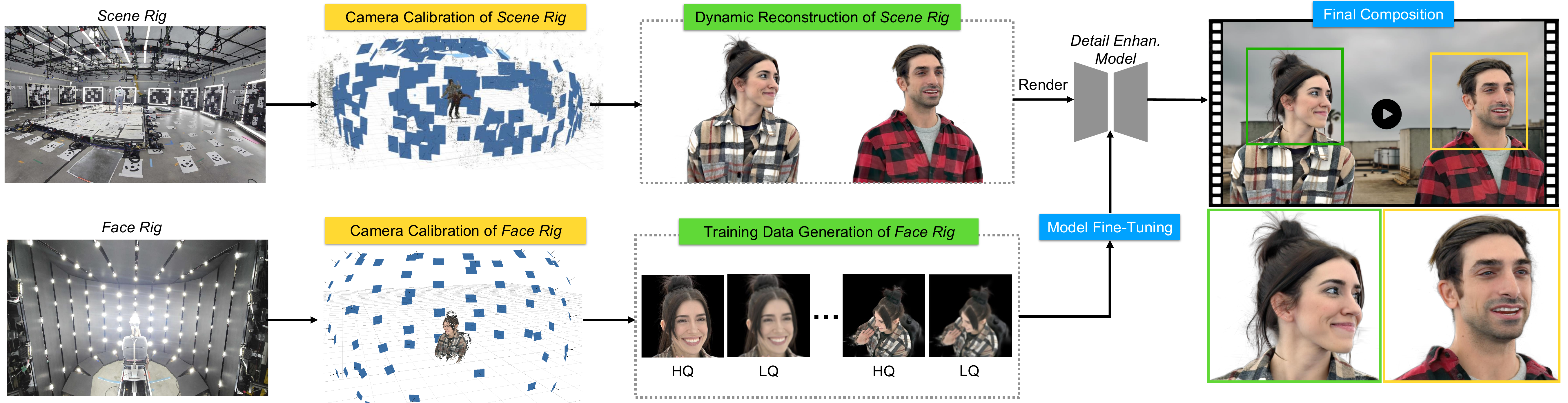} 
    \caption{
        Pipeline Overview. Actors perform in the \scenerig, where full-body performances are captured. Using our Poly4DGS framework, we reconstruct the \textbf{Performance}.
        The same actors are then captured in the \facerig. We generate Poly4DGS models for a portion of their facial performance: a high-quality model (\textbf{HQGS}, 4M Gaussians) and a low-quality model (\textbf{LQGS}, 50K-200K Gaussians). These reconstructions are used to train an \textbf{Image Enhancement Module} which refines the renderings of the low-quality GS to be like the high quality one. Finally, the trained model is used to enhance renderings from the 4DGS performance. Please refer to our supplementary video for the final composition of 4K render results.
    }
    \label{fig:overview}
\end{figure*}

\subsection{Image and Video Enhancement}

Image and video enhancement can be achieved through super-resolution. Early video super-resolution \SA{approaches include} recurrent-based~\cite{huang2017video, sajjadi2018frame, haris2019recurrent, liang2022recurrent, shi2022rethinking} and sliding-window-based~\cite{caballero2017real, yi2019progressive, li2020mucan, xu2021temporal, liang2024vrt} methods.
Although both enhance their input, they tend to produce artifacts with real-world videos. \SA{RealBasicVSR~\cite{chan2021basicvsr, chan2022investigating} appends a pre-cleaning stage for artifact removal.
RealViformer~\cite{zhang2024realviformer} uses covariance-based re-scaling for details. But they still face over-smoothing and temporal inconsistency.}
\nopagebreak
With the emergence of Diffusion models, recent efforts integrate diffusion priors \SA{with super-resolution}~\cite{wang2024exploiting, lin2023diffbir, yang2023pixel, wu2024seesr, zhao2024wavelet}.
StableSR~\cite{wang2024exploiting} couples a time-aware encoder and warping module with Stable Diffusion, while DiffBIR~\cite{lin2023diffbir} unifies generation and restoration via ControlNet.
\SA{Some methods~\cite{yang2023pixel, wu2024seesr}} infuse semantic cues to refine details.
Others~\cite{zhou2024upscale, yang2023mgldvsr, yuan2024inflation, chen2024learning} augment text-to-image backbones~\cite{rombach2022high, ho2022imagen} with temporal layers or adapters, yet enforcing consistent details over long sequences remains difficult.


Text-to-video (T2V) priors also inspire \SA{video super-resolution}.~\citet{xie2025star} propose a spatial-temporal augmentation strategy, a \SA{local information enhancement module, and a dynamic frequency loss for artifact correction. Several works~\cite{he2024venhancer,wang2023lavie} apply T2V to AI-generated videos.} Upscale-A-Video~\cite{zhou2024upscale} relies on global latent propagation for longer sequences, while \SA{other methods~\cite{chen2024learning,yuan2024inflation}} embed U-Nets within a VAE pipeline. Despite gains in fidelity, artifact handling and long-range temporal smoothness remain open problems.


\SA{
A line of work focuses on mitigating reconstruction artifacts in NeRFs or 3DGS by training restoration models.
\citet{NeRFLix} introduce a dedicated degradation pipeline to simulate NeRF-specific artifacts.
\citet{RoGUENeRF} pretrain on a small NeRF dataset and fine-tune per-scene to improve view consistency.
\citet{GANeRF} train a per-scene discriminator to enhance reconstructions.
\FI{However, these models primarily refine input signals and can't generate new details beyond the input images which we require.}
\citet{wu2025difix3d} trains a diffusion model on large-scale 3D reconstructions for generalizable artifact removal. \FI{Because of the generality of this approach, the added details are not based on specific subject appearance, but on general priors.}

}


For production-quality \SA{high-resolution} generation, \SA{tuning-based methods}~\cite{teng2023relay, hoogeboom2023simple, ren2024ultrapixel, liu2024linfusion, guo2024make, zheng2024any} fine-tune large models on scarce high-resolution data, while \SA{tuning-free pipelines}~\cite{he2023scalecrafter, du2024demofusion, haji2024elasticdiffusion, lin2024cutdiffusion, lee2023syncdiffusion, jin2023training, hwang2024upsample, cao2024ap} perform patch-wise inference. \SA{Some methods mitigate block artifacts~\cite{bar2023multidiffusion, du2024demofusion}, expand receptive fields~\cite{he2023scalecrafter}, and manipulate latents across scales~\cite{huang2024fouriscale}}. Pivoting strategies~\cite{guo2025make, qiu2024freescale} fuse stable \SA{low-resolution} semantics with multi-scale upsamplers, but balancing local detail with global consistency \SA{remains a} challenge in super-resolution. \SA{Moreover, to the best of our knowledge, no video-based super-resolution method supports 4K. Instead, we modify an image-based model \cite{flux} with a frame warping scheme and low frequency swapping for temporal stability}.

\section{Overview}
The workflow of our method, from volumetric data capture to dynamic scene reconstruction and enhancement, is illustrated in Fig.~\ref{fig:overview}.
We use both a \scenerig for multi-actor dynamic scenes and a \facerig to capturing high-resolution facial details.
First, the \scenerig captures a multi-view performance, allowing actors to move freely. We adopt a \SA{variation of the} 4DGS method, \emph{Poly4DGS}, to reconstruct the dynamic scene with globally and temporally consistent structure but limited facial details.

Next, the \facerig captures high-resolution facial data of the same actors within a small volume.
This data is used to train a diffusion-based detail enhancement model, \SA{which we use to enhance the \scenerig renderings}.

\section{Dynamic Reconstruction and Rendering}
\label{sec:recon}
Our first stage aims at reconstructing and rendering actors' free-form performances using 4DGS.
The quality and temporal consistency of 4DGS is highly correlated with the quality of the input data.
Unfortunately, capturing fast-moving actors in a multi-view setup is challenging due to motion blur, defocus, and actors moving out of frame.
To tackle these challenges, we use a \scenerig combining static and dynamically aimed cameras to record the actors in the highest resolution possible.
Given the acquired sequences, we introduce a new camera calibration method that leverages a clean background to ensure accurate and consistent camera parameters.
\SA{We then reconstruct the performance using \emph{Poly4DGS} allowing to render dynamic scenes from arbitrary viewpoints.}

\subsection{Hardware Setup of the \scenerig}

The \scenerig features 180 synchronized
4K Z-CAM e2 cinema cameras
30 of these cameras are mounted on the ceiling and 25 cameras are placed on the floor, providing top-down and bottom-up perspectives to minimize occlusions.

The remaining cameras are mounted on 12 mobile carts arranged circularly around the \SA{$6m \times 8m$} stage, allowing our system to adapt to different configurations.
The cameras on each cart follow a structured zig-zag pattern, with slight vertical offsets between adjacent units.
Calibration patterns are placed throughout the room and stage to ease camera alignment.

Our stage features two types of cameras: static wide field-of-view (FoV) cameras and tracking cameras with instrumented pan, tilt, zoom, and focus.
The 90 static cameras are equipped with 14-42mm lenses, providing broad coverage of the entire stage.
Additionally, 40 landscape tracking cameras with 45-175mm zoom lenses dynamically adjust the camera pan, tilt, zoom, and focus to track the actors' faces. 
\SA{Tracking is performed using unobtrusive OptiTrack\textsuperscript{\texttrademark} markers placed on the clothing below the back of the neck of actors.}
Finally, 50 additional portrait orientation cameras track and frame full-body, upper-body, and lower-body movements.

The stage is uniformly lit by white LED strips mounted on the carts and ceiling, ensuring consistent illumination across the stage. 
The LED strips emit 1 ms flashes in sync with the 24 fps camera shutters to reduce motion blur, strobing at 72 Hz to exceed the flicker fusion frequency for actor comfort.

\subsection{Volumetric Data Capture and Processing}

\SA{We direct three groups of actors} to perform diverse \SA{actions, including running, walking, talking, and playing basketball}. 
The footage is recorded at 4K, 24 fps. We also capture a color chart to calibrate color between the \scenerig and the \facerig.


\paragraph{Camera alignment and calibration.}
Camera alignment of our volumetric capture system is challenging due to dynamic cameras with varying zoom levels.
The shallow depth of field in zoomed-in views blurs the background, making these views difficult to calibrate.
Additionally, focal length changes can be confused with camera movement, leading to ambiguities between intrinsics and extrinsics.
Moving cameras further complicate the process, requiring frame-by-frame re-calibration.
Without proper constraints, this can lead to alignment inconsistencies, such as drift and stutter.

To align both static and dynamic cameras, we use a two-step process.
First, we calibrate only the static wide-FoV cameras, using lidar scans and the middle frame of the sequence while keeping their focal lengths fixed.  
Once calibrated, we fix the static cameras' position and then calibrate the tracking cameras separately.
We ensure that zoom changes are properly accounted for rather than freely estimated per frame.
To do so, we fit a smooth function to the changes in focal length provided by the cameras and use it as a regularization across frames, preventing sudden jumps or inconsistencies.
We find that by doing so, the number of identified points in the initial point cloud increases by nearly 50\%.



\subsection{Dynamic Scene Reconstruction Using \emph{Poly4DGS}}


Given multi-view video data as input, we design the dynamic scene reconstruction method, \emph{Poly4DGS}, based on advanced 3DGS~\cite{kerbl3Dgaussians} and 4DGS~\cite{yang2023gs4d,4drotor} techniques. We also incorporate new training strategies, including color space adjustment, exposure, and black-level optimization, to maximize reconstruction quality while preserving temporal consistency.

\subsubsection{\emph{Poly4DGS}}
\hfill\\
We represent dynamic 3D scenes using a set of 4D primitives. Unlike previous approaches that parameterize each 4D primitive as Gaussian functions in 4D space using 4D-Rotor~\cite{4drotor} or dual-quaternion~\cite{yang2023gs4d}, we adopt a simplified representation \SA{that is easy to implement based on existing 3DGS rasterizer \cite{ye2025gsplat}}. Our method directly parametrizes the motion of each Gaussian as polynomial expansions over time for each property. Formally, we define:
\begin{align*}
    \mu_t &= \Sigma_{i=0}^{n_\mu} \mu_i (t - t_0)^i \text{,} \quad
    \mathbf{q}_t = \Sigma_{i=0}^{n_\mathbf{q}} \mathbf{q}_i (t - t_0)^i \text{,}\\
    \mathbf{s}_t &= \Sigma_{i=0}^{n_\mathbf{s}} \mathbf{s}_i (t-t_0)^i \text{,} \quad
    \mathbf{o}_t = {\mathbf{o}_0} e^{-\frac{1}{2} \Sigma_{i=1}^{n_\mathbf{o}}\lambda_i (t - t_0)^{2i} },
\end{align*}

where $\mu_t, \mathbf{q}_t$, $\mathbf{s}_t$, and $\mathbf{o}_t$ represent the mean, quaternion rotation, scaling, and opacity of a 3D Gaussian at timestamp $t$, respectively.
Intuitively, the parameter $t_0$ denotes the temporal center of the Gaussian, and $\lambda_i$ controls the length of the Gaussian lifespan.
Notably, \citet{4drotor} have shown that when $n_\mu=1, n_\mathbf{q}=0, n_\mathbf{s}=0$ and $n_\mathbf{o}=1$, this parameterization corresponds to slicing a 4D Gaussian at time $t$.
By introducing higher-order terms for different properties, our approach allows more flexibility, thus enabling each primitive to capture more complex motions. Empirically, we find $n_\mu = 2$, $n_\mathbf{q} = 1$, $n_\mathbf{s}=0$, and $n_\mathbf{o}=2$ result in \SA{a slight improvement in PSNR (+0.2dB) without introducing visible overhead.}
Since performance sequences exhibit minimal texture changes, we use time-constant spherical harmonics.
We use the standard Gaussian splatting rendering \cite{kerbl3Dgaussians} with antialiasing \cite{Yu2023MipSplatting} to rasterize the 3D Gaussians at time $t$ and we use ~\citet{kheradmand20243d} to handle pruning and relocation.

With \emph{Poly4DGS}, the number of primitives required to represent a 4D sequence with comparable quality scales approximately \SA{linearly} with the length of the sequence.
To balance representation quality and computational feasibility, we divide the whole sequence into smaller segments and train each segment independently.
The segmentation ensures high-quality results but also enhances the scalability, as the training of segments can be parallelized efficiently on GPU clusters. 

We initialize 4D primitives from a sequence of dense point clouds reconstructed per frame.
We observe that using dense point clouds of approximately 200k points leads to much faster convergence compared to sparse point clouds.

\subsubsection{Training Color Space}  
\hfill\\
We aim to train our 4DGS with linear OpenEXR files to maintain the input's dynamic range while ensuring proper control over the training color space.
The direct solution of training Gaussian Splatting in linear color space results in poorer outcomes, \SA{typically we observed a 5dB decrease in PSNR as detailed in supplemental materials.
We hypothesize this is due to a conditioning issue in the optimization process, similar in spirit to findings for camera optimization and {\em floaters} \cite{park2023camp,10.2312:sr.20231122}}.  
To address this, we propose storing \SA{color parameters in an unbounded tone-mapped space, but rasterizing using linearized colors.
We experimented with several color spaces such as $log(1+x)$ and $sRGB$ without clamping, and found they all performed equally. To maintain a single tone-mapping and inverse operator we used the unbounded $sRGB$ color space.}
Specifically, the Gaussian colors, \( c_\text{G} \), obtained after evaluating the spherical harmonics, are transformed to linear space using the inverse sRGB mapping \( f_{\text{srgb}}^{-1} \), before being rasterized:  
\begin{equation}
c_{\text{linear}} = f_{\text{srgb}}^{-1}(c_\text{G}) \quad \textrm{and} \quad
C_{\text{rast}} = \mathcal{R}(c_{\text{linear}}, \ldots),
\end{equation}
where \( \mathcal{R} \) represents the rasterization. \SA{\( f_{\text{srgb}}^{-1} \) operates in $\mathbb{R}^{+}$ without clamping.} We leave out other Gaussian attributes for simplicity.


\begin{figure}[!htbp]
    \centering
    \includegraphics[width=0.49\linewidth]{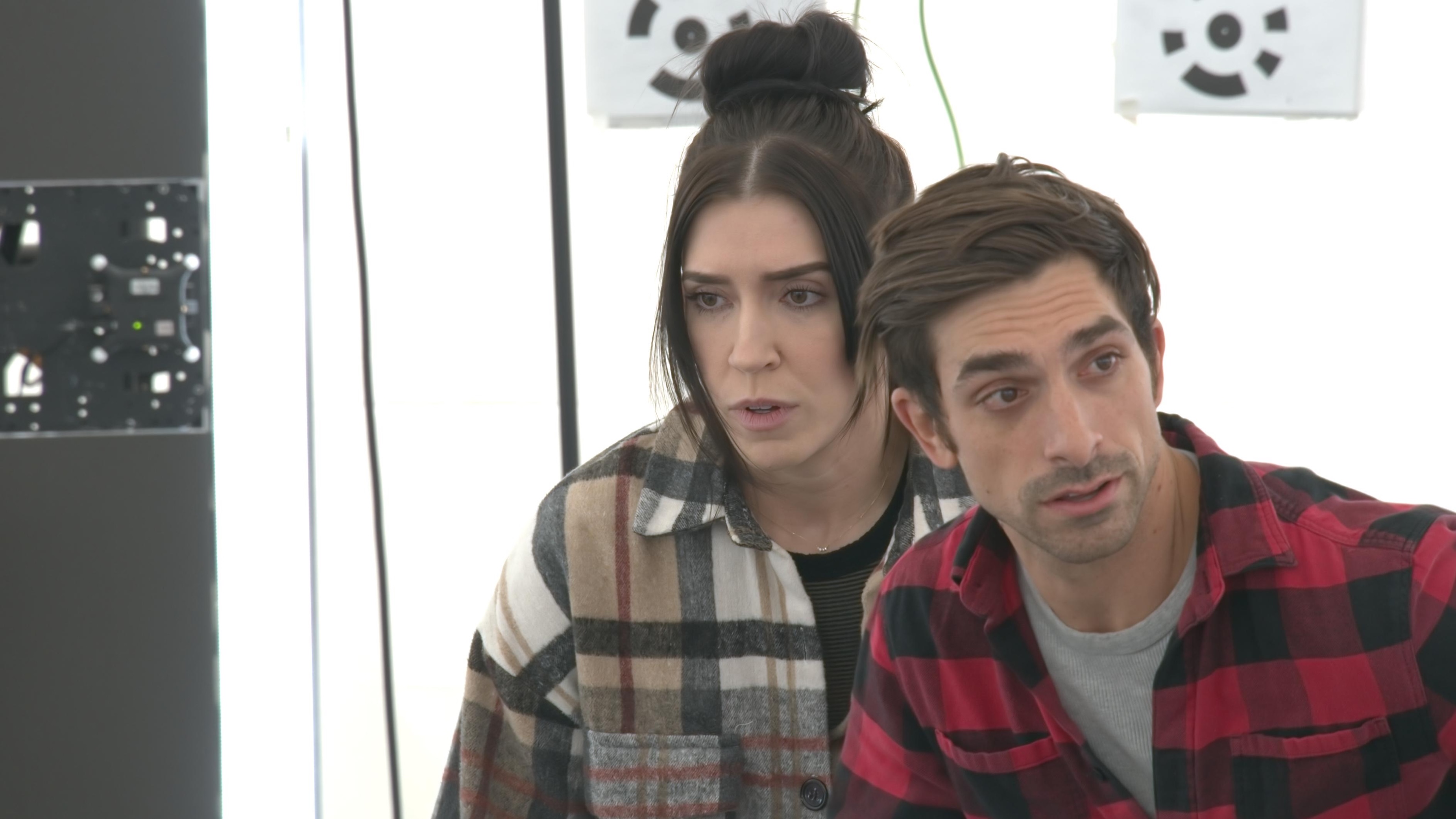}
    \hfill
    \includegraphics[width=0.49\linewidth]{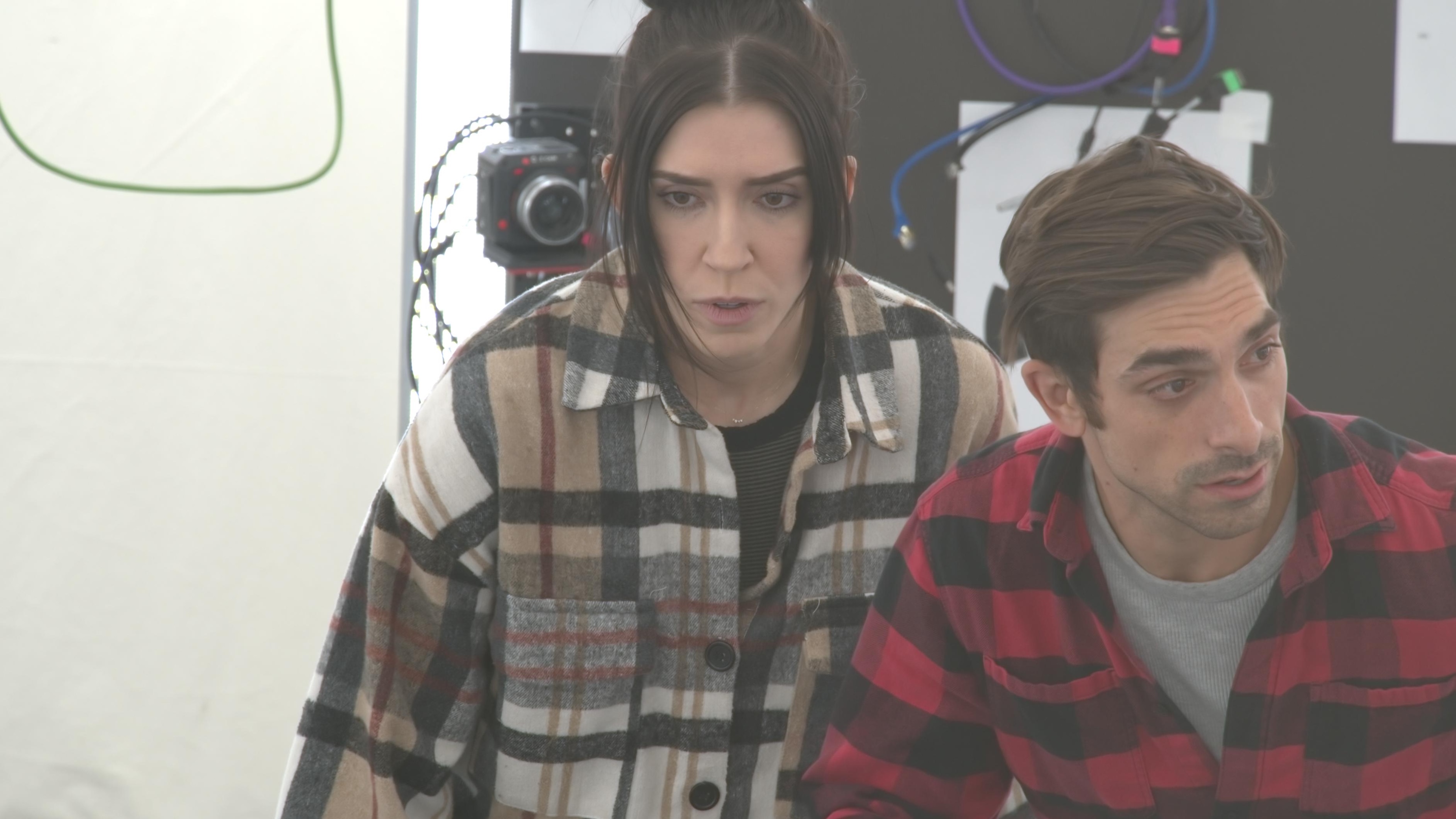}
    \hfill
    \makebox[0.49\linewidth]{\centering \small Camera 1}
    \makebox[0.49\linewidth]{\centering \small Camera 2}
    \caption{The \scenerig captures suffer from severe lens glare for some cameras.}
    \label{fig:lensglare}
\end{figure}

\subsubsection{Exposure and Black-Level Optimization}
\hfill\\
The zoom lens optics produce a noticeable amount of veiling glare and lens flares in our input frames, as visible in Fig.~\ref{fig:lensglare}. We propose to correct for this using \SA{$32 \times 32$} spatially-varying exposure and black-level adjustments \SA{grids}, also accounting for sensor variations:  
\begin{equation}
C_{\text{adjusted}} = (C_{\text{rast}} + B^c(x, y)) \cdot E^c(x, y),
\end{equation} 
where \( B^c(x, y) \) and \( E^c(x, y) \) indicate the black level and exposure at pixel position $(x,y)$, \FI{and $\cdot$ is the Hadamard product}. We \SA{upsample the grid, by applying a FFT to it, zero-padding to $C_{\text{rast}}$ size and applying the IFFT. The backward pass of this process is} orders of magnitude faster than the slicing operation used by bilateral grids~\cite{wang2024bilateral} at high resolution.  

The adjusted color can then be transformed to a display color such as sRGB; combining all the steps, the final output color \( C_{\text{final}} \) is:  
\begin{equation}
C_{\text{final}} = f_{\text{srgb}}(C_{\text{adjusted}}) =f_{\text{srgb}} \left( \left( \mathcal{R}\left( f_{\text{srgb}}^{-1}(c_{G}) \right)+ B^c \right) \cdot E^c \right).
\end{equation} 

\SA{More details about the tone-mapping, and the impact of the GS storage color space are provided in the supplemental document}


\subsubsection{Training Objective}
\hfill\\
Our training objective builds upon the standard 3DGS reconstruction loss, \( \mathcal{L}_{\text{recon}} \), which combines \( L_1 \) and \( L_\text{SSIM} \). The reconstruction loss \( \mathcal{L}_{\text{recon}} \) is computed in unbounded sRGB space. Additionally, we include the following regularization terms:  

\textbf{Exposure \SA{and black-level} regularization:}  
   \SA{We} encourage the exposure maps \( E^c(x, y) \) to have an average value of 1 across all spatial and camera dimensions, to ensure stability. We also encourage higher black-level offsets \( B^c(x, y) \) for the current frame, to reduce glare:  
   \begin{equation}
   \mathcal{L}_{\text{exposure}} = \left( \sum_{x, y, c} \frac{E^c(x, y) - 1}{N_x N_y N_c} \right)^2 \quad \mathcal{L}_{\text{black}} = -\frac{1}{N}  \sum_{pixels} B^c(x, y)
   \end{equation}  




Our final training objective is a weighted sum of all components:  
\begin{equation}
\mathcal{L}_{\text{total}} = \mathcal{L}_{\text{recon}} + \lambda_e \mathcal{L}_{\text{exposure}} + \lambda_b \mathcal{L}_{\text{black}} 
\end{equation}
where \( \lambda_e = 10 \), and \( \lambda_b = 0.05 \) are hyperparameters controlling the contributions of each regularization term.

\section{Detail Enhancement}
\label{sec:sr}
Although we aim to maximize the quality of novel-view synthesis in the first stage, the level of detail achieved remains limited by the large scale of the capture area, the finite number of cameras, and imperfections in lens optics.
To obtain production-level details, particularly for facial close-ups, we use a smaller-scale multi-view capture stage, the \facerig.
From the \facerig data, we generate a paired training dataset \SA{of sequences using low- and high-quality Dynamic GS reconstructions}.
This dataset is used to train a \SA{Detail Enhancement model finetuned from an Image Diffusion model. Its role is to remove Gaussian artifacts from lower-quality \scenerig renderings and generate fine details, enhancing visual photorealism.

To improve the model's temporal stability, we condition it on an Optical Flow reprojection of the preceding generated frame.
Finally, to help compositing, we modify the model architecture to jointly predict the detailed image and its corresponding alpha.}

\begin{figure*}
\centering
\begin{tabular}{c|c}

\resizebox{6.5cm}{!}{
\begin{tikzpicture}[node distance=1.6cm and 1.8cm, on grid, font=\small]


\node[draw, align=center, fill=gray!10, minimum width=1cm, minimum height=1cm,inner sep=0] (noise) at (0,0) {Latent\\Noise};

\node[draw, align=center, fill=red!20, rounded corners=5pt, minimum width=1.5cm, minimum height=1cm] (trans1) at (1.5,0) {Transformer\\ $\circlearrowright$ x N};

\node[draw, align=center, fill=gray!10, minimum width=1cm, minimum height=1cm] (latent2) at (3,0) {Output\\Latent};

\node[draw, trapezium, trapezium left angle=75, trapezium right angle=75, trapezium stretches body, fill=blue!20, minimum width=1.5cm, minimum height=1cm, shape border rotate=90] (dec1) at (4.25,0) {Decoder};

\node[align=center] (label) at (5.75,1) {Output Image};
\node[draw, align=center, fill=gray!10, minimum width=1.5cm, minimum height=1.5cm,inner sep=0] (output1) at (5.75,0) {\includegraphics[width=0.085\linewidth]{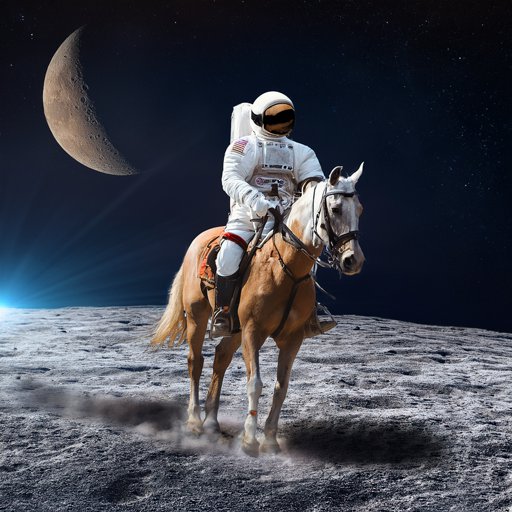}};

\draw[->] (noise) -- (trans1);
\draw[->] (trans1) -- (latent2);
\draw[->] (latent2) -- (dec1);
\draw[->] (dec1) -- (output1);

\end{tikzpicture}
}

&

\multirow{2}{*}[2.5em]{
\resizebox{11cm}{!}{
\begin{tikzpicture}[node distance=1.6cm and 1.8cm, on grid, font=\small]

\node[draw, trapezium, trapezium left angle=75, trapezium right angle=75, trapezium stretches body, fill=blue!20, minimum width=1.5cm, minimum height=1cm, shape border rotate=270] (enc1) at (2.0,0) {Encoder};

\node[align=center] (label3) at (-1,1.55) {Warped prev.};
\node[draw, align=center,inner sep=0] (input3) at (-1,0.75) {\includegraphics[width=0.07\linewidth]{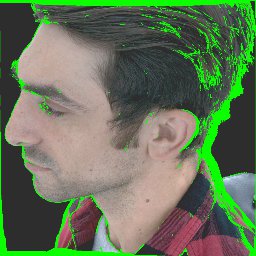}};

\node[align=center] (label2) at (-1,-1.6) {Warp mask};
\node[draw, align=center,inner sep=0] (input2) at (-1,-0.75) {\includegraphics[width=0.07\linewidth]{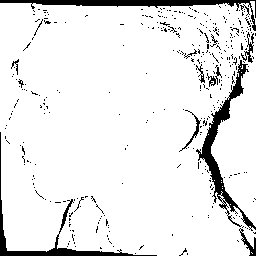}};

\draw[->] (input2) -- (enc1);
\draw[->] (input3) -- (enc1);

\node[align=center] (label1) at (0.5,1.55) {LQ RGB};
\node[draw, align=center,inner sep=0] (input1) at (0.5,0.75) {\includegraphics[width=0.07\linewidth]{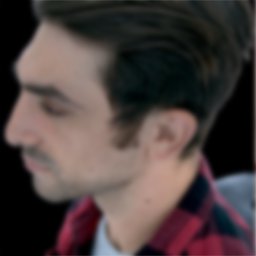}};

\node[align=center] (label0) at (0.5,-1.6) {LQ Alpha};
\node[draw, align=center,inner sep=0] (input0) at (0.5,-0.75) {\includegraphics[width=0.07\linewidth]{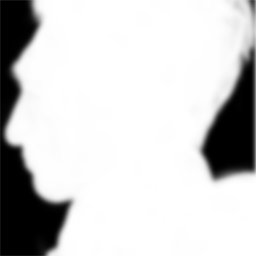}};

\draw[->] (input0) -- (enc1);
\draw[->] (input1) -- (enc1);

\node[draw, align=center, fill=gray!10, minimum width=1cm, minimum height=1cm] (latent3) at (3.6,0.075) {Cond.\\Latents};

\node[draw, align=center, fill=gray!10, minimum width=1cm, minimum height=1cm] (latent2) at (3.55,0.025) {Cond.\\Latents};

\node[draw, align=center, fill=gray!10, minimum width=1cm, minimum height=1cm] (latent1) at (3.5,-0.025) {Cond.\\Latents};

\node[draw, align=center, fill=gray!10, minimum width=1cm, minimum height=1cm] (latent0) at (3.45,-0.075) {Cond.\\Latents};

\node[draw, align=center, fill=gray!10, minimum width=1cm, minimum height=1cm,font = {\scriptsize}] (noise1) at (5.0,1.0) {Latent\\Noise\\RGB};

\node[draw, align=center, fill=gray!10, minimum width=1cm, minimum height=1cm,font = {\scriptsize}] (noise0) at (5.0,-1.0) {Latent\\Noise\\Alpha};

\node [draw,circle,minimum width=0.5 cm] (add) at (5.0,0){Cat};

\node[draw, align=center, fill=red!20, rounded corners=5pt, minimum width=1.5cm, minimum height=1cm] (trans1) at (6.6,0) {Transformer\\ $\circlearrowright$ x N};

\node[draw,circle,minimum width=0.5 cm] (split0) at (8.1,0) {Split};

\node[draw, align=center, fill=gray!10, minimum width=1cm, minimum height=1cm,font = {\scriptsize}] (latentout0) at (9.2,0.55) {Output\\Latent\\RGB};

\node[draw, align=center, fill=gray!10, minimum width=1cm, minimum height=1cm,font = {\scriptsize}] (latentout1) at (9.2,-0.55) {Output\\Latent\\Alpha};

\node[draw, trapezium, trapezium left angle=75, trapezium right angle=75, trapezium stretches body, fill=blue!20, minimum width=1.5cm, minimum height=1cm, shape border rotate=90] (dec1) at (10.55,0) {Decoder};

\node[align=center] (label4) at (12,-1.6) {DE Alpha};
\node[draw, align=center, inner sep=0, minimum height=1.25cm] (output0) at (12,-0.75) {\includegraphics[width=0.07\linewidth]{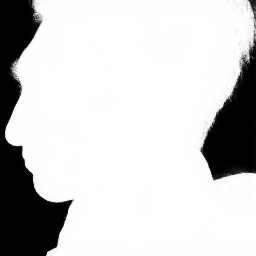}};

\node[align=center] (label5) at (12,1.55) {DE RGB};
\node[draw, align=center, inner sep=0, minimum height=1.25cm] (output1) at (12,0.75) {\includegraphics[width=0.07\linewidth]{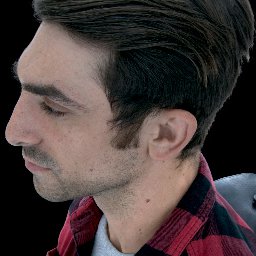}};

\draw[->] (enc1) -- (latent1);
\draw[-] (latent3.20) -- (add);
\draw[-] (latent2.10) -- (add);
\draw[-] (latent1.0) -- (add);
\draw[-] (latent0.-10) -- (add);
\draw[-] (noise0) -- (add);
\draw[-] (noise1) -- (add);
\draw[->] (add) -- (trans1);
\draw[->] (trans1) -- (split0);
\draw[->] (split0) -- (latentout0);
\draw[->] (split0) -- (latentout1);
\draw[->] (latentout0) -- (dec1);
\draw[->] (latentout1) -- (dec1);
\draw[->] (dec1) -- (output0);
\draw[->] (dec1) -- (output1);

\end{tikzpicture}
}
}

\\

\resizebox{6.5cm}{!}{
\begin{tikzpicture}[node distance=1.6cm and 1.8cm, on grid, font=\small]


\node[draw, align=center, fill=yellow!30, minimum size=1.5cm] (input2) at (0.1,-0.45) {Input\\Conditions};

\node[draw, align=center, fill=yellow!30, minimum size=1.5cm] (input1) at (0.05,-0.5) {Input\\Conditions};

\node[draw, align=center, fill=yellow!30, minimum size=1.5cm] (input0) at (0,-0.55) {Input\\Conditions};

\node[draw, trapezium, trapezium left angle=75, trapezium right angle=75, trapezium stretches body, fill=blue!20, minimum width=1.5cm, minimum height=1cm, shape border rotate=270] (enc1) at (1.625,-0.5) {Encoder};

\node[draw, align=center, fill=gray!10, minimum width=1cm, minimum height=1cm] (latent2) at (3.05,-0.45) {Cond.\\Latents};

\node[draw, align=center, fill=gray!10, minimum width=1cm, minimum height=1cm] (latent1) at (3,-0.5) {Cond.\\Latents};

\node[draw, align=center, fill=gray!10, minimum width=1cm, minimum height=1cm] (latent0) at (2.95,-0.55) {Cond.\\Latents};

\node[draw, align=center, fill=gray!10, minimum width=1cm, minimum height=1cm] (noise) at (3,0.6) {Latent\\Noise};

\node [draw,circle,minimum width=0.5 cm] (add) at (4.35,0){Concat};

\node[draw, align=center, fill=red!20, rounded corners=5pt, minimum width=1.5cm, minimum height=1cm] (trans1) at (6,0) {Transformer\\ $\circlearrowright$ x N};

\node[draw, align=center, fill=gray!10, minimum width=1cm, minimum height=1cm] (latentout) at (7.75,0) {Output\\Latent};

\node[draw, trapezium, trapezium left angle=75, trapezium right angle=75, trapezium stretches body, fill=blue!20, minimum width=1.5cm, minimum height=1cm, shape border rotate=90] (dec1) at (9.25,0) {Decoder};

\node[draw, align=center, fill=gray!10, minimum width=1.5cm, minimum height=1.5cm] (output1) at (11,0) {Output\\Image};

\draw[->] (input1) -- (enc1);
\draw[->] (enc1) -- (latent1);
\draw[-] (latent2.20) -- (add);
\draw[-] (latent1.10) -- (add);
\draw[-] (latent0.0) -- (add);
\draw[-] (noise) -- (add);
\draw[->] (add) -- (trans1);
\draw[->] (trans1) -- (latentout);
\draw[->] (latentout) -- (dec1);
\draw[->] (dec1) -- (output1);

\end{tikzpicture}
}

&

\end{tabular}
\caption{\SA{Architectural changes made to the base Flux model~\cite{flux}. Starting from the Latent Diffusion Architecture (top left), we add input channels to condition the network (bottom left). To improve temporal stability and generate an alpha channel, we condition our model on the previous warped output, a validity mask, and both the LQ RGB and Alpha. We also double the size of the latent space to predict RGB and Alpha jointly (right).}}
\label{fig:DE_arch}
\end{figure*}
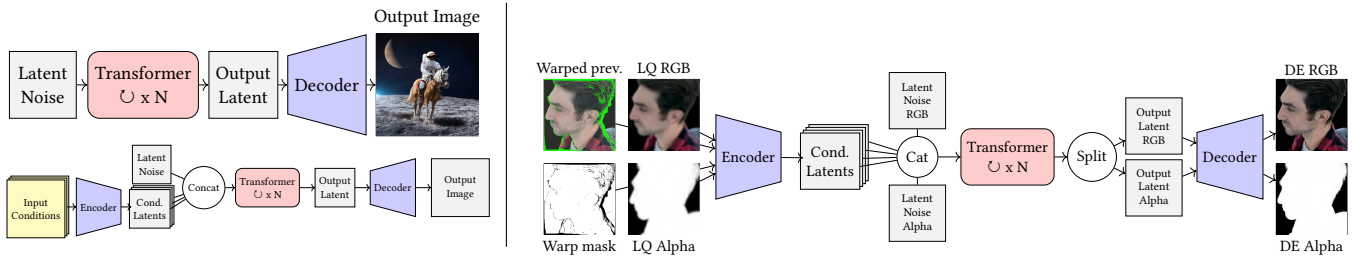

\begin{figure}[!htbp]
    \begin{minipage}{0.21\linewidth}
    \centering
    \includegraphics[trim=0 0 0 20pc, clip, width=1.0\linewidth]{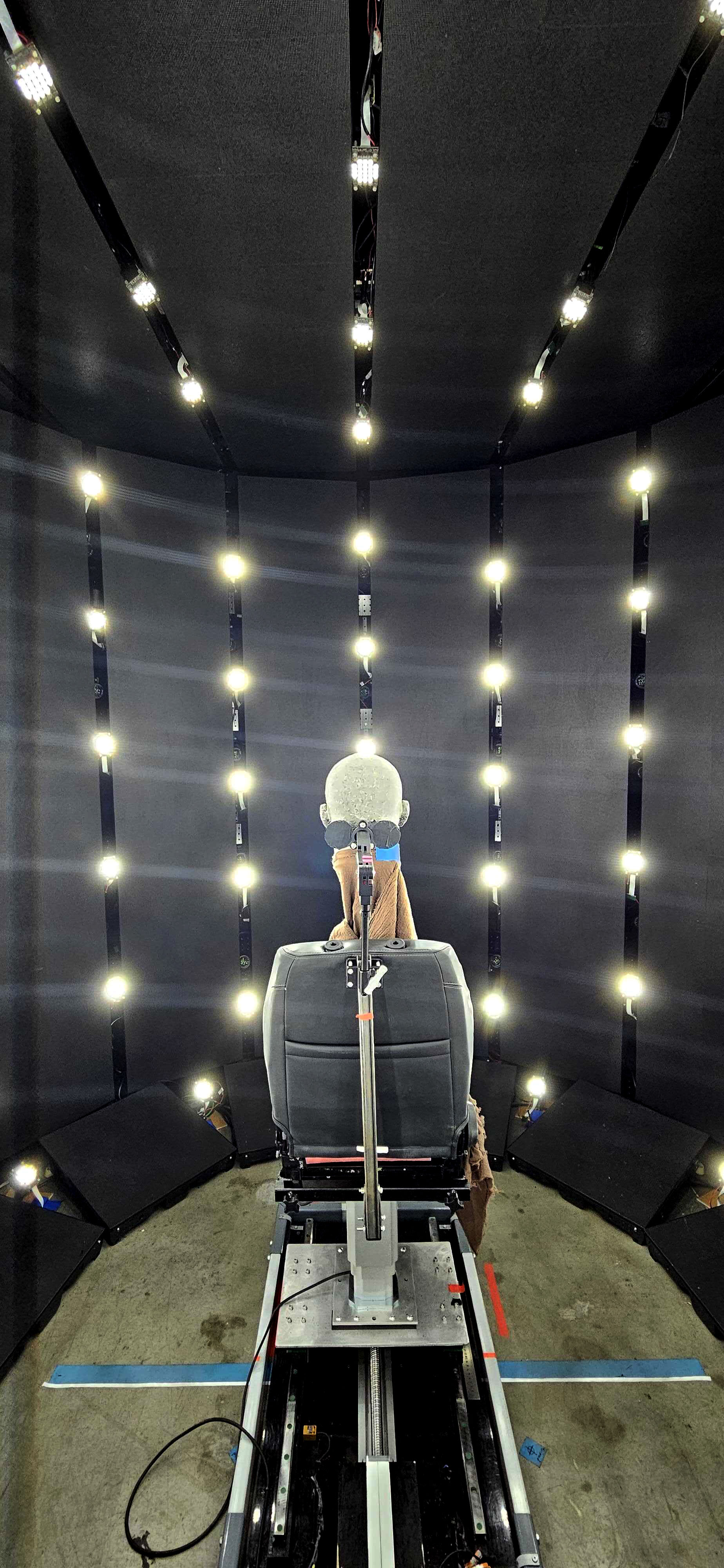}
    \end{minipage}
    \begin{minipage}{0.275\linewidth}
    \centering
    \includegraphics[trim=0 40pc 0 0, clip, width=1.0\linewidth]{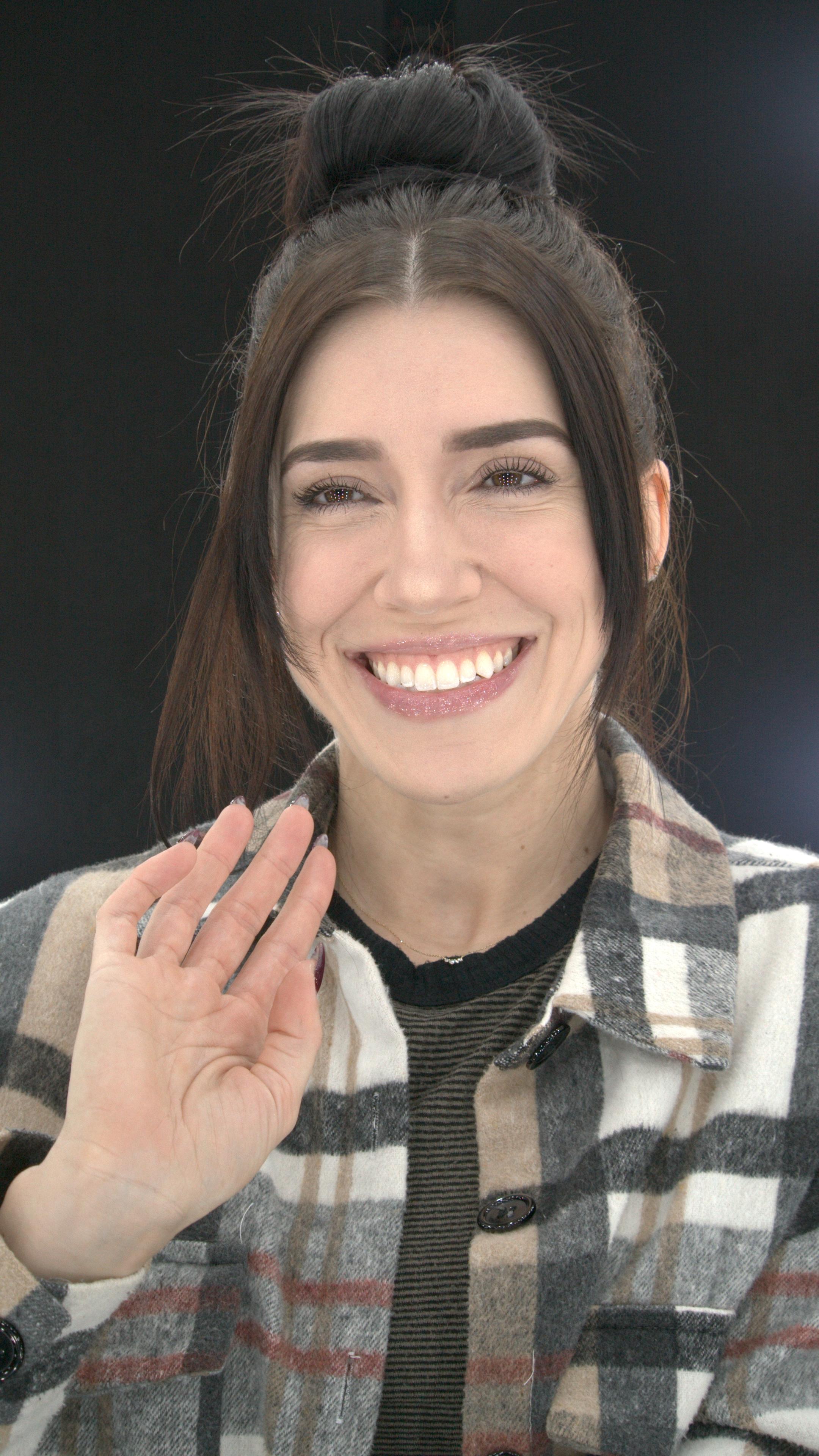}
    \end{minipage}
    \begin{minipage}{0.242\linewidth}
    \centering
    \includegraphics[trim=100pc 0 30pc 60pc, clip, width=1.0\linewidth]{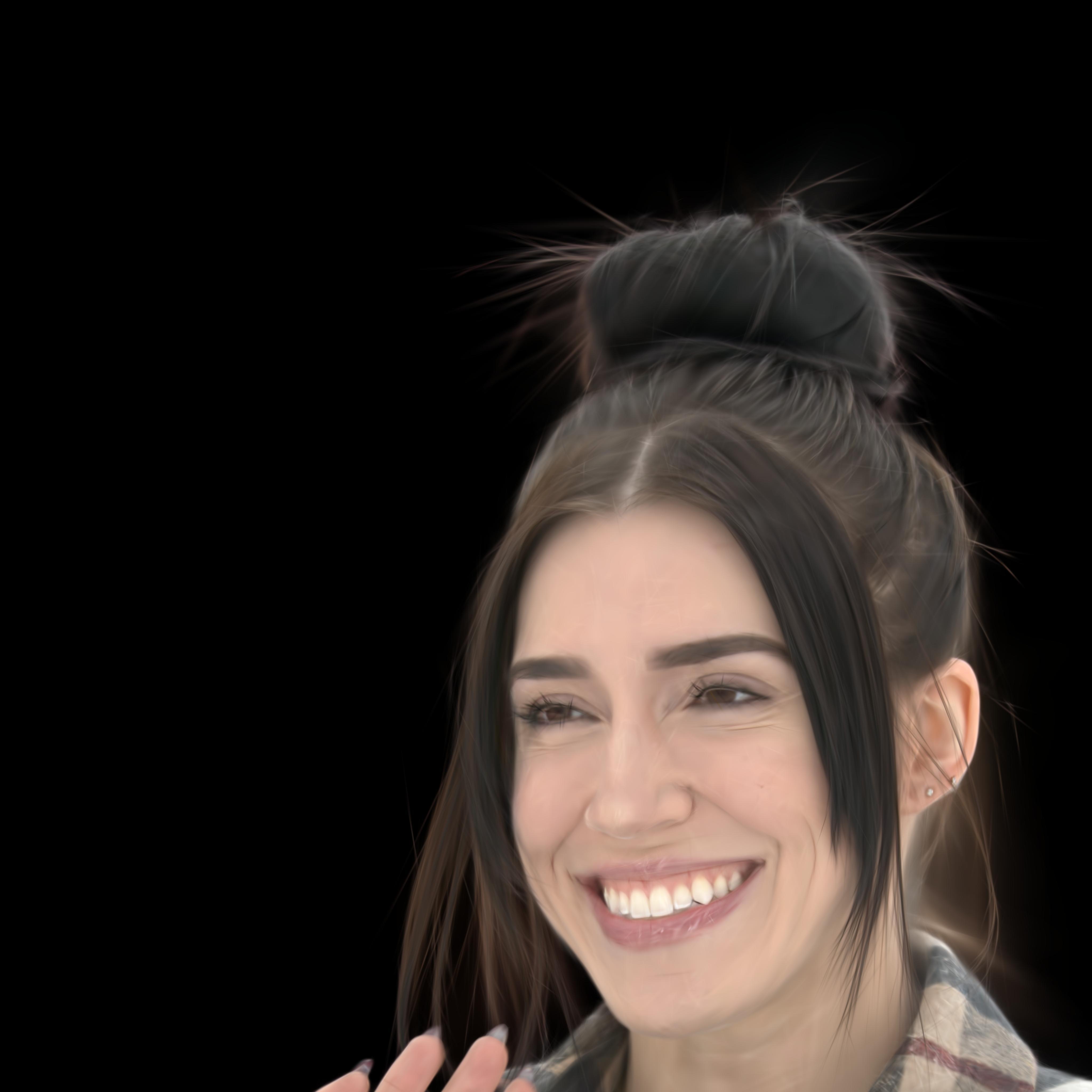}
    \includegraphics[trim=150pc 100pc 90pc 200pc, clip, width=1.0\linewidth]{images/face_rig/low_quality_reconstruction.jpg}
    \end{minipage}
    \begin{minipage}{0.242\linewidth}
    \centering
    \includegraphics[trim=100pc 0 30pc 60pc, clip, width=1.0\linewidth]{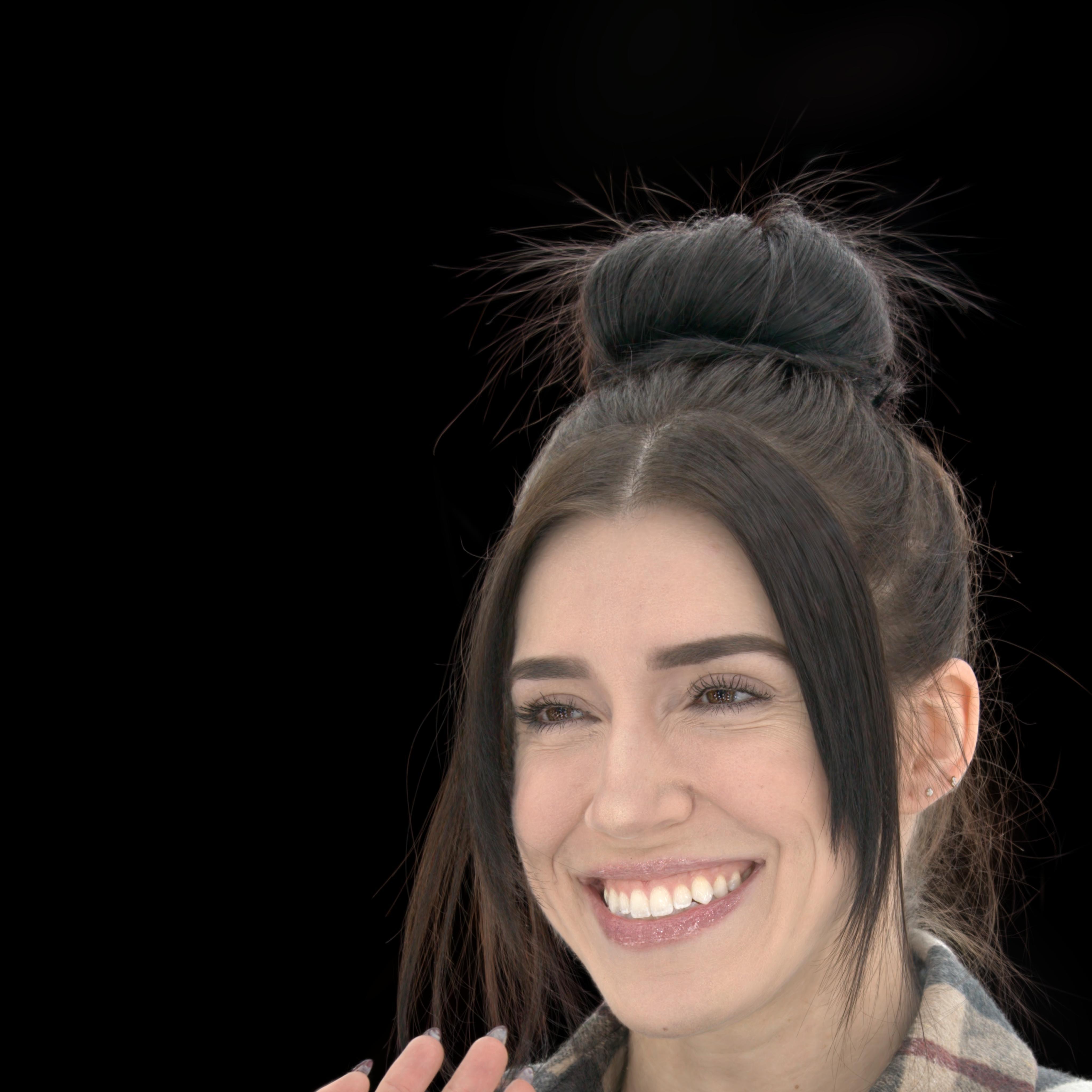}
    \includegraphics[trim=150pc 100pc 90pc 200pc, clip, width=1.0\linewidth]{images/face_rig/high_quality_reconstruction.jpg}
    \end{minipage}
    \caption{Illustration of the \facerig and corresponding reconstructions. From left to right: the \facerig hardware, an input image, a low-quality GS render and the corresponding high-quality GS render used for supervision.}
    \label{fig:face_rig}
\end{figure}

\subsection{Hardware Setup of the \facerig}
\label{sec:facerig_dta}
The \facerig is a volumetric capture stage with a multi-view camera array placed within a capped cylinder of black LED panels and white LED lights as shown in Fig.~\ref{fig:face_rig}, left.
The 80 LED panels (16 columns, each consisting of 5 panels) form a capture volume approximately 250cm tall and 276cm in diameter, with the ceiling and floor covered with 30 and 10 panels, respectively.
This stage features 75 synchronized 4K Z-CAM E2 cinema cameras, evenly installed and peering through 5cm gaps in the walls and ceiling. To ensure consistent flat lighting for facial data capture, 110 white LED lights are evenly distributed throughout the stage.

\subsection{Volumetric Data Capture and Processing}
\SA{After performing in the \scenerig,} we ask the actors to individually perform a short sequence of diverse facial expressions in the \facerig~.
This footage is captured at 24fps 4K resolution.
We also capture a color chart for color calibration against the \scenerig.

From each actor's performance, we uniformly sample \SA{60 sub-sequences of 8 frames}, ensuring a diverse range of expressions.
For each selected \SA{sub-sequence, we reconstruct two \emph{Poly4DGS} models: a high-quality model with 1 million Gaussians per-frame and a lower-quality model with a number of Gaussians sampled from 50,000 to 200,000.
This range corresponds to the minimum and maximum number of Gaussians left when cropping the upper-body of actors in the \scenerig reconstructions.
This ensures that the low quality reconstructions match the quality expected at test time.}

\subsection{\SA{Training Data Generation}}
\SA{
To obtain training data for our detail enhancement model, we render paired sequences of low- and high-quality GS in a 16-bit RGBA EXR format at $4096 \times 4096$ resolution.
For each reconstructed subsequence, we generate 60 camera paths each spanning 8 frames, with a moving camera focusing on the face.
In total, this results in approximately 3600 paired sequences of low- and high-quality renderings per actor, covering a diverse range of Gaussians artifacts, facial expressions, camera motions, and focal lengths.
After rendering, the dataset contains approximately 3600 paired sequences of low- and high-quality renderings per actor.
Additional details about the camera paths and data generation are included in the supplemental material.}



\subsection{\SA{Detail Enhancement Model}}
\SA{The goal of our detail enhancement model is to remove the Gaussian-like artifacts from the \scenerig renderings and restore fine details, especially useful for rendering close-up shots.

One promising approach for this task would be to use a conditioned video diffusion model \cite{blattmann2023stable,jin2024pyramidal,yang2024cogvideox}.
To meet the production quality requirements, we aim to enhance our images at 4K resolution.
However, video diffusion models are memory-intensive and resolution-limited, typically ranging from 480p to 768p, well below our target of 2160p.

We thus leverage an image diffusion model, Flux~\cite{flux}, which natively supports resolutions from 1080p to 1440p and incorporates strong natural image priors.
We introduce key architectural changes to this pretrained model to address the main challenges we face. 
First, we need to condition the model on the renderings from the \scenerig.
Second, we aim to obtain a temporally stable output, as generating frames independently causes temporal flickering.
Finally, as the model will be used for virtual production, we seek to enhance the Alpha channel together with RGB to ease the composition with backgrounds.

\subsubsection{Image-Conditioning}
As shown in the inference generation scheme of Flux \cite{flux} in Fig.~\ref{fig:DE_arch} (top left), 
latent noise is sampled from a normal distribution and then denoised multiple times by the diffusion transformer, to produce an output latent, which is converted to an image using the VAE Decoder.
Following the existing works \cite{zeng2024rgb,Luo2024IntrinsicDiffusion}, we modify the Diffusion Transformers (DiT) to condition it on multiple images.
We extend the weight matrix of the first linear layer so it takes as input, $(N+1)\mathcal{L}_{ch}$ channels instead of $\mathcal{L}_{ch}$, where $N$ represents the number of conditioning images added.

The inference scheme for this model is shown in Fig.~\ref{fig:DE_arch} (bottom left).
The input conditions are encoded separately with the VAE and concatenated to the latent noise, before being processed.

\subsubsection{Temporal Stability}
We first explore training a model conditioned on low-quality GS to predict high-quality renderings. While this yields high-quality results for single frames, as shown in our supplemental video, the output lacks temporal stability.

\paragraph{Optical Flow Warping}
To make the output of our model temporally consistent, especially for fine details, we propose conditioning it on the previous output frame, if available.
Therefore, we add two additional conditions to the model: a warped version of the previous output frame and a validity mask.
We compute the optical flow \cite{raft} between the current and previous input low-quality (LQ) renderings.
We then resample both the previous LQ and enhanced renderings.
The resampled LQ rendering is compared to the current one to generate the validity mask.
For the first frame, both conditions are replaced with zeros, and the network is trained with $50\%$ dropout on these inputs.

\paragraph{Low-Frequency Stabilization}
Even with this guidance, the output suffers from low-frequency flicker, such as global intensity shifts.
As we show in the supplemental material, this is partially caused by the VAE’s inability to preserve low-frequency information through encoding and decoding.
To alleviate this, we compute a 5-level Laplacian pyramid for both the LQ input and detail-enhanced result and then swap the lowest level ($120 \times 68$ pixels) of the output with that of the input. Experiments show this simple approach is highly effective in terms of improving temporal stability. 

\subsubsection{Alpha Channel Enhancement}
Since we aim to composite the final frames onto virtual backgrounds, an accurate Alpha channel is essential.
To enhance the Alpha channel of the LQ input, we modify the model to additionally condition on this channel.
We also double the number of input latent channels, similar to adding an extra input condition.
Finally, we adjust the output linear layer to predict twice the number of channels.
Our final architecture, shown in Fig.~\ref{fig:DE_arch} (right), takes four input conditions - LQ RGB, LQ Alpha, warped previous output, and warped validity mask - along with two latent noise inputs, resulting in $6 \times \mathcal{L}_{ch}$ input channels.
It outputs two latents, \textit{i.e.}, $2 \times \mathcal{L}_{ch}$ channels. The Alpha and RGB channels are decoded independently by the decoder.

The model is trained simultaneously on all seven actors, with a different text prompt for each subgroup. Additional implementation details are provided in the supplemental materials.
}

\begin{figure*}
\centering
Example input frames (cropped) with insets.
\addtolength{\tabcolsep}{-0.4em}
\begin{tabular}{cccc}
     \adjincludegraphics[trim= {0.0\width} {0.03\height} {0.0\width} {0.10\height}, clip, width=0.247\linewidth]{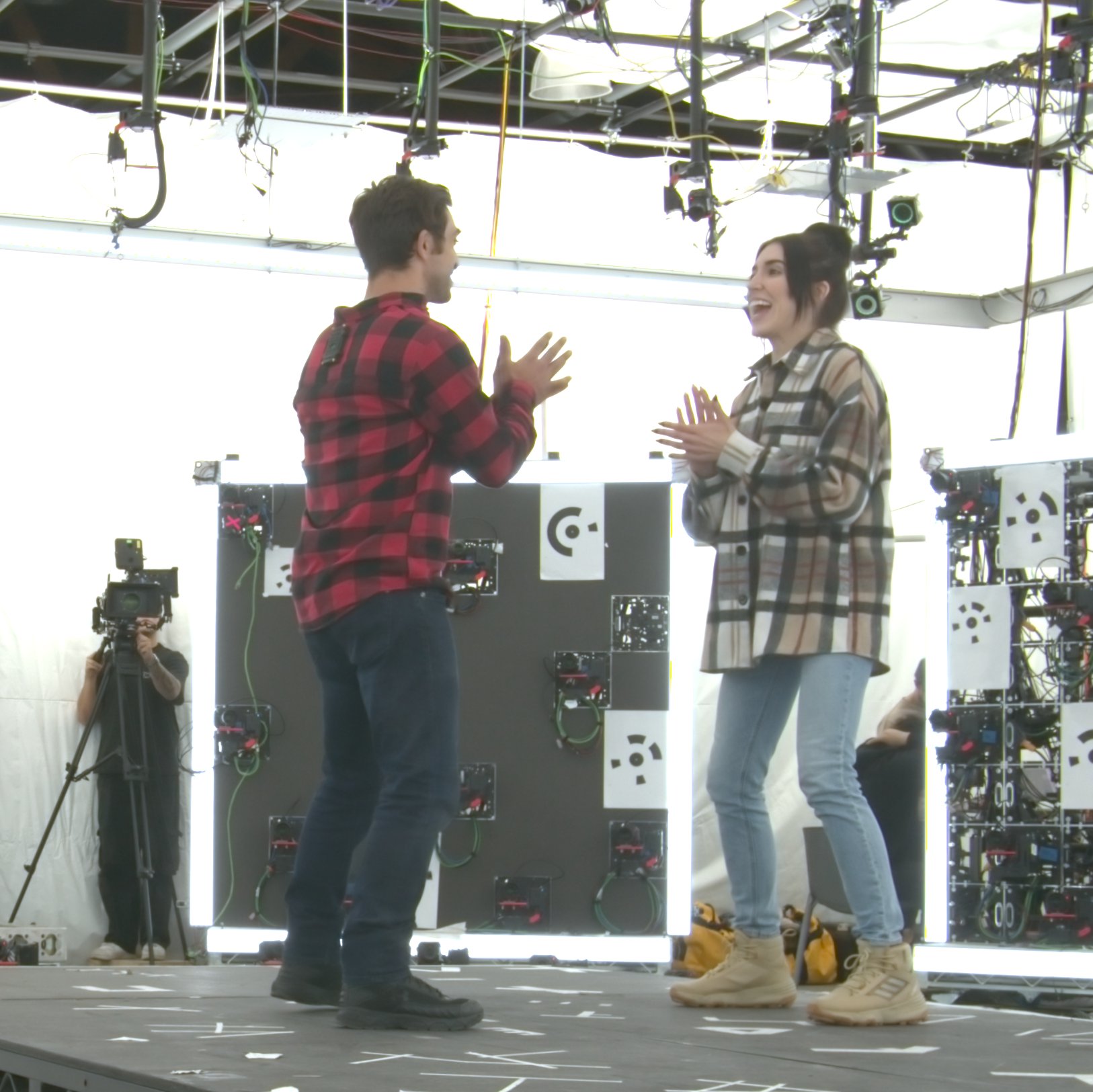}&
    \adjincludegraphics[trim= {0.0\width} {0.03\height} {0.0\width} {0.10\height}, clip,width=0.247\linewidth]{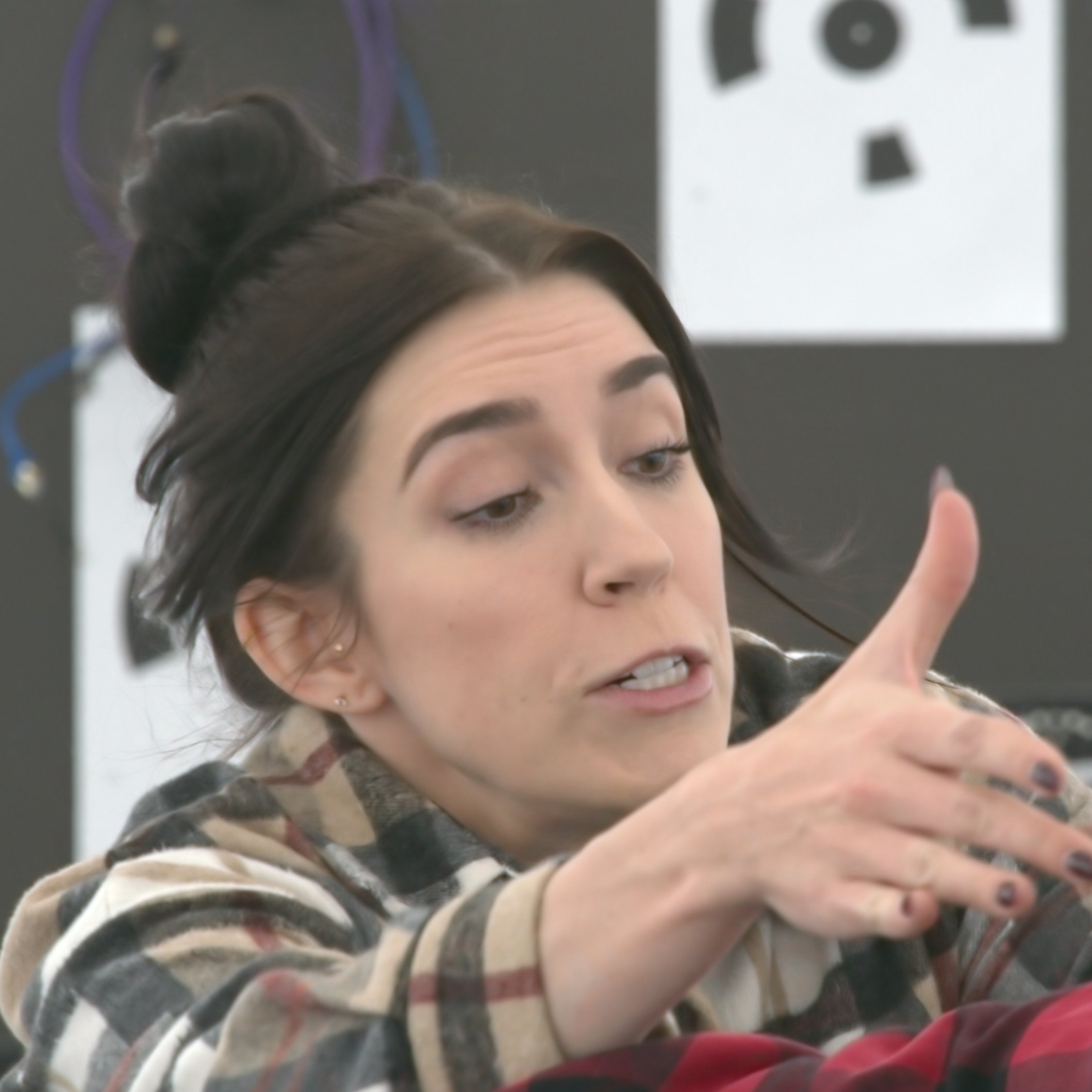}&
    \adjincludegraphics[trim= {0.5\width} {0.225\height} {0.0\width} {0.0\height}, clip,width=0.247\linewidth]{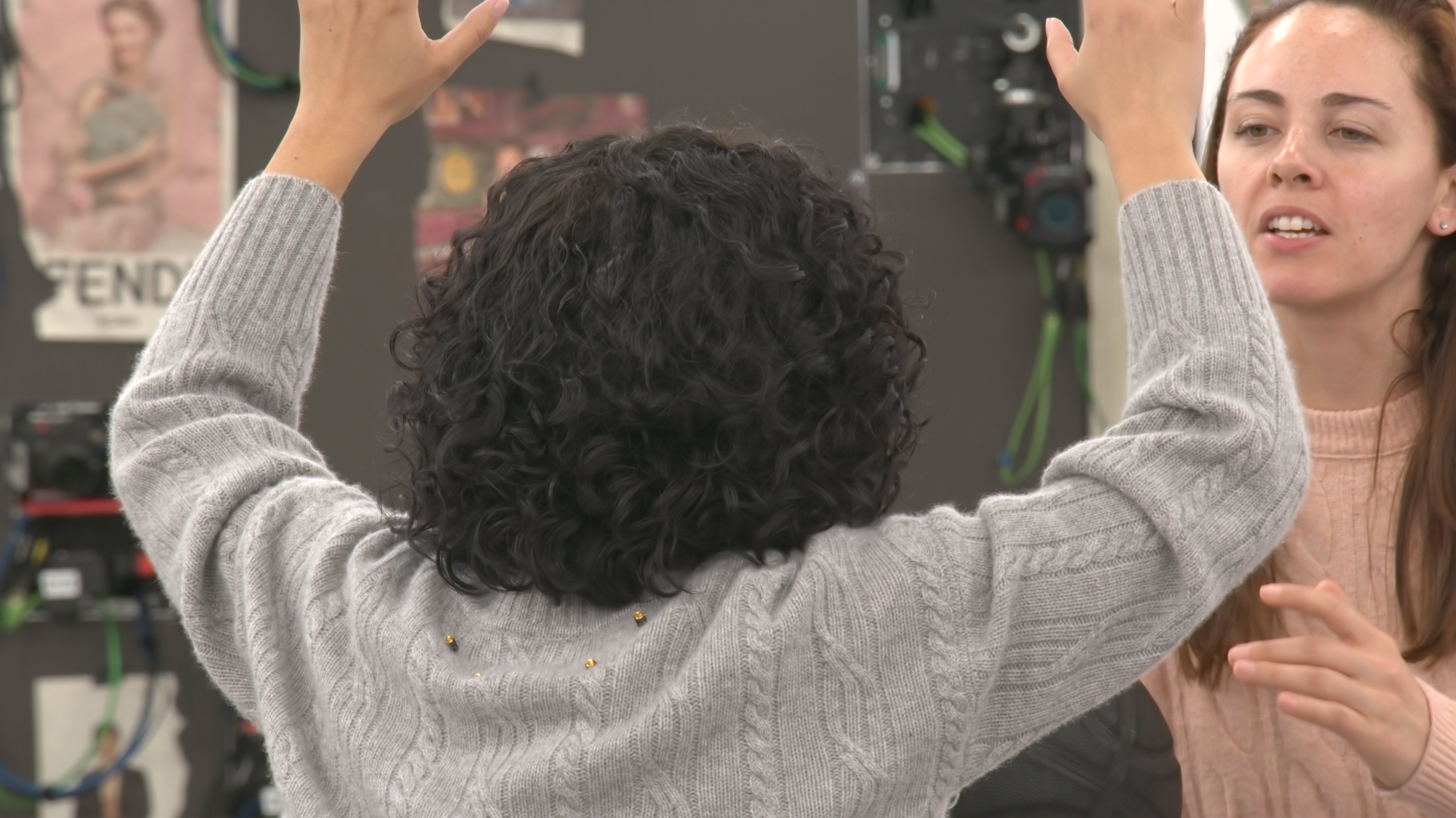}&
    \adjincludegraphics[trim= {0.1\width} {0.0\height} {0.3\width} {0.07\height}, clip,width=0.247\linewidth]{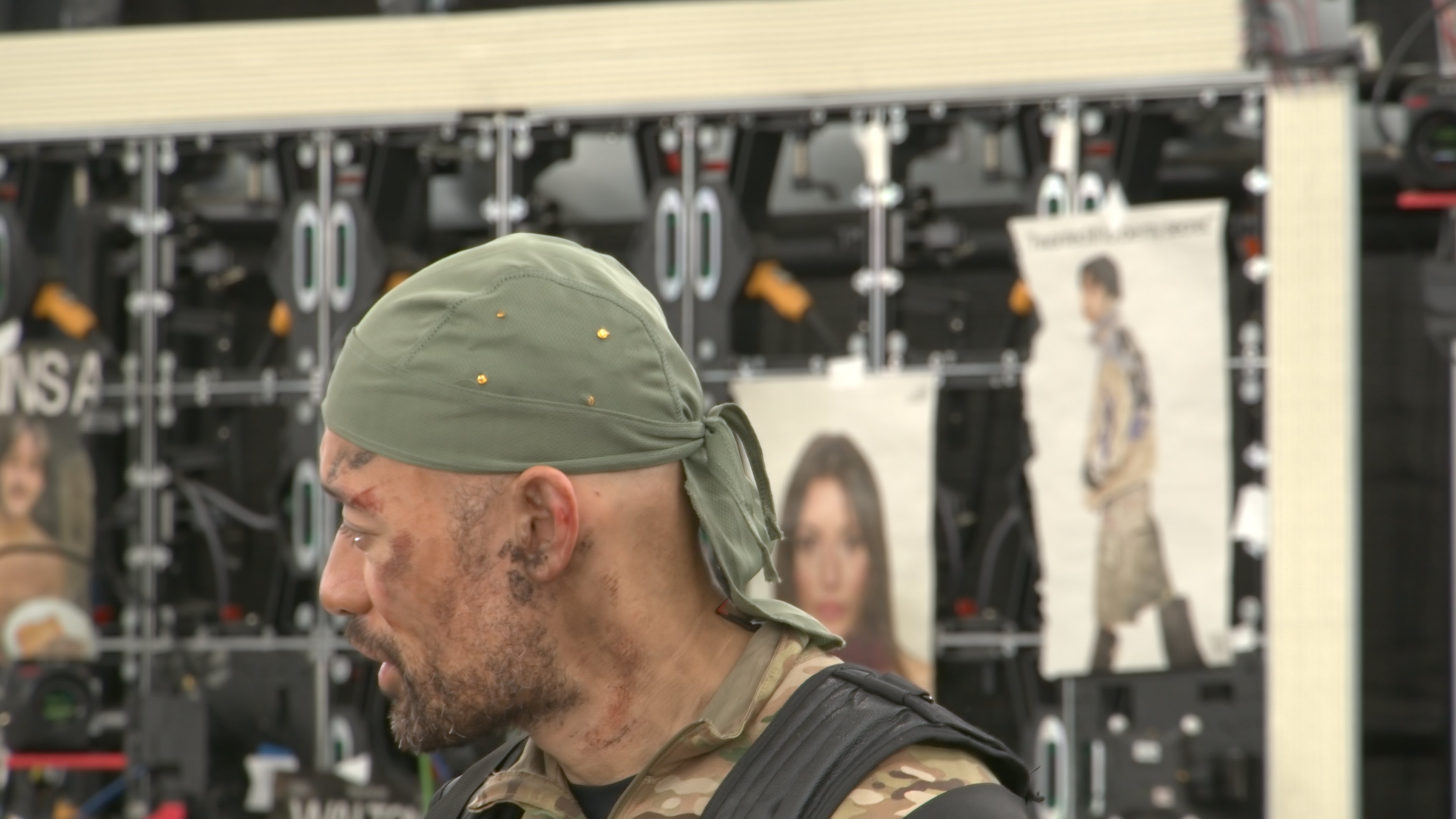} \\
    \adjincludegraphics[trim= {0.3\width} {0.64\height} {0.21\width} {0.15\height}, clip,width=0.247\linewidth]{images/main_results/VPS07-zcam-e2-017.1017.jpg}&
    \adjincludegraphics[trim= {0.38\width} {0.49\height} {0.34\width} {0.39\height}, clip,width=0.247\linewidth]{images/main_results/VPS07-zcam-e2-019.1051.jpg}&
    \adjincludegraphics[trim= {0.7\width} {0.67\height} {0.0\width} {0.1\height}, clip,width=0.247\linewidth]{images/main_results/frames/VPS07-zcam-e2-004.0198.jpg}&
    \adjincludegraphics[trim= {0.15\width} {0.22\height} {0.55\width} {0.55\height}, clip,width=0.247\linewidth]{images/main_results/frames/VPS07-zcam-e2-025.0100.jpg} 
\end{tabular}

\begin{tabular}{cc}
     Our reconstruction using \emph{Poly4DGS} & Our Detail Enhanced Results  \\

    \begin{minipage}{0.5\linewidth}
    \centering
    \adjincludegraphics[trim= {0.25\width} {0.0\height} {0.15\width} {0.0\height}, clip, clip, width=0.495\linewidth]{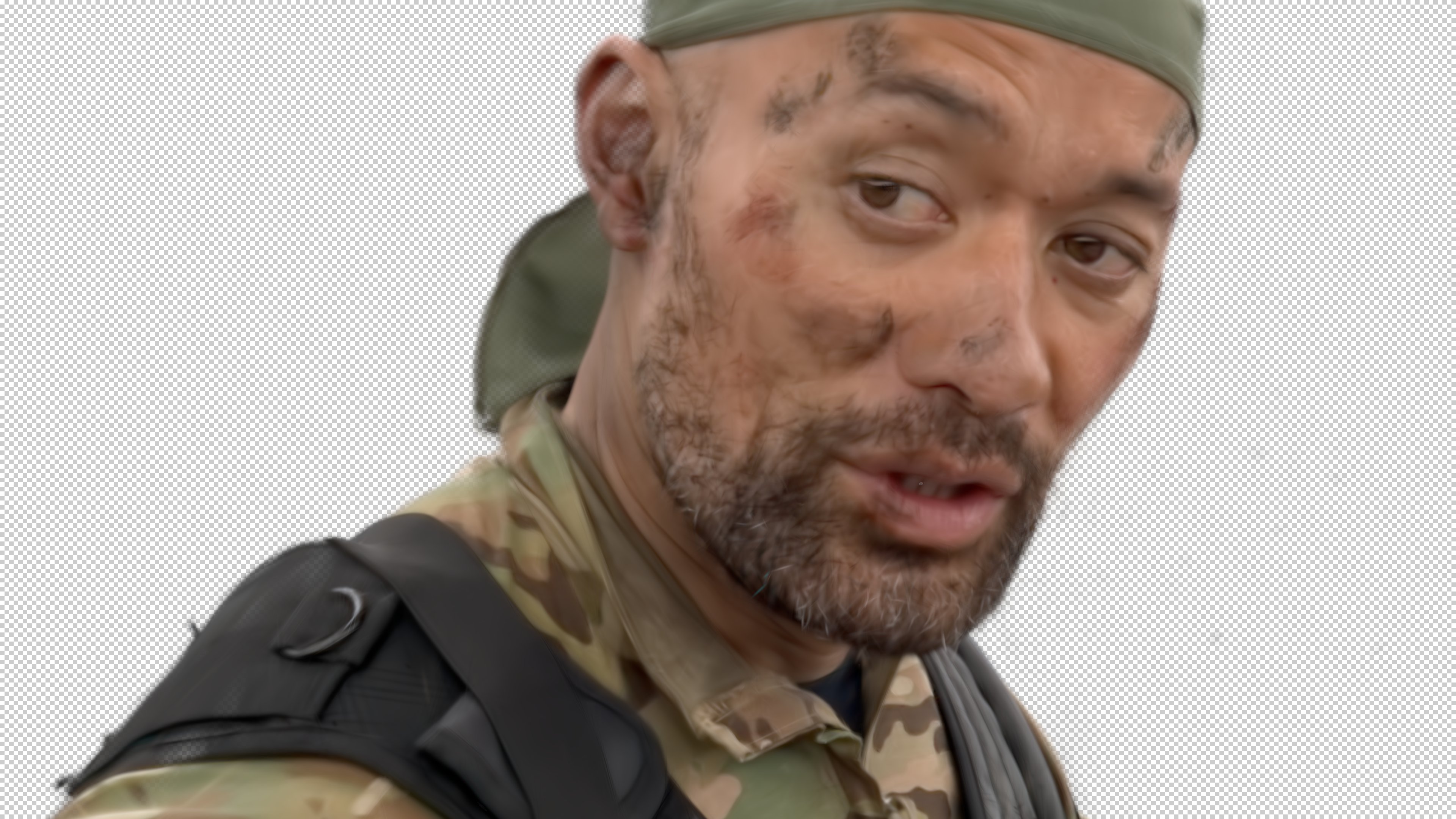}
    \adjincludegraphics[trim= {0.2\width} {0.0\height} {0.2\width} {0.0\height}, clip, clip, width=0.495\linewidth]{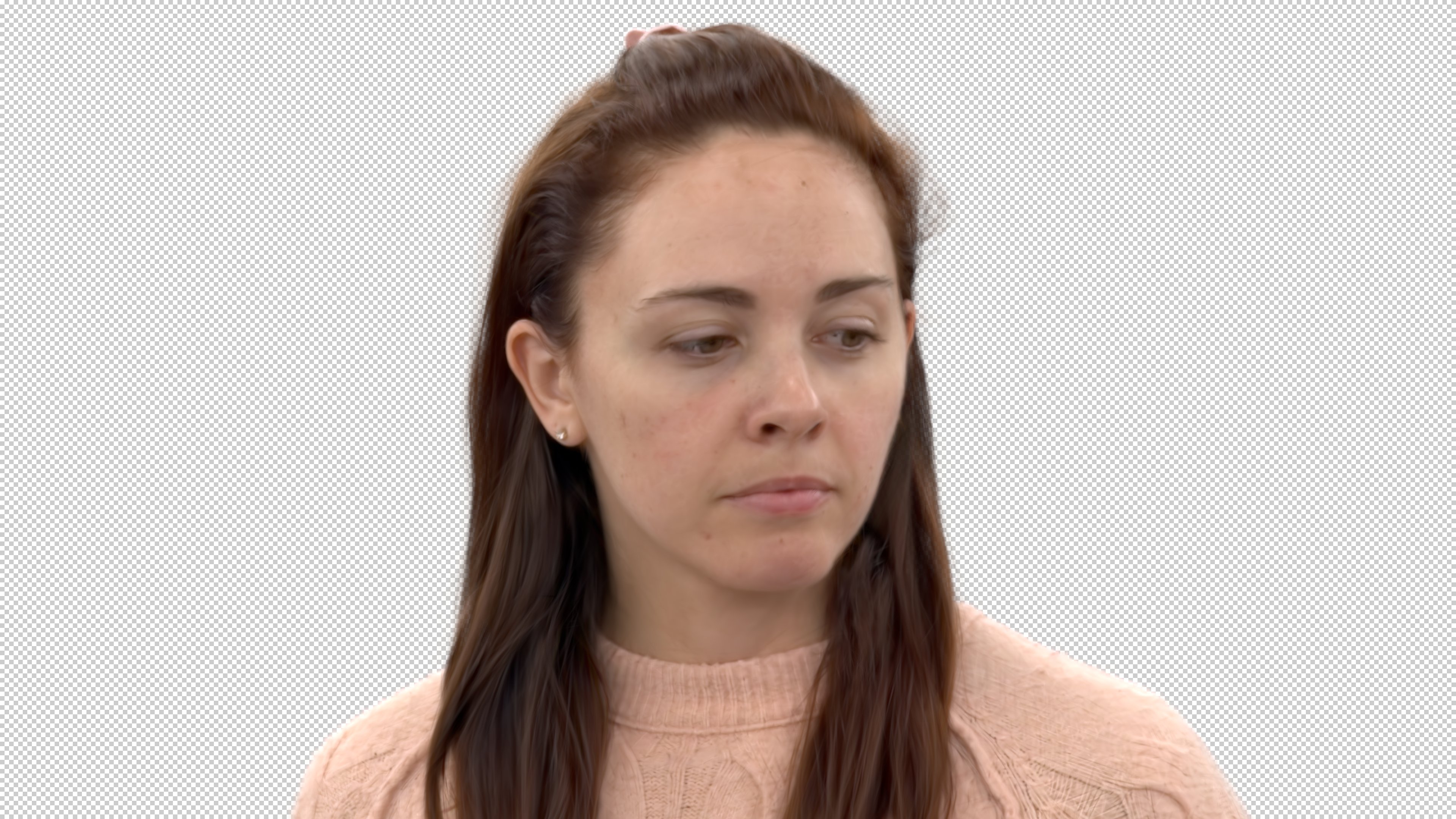}
    \\
    \adjincludegraphics[trim= {0.55\width} {0.3\height} {0.25\width} {0.5\height}, clip, width=0.495\linewidth]{images/main_results/input/ET_cam0000_1001.jpg}
    \adjincludegraphics[trim= {0.425\width} {0.425\height} {0.375\width} {0.375\height}, clip, width=0.495\linewidth]{images/main_results/input/TK_cam0000_1001.jpg}
    \end{minipage}

&
    
    \begin{minipage}{0.5\linewidth}
    \centering
    \adjincludegraphics[trim= {0.25\width} {0.0\height} {0.15\width} {0.0\height}, clip, clip, width=0.495\linewidth]{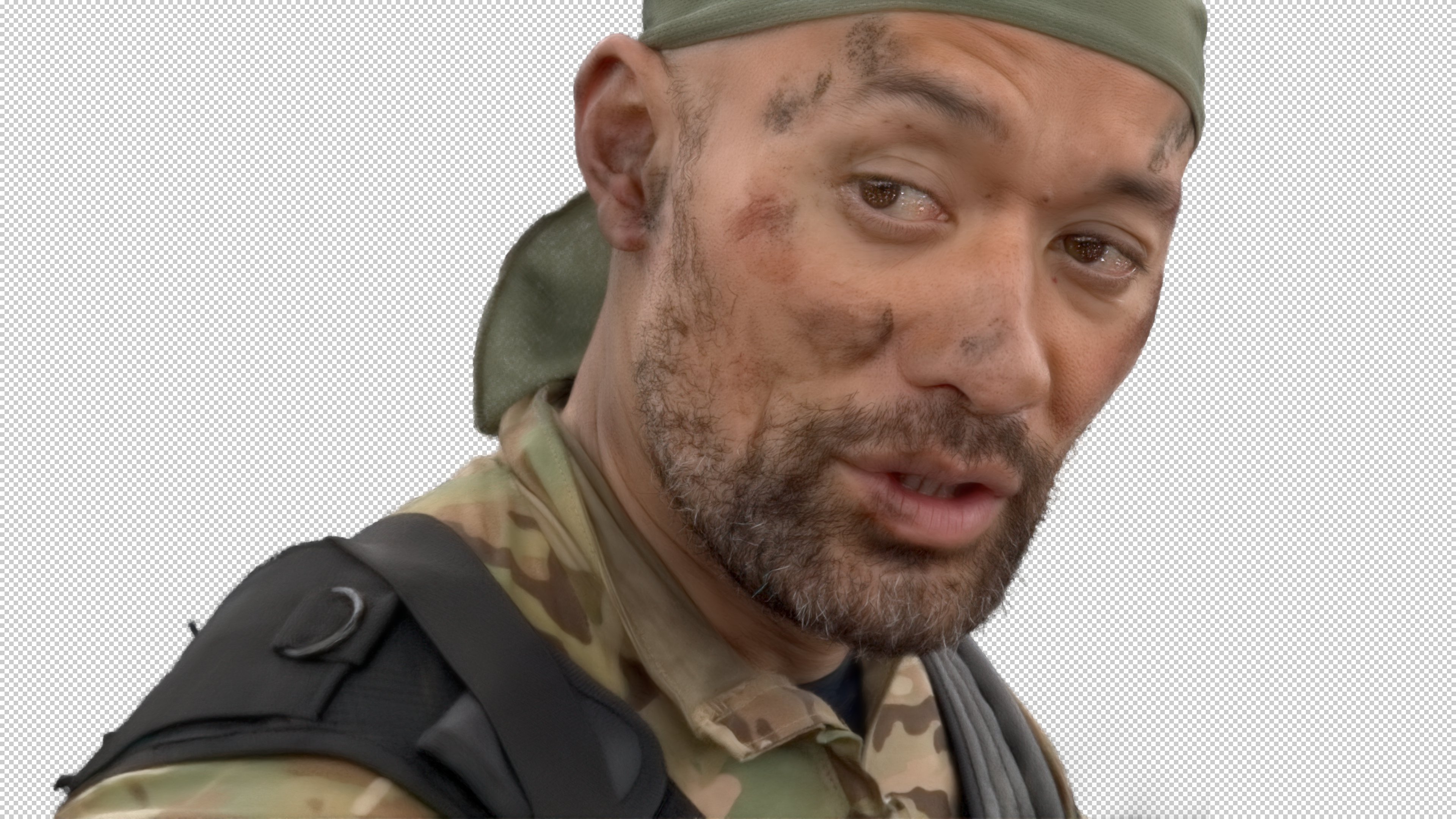}
    \adjincludegraphics[trim= {0.2\width} {0.0\height} {0.2\width} {0.0\height}, clip, clip, width=0.495\linewidth]{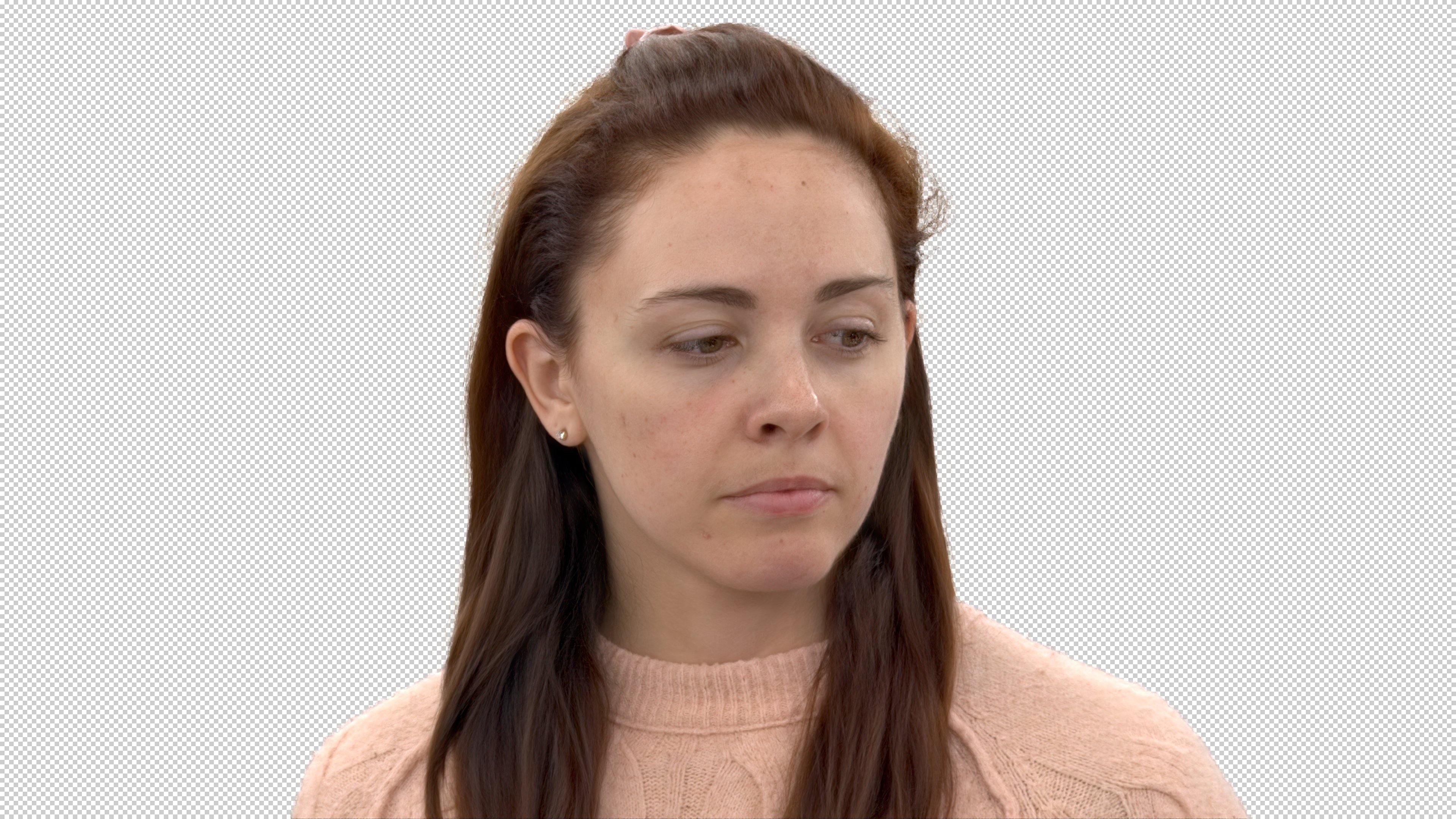}
    \\
    \adjincludegraphics[trim= {0.55\width} {0.3\height} {0.25\width} {0.5\height}, clip, width=0.495\linewidth]{images/main_results/output/ET_cam0000_1001.jpg}
    \adjincludegraphics[trim={0.425\width} {0.425\height} {0.375\width} {0.375\height}, clip, width=0.495\linewidth]{images/main_results/output/TK_cam0000_1001.jpg}
    \end{minipage} \\

    \noalign{\vskip 1mm}     

    \begin{minipage}{0.5\linewidth}
    \centering
    \adjincludegraphics[trim= {0.3\width} {0.0\height} {0.1\width} {0.0\height}, clip, clip, width=0.495\linewidth]{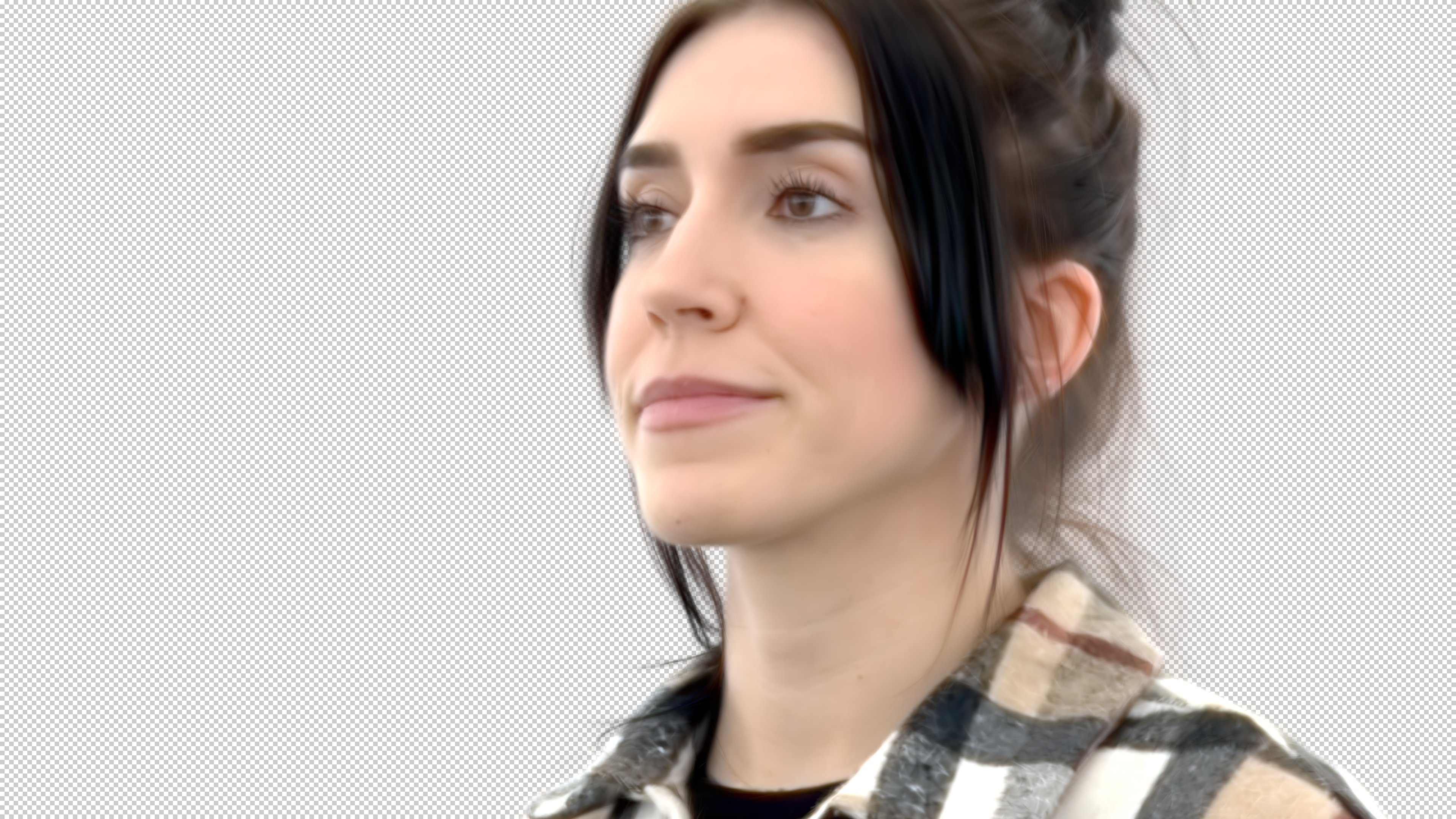}
    \adjincludegraphics[trim= {0.2\width} {0.0\height} {0.2\width} {0.0\height}, clip, clip, width=0.495\linewidth]{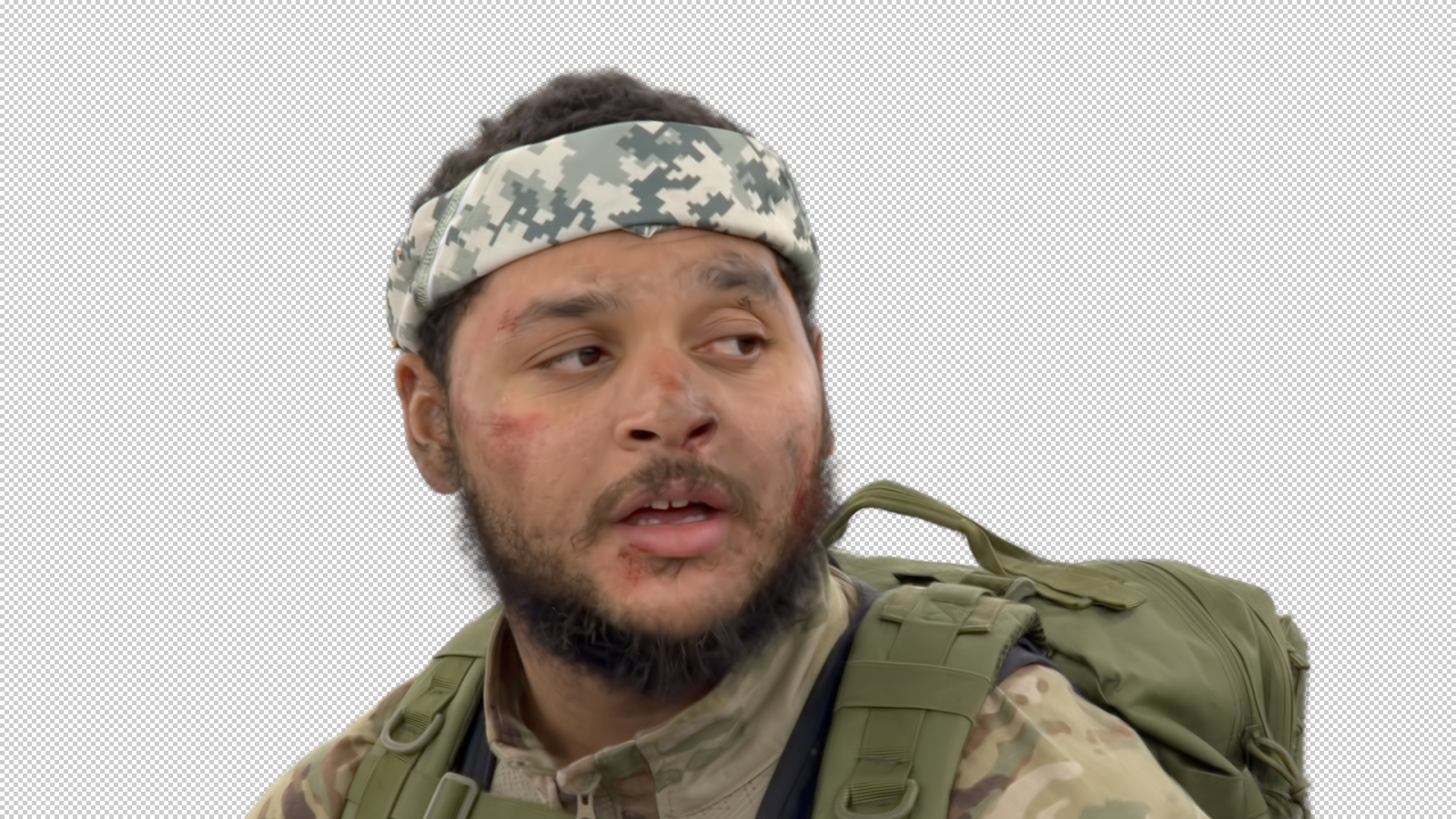}
    \\
    \adjincludegraphics[trim= {0.55\width} {0.4\height} {0.25\width} {0.4\height}, clip, width=0.495\linewidth]{images/main_results/input/EW_cam0000_1001.jpg}
    \adjincludegraphics[trim={0.275\width} {0.35\height} {0.525\width} {0.45\height}, clip, width=0.495\linewidth]{images/main_results/input/MH_cam0000_1001.jpg}
    \end{minipage} &
    
    \begin{minipage}{0.5\linewidth}
    \centering
    \adjincludegraphics[trim= {0.3\width} {0.0\height} {0.1\width} {0.0\height}, clip, clip, width=0.495\linewidth]{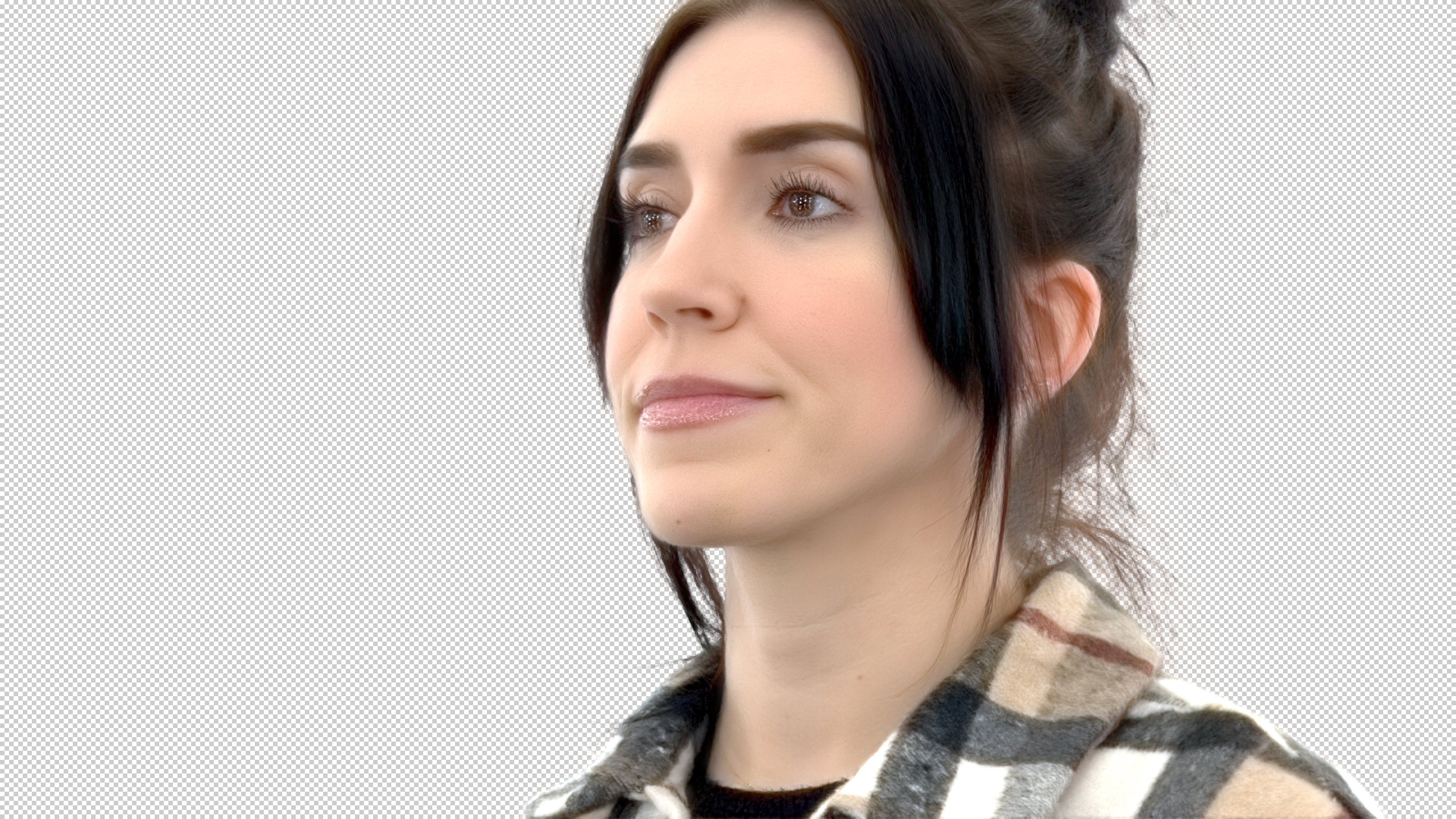}
    \adjincludegraphics[trim= {0.2\width} {0.0\height} {0.2\width} {0.0\height}, clip, clip, width=0.495\linewidth]{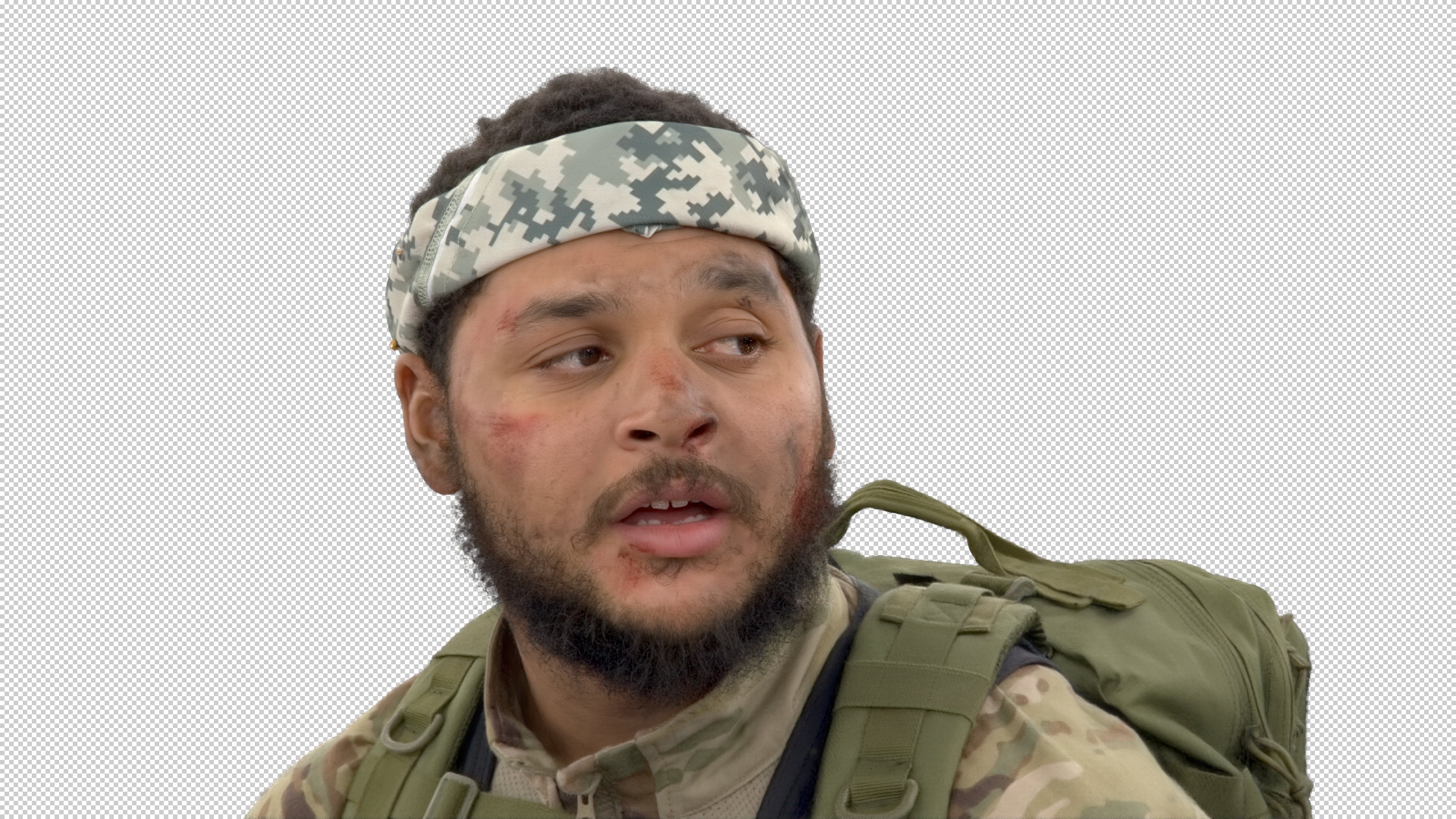}
    \\
    \adjincludegraphics[trim= {0.55\width} {0.4\height} {0.25\width} {0.4\height}, clip, width=0.495\linewidth]{images/main_results/output/EW_cam0000_1001.jpg}
    \adjincludegraphics[trim={0.275\width} {0.35\height} {0.525\width} {0.45\height}, clip, width=0.495\linewidth]{images/main_results/output/MH_cam0000_1001.jpg}
    \end{minipage}
    
\end{tabular}
\caption{Top: input training views captured in our \scenerig. Bottom-left: our \emph{Poly4DGS} reconstruction with insets. Bottom-right: final results using our super-resolution module and compositing. We can observe that the Gaussian artifacts present at extreme zoom levels are effectively removed and new details added.}
\label{fig:main_results}
\end{figure*}

\begin{figure*}
\centering

\begin{tabular}{*{5}{P{0.15\linewidth}}}
    GT & Full & w/o dense pt. & w/o black level & w/o expo. opt. \\
    \multicolumn{5}{c}{
\includegraphics[width=0.8\linewidth]{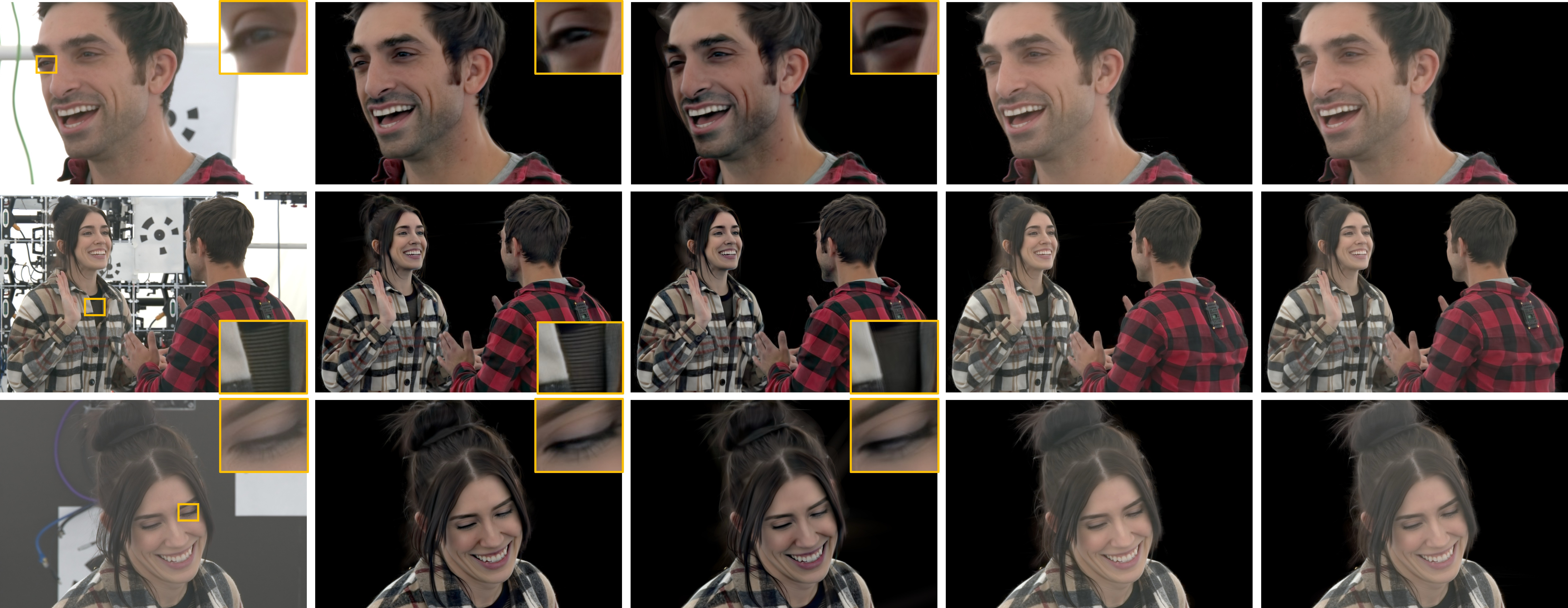}
    }
\end{tabular}
\caption{Qualitative results for several ablation studies. Our full method achieves the highest quality, delivering good color contrast and sharp details. Without initializing from dense point cloud (w/o dense pt.), the method fails to reconstruct fine details like the stripe on the girl's shirt. Without black-level and exposure optimization (w/o black level and w/o expo. opt.), view inconsistencies caused by camera exposure variations and lens glare are baked in the GS. }
\label{fig:ablations_recon}
\end{figure*}

\begin{figure*}
\centering
\addtolength{\tabcolsep}{-0.5em}
\begin{tabular}{ccccccc}

HQ & LQ & Ours & UpscaleAVideo & KEEP & ResShift & SinSR \\

\begin{minipage}{0.128\linewidth}
\centering
\adjincludegraphics[trim= {0.0\width} {0.0\height} {0.0\width} {0.0\height}, clip, width=1.0\linewidth]{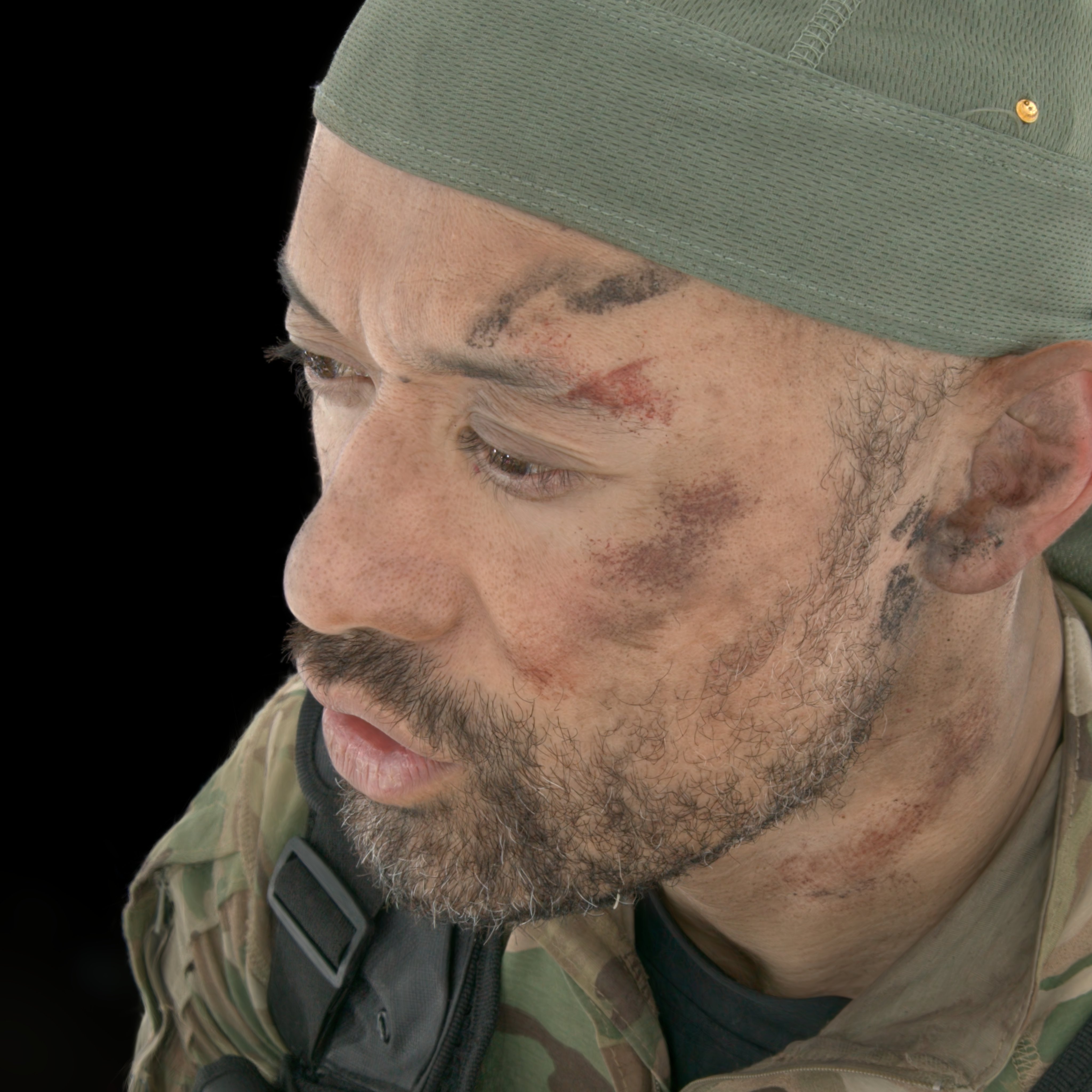} \\
\adjincludegraphics[trim= {0.43\width} {0.52\height} {0.47\width} {0.38\height}, clip, width=0.475\linewidth]{images/comparisons/seq09/HQ/cam0000_0012.jpg}
\hfill
\adjincludegraphics[trim= {0.29\width} {0.31\height} {0.61\width} {0.59\height}, clip, width=0.475\linewidth]{images/comparisons/seq09/HQ/cam0000_0012.jpg}
\end{minipage} &
\begin{minipage}{0.128\linewidth}
\centering
\adjincludegraphics[trim= {0.0\width} {0.0\height} {0.0\width} {0.0\height}, clip, width=1.0\linewidth]{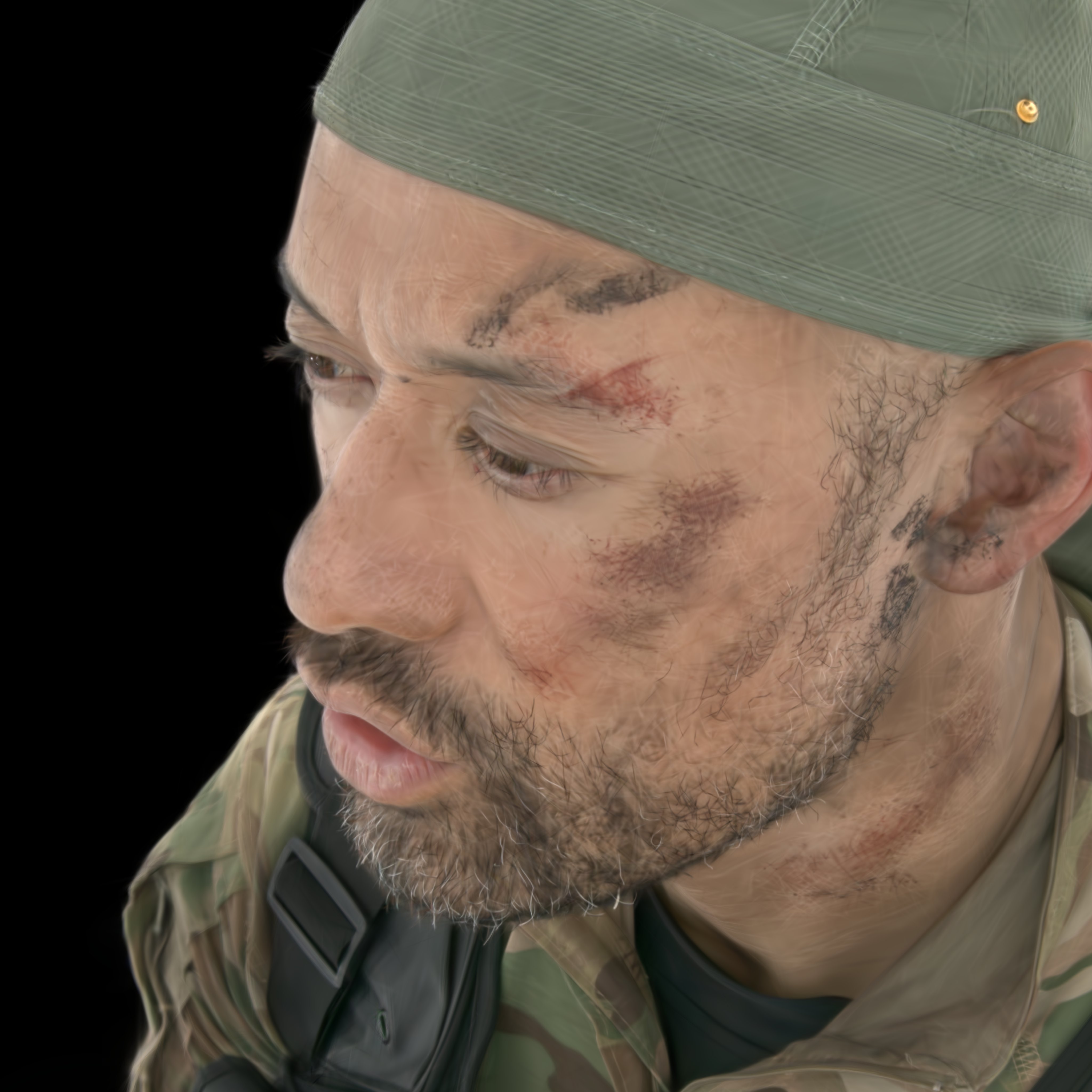} \\
\adjincludegraphics[trim= {0.43\width} {0.52\height} {0.47\width} {0.38\height}, clip, width=0.47\linewidth]{images/comparisons/seq09/LQ/cam0000_0012.jpg}
\hfill
\adjincludegraphics[trim= {0.29\width} {0.31\height} {0.61\width} {0.59\height}, clip, width=0.47\linewidth]{images/comparisons/seq09/LQ/cam0000_0012.jpg}
\end{minipage} &
\begin{minipage}{0.128\linewidth}
\centering
\adjincludegraphics[trim= {0.0\width} {0.0\height} {0.0\width} {0.0\height}, clip, width=1.0\linewidth]{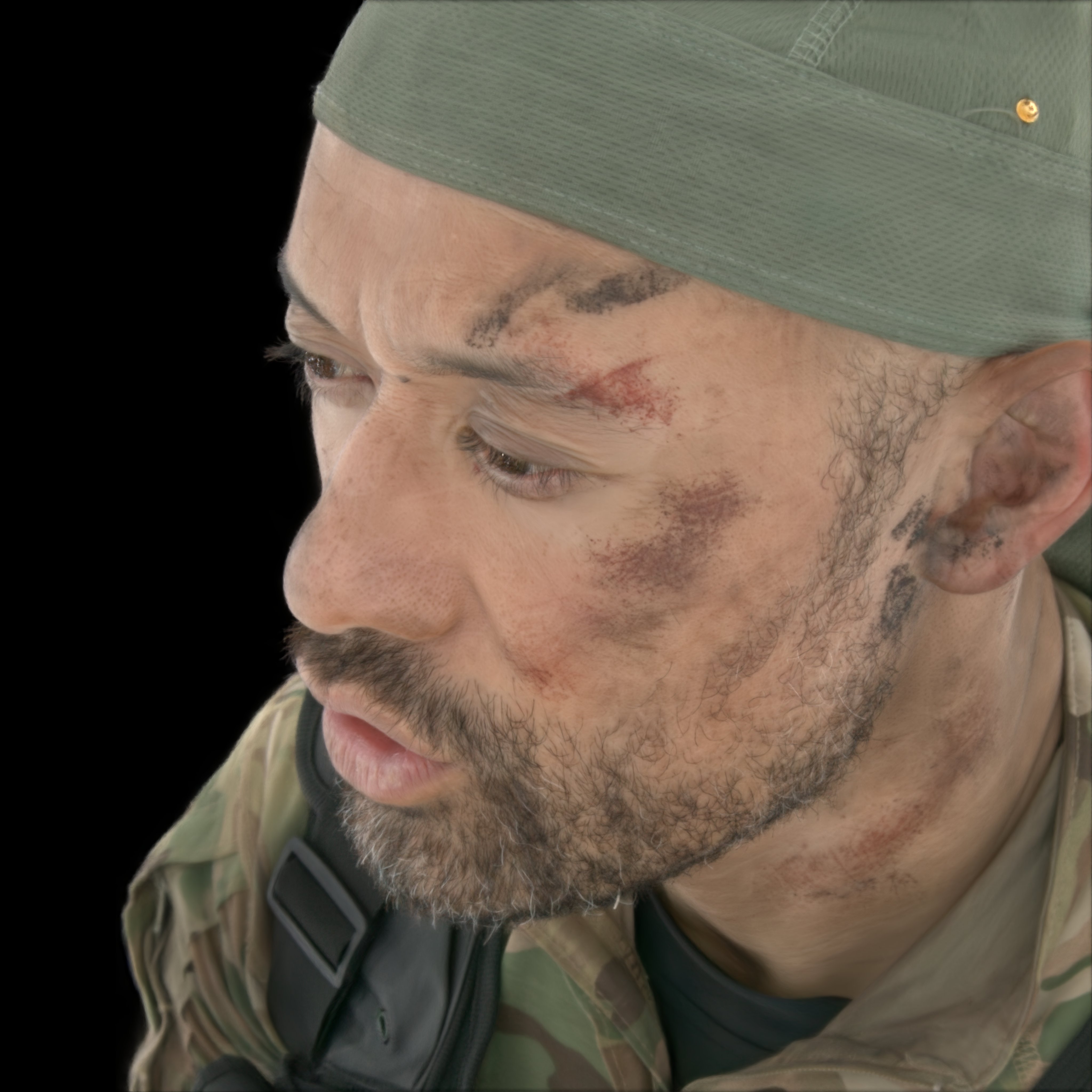} \\
\adjincludegraphics[trim= {0.43\width} {0.52\height} {0.47\width} {0.38\height}, clip, width=0.47\linewidth]{images/comparisons/seq09/Ours/cam0000_0012.jpg}
\hfill
\adjincludegraphics[trim= {0.29\width} {0.31\height} {0.61\width} {0.59\height}, clip, width=0.47\linewidth]{images/comparisons/seq09/Ours/cam0000_0012.jpg}
\end{minipage} &
\begin{minipage}{0.128\linewidth}
\centering
\adjincludegraphics[trim= {0.0\width} {0.0\height} {0.0\width} {0.0\height}, clip, width=1.0\linewidth]{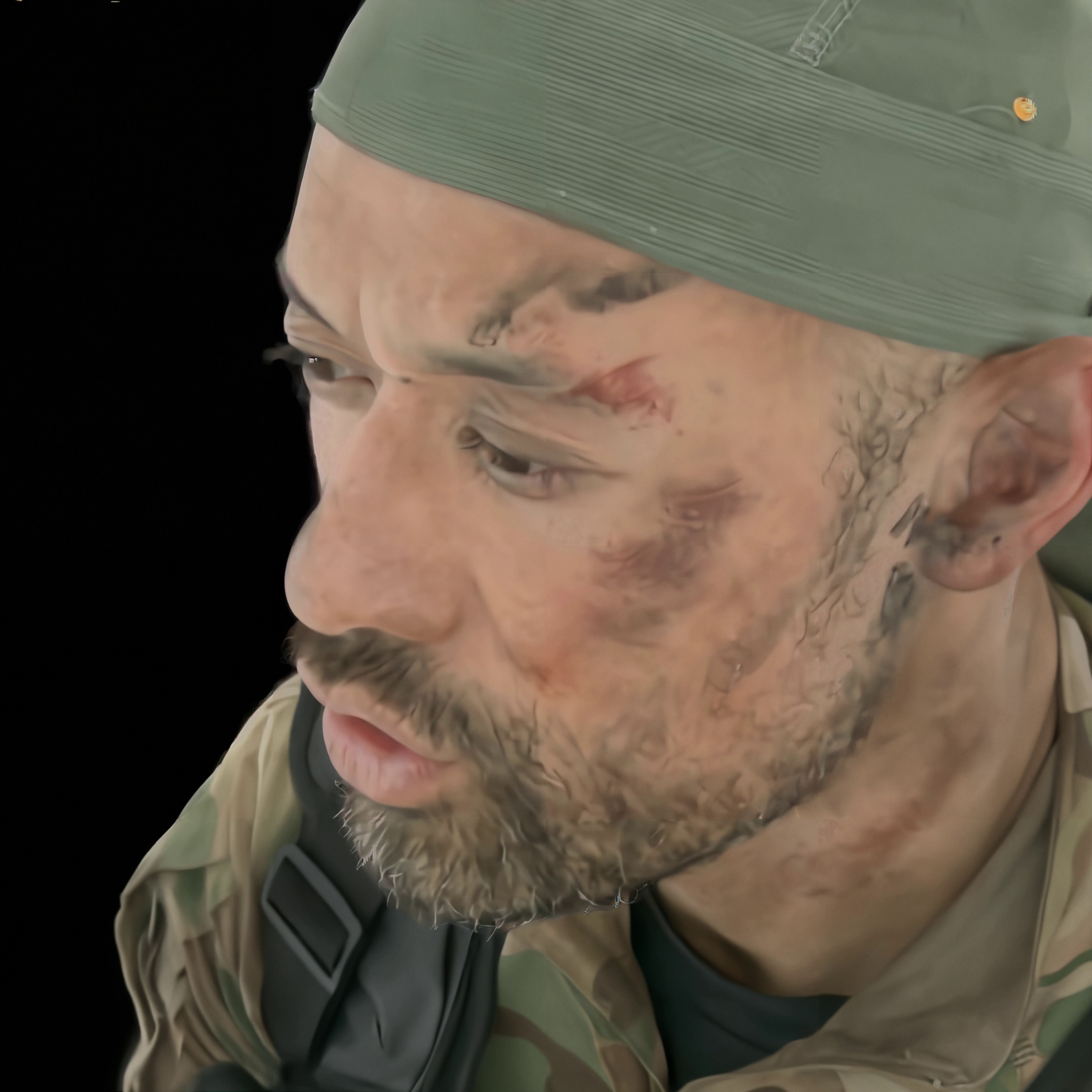} \\
\adjincludegraphics[trim= {0.43\width} {0.52\height} {0.47\width} {0.38\height}, clip, width=0.47\linewidth]{images/comparisons/seq09/UpscaleAVideo/cam0000_0012.jpg}
\hfill
\adjincludegraphics[trim= {0.29\width} {0.31\height} {0.61\width} {0.59\height}, clip, width=0.47\linewidth]{images/comparisons/seq09/UpscaleAVideo/cam0000_0012.jpg}
\end{minipage} &
\begin{minipage}{0.128\linewidth}
\centering
\placeholderFailed \\
\hfill
\end{minipage} &
\begin{minipage}{0.128\linewidth}
\centering
\adjincludegraphics[trim= {0.0\width} {0.0\height} {0.0\width} {0.0\height}, clip, width=1.0\linewidth]{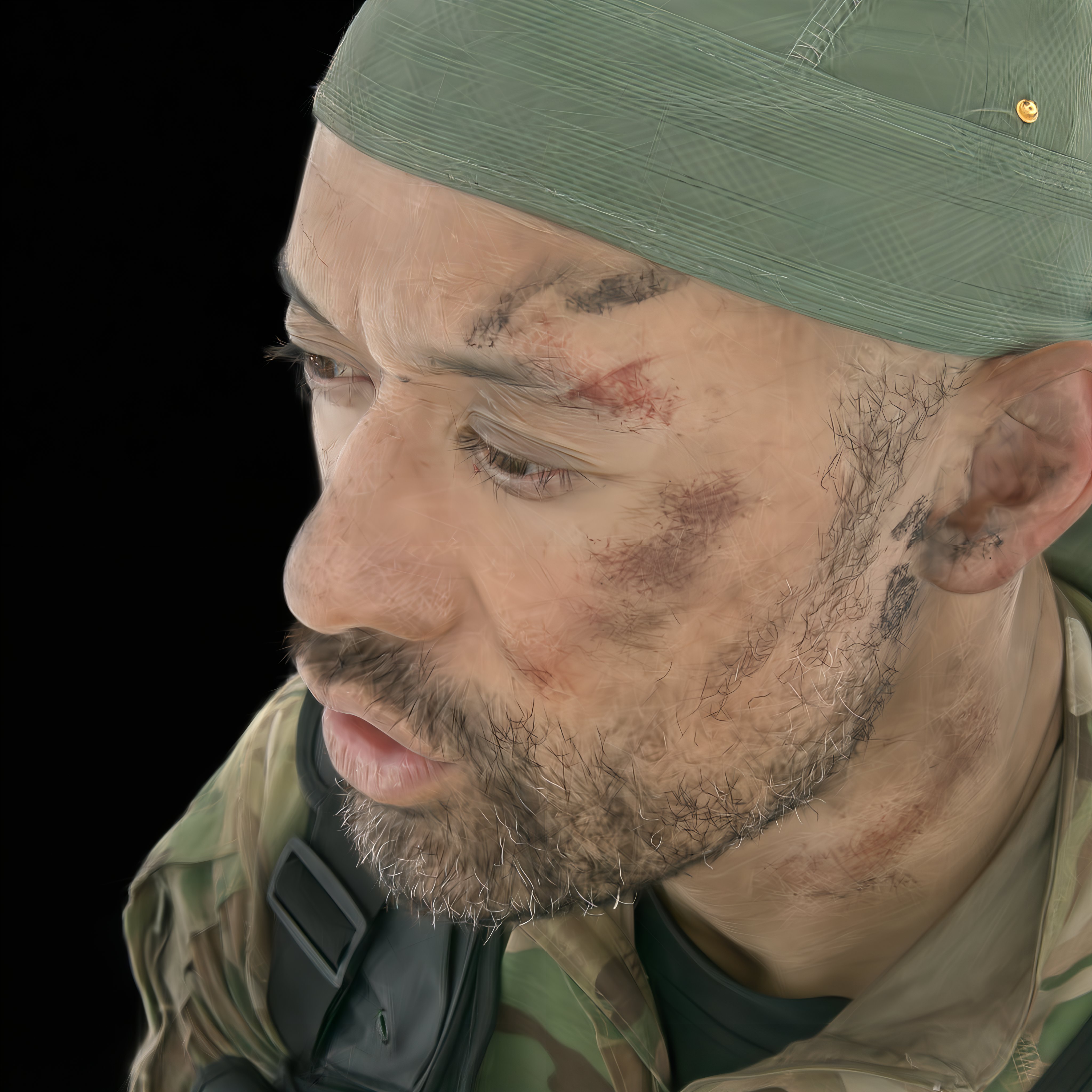} \\
\adjincludegraphics[trim= {0.43\width} {0.52\height} {0.47\width} {0.38\height}, clip, width=0.47\linewidth]{images/comparisons/seq09/ResShift/cam0000_0012.jpg}
\hfill
\adjincludegraphics[trim= {0.29\width} {0.31\height} {0.61\width} {0.59\height}, clip, width=0.47\linewidth]{images/comparisons/seq09/ResShift/cam0000_0012.jpg}
\end{minipage} &
\begin{minipage}{0.128\linewidth}
\centering
\adjincludegraphics[trim= {0.0\width} {0.0\height} {0.0\width} {0.0\height}, clip, width=1.0\linewidth]{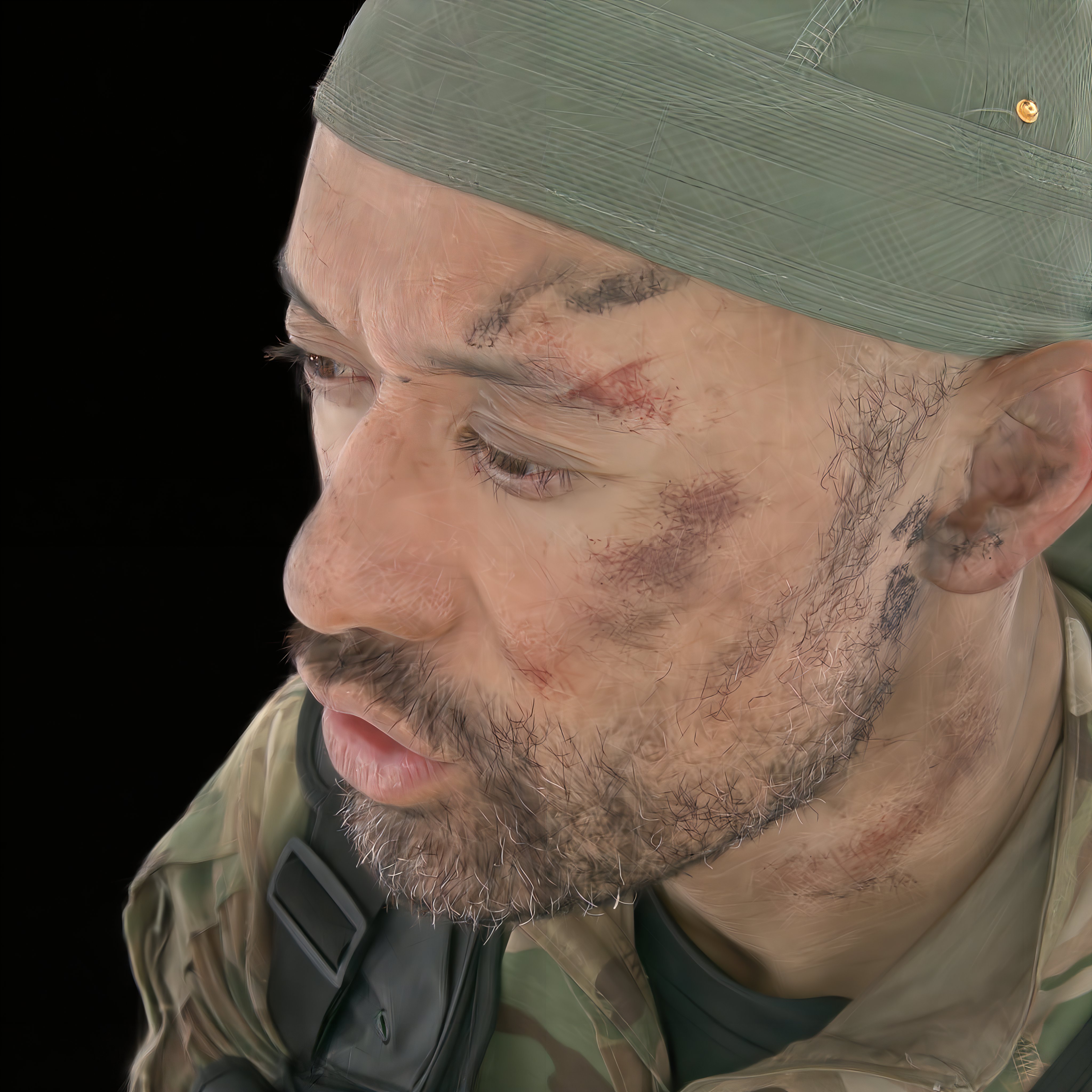} \\
\adjincludegraphics[trim= {0.43\width} {0.52\height} {0.47\width} {0.38\height}, clip, width=0.47\linewidth]{images/comparisons/seq09/SinSR/cam0000_0012.jpg}
\hfill
\adjincludegraphics[trim= {0.29\width} {0.31\height} {0.61\width} {0.59\height}, clip, width=0.47\linewidth]{images/comparisons/seq09/SinSR/cam0000_0012.jpg}
\end{minipage} \\


\begin{minipage}{0.128\linewidth}
\centering
\adjincludegraphics[trim= {0.0\width} {0.0\height} {0.1\width} {0.0\height}, clip, width=1.0\linewidth]{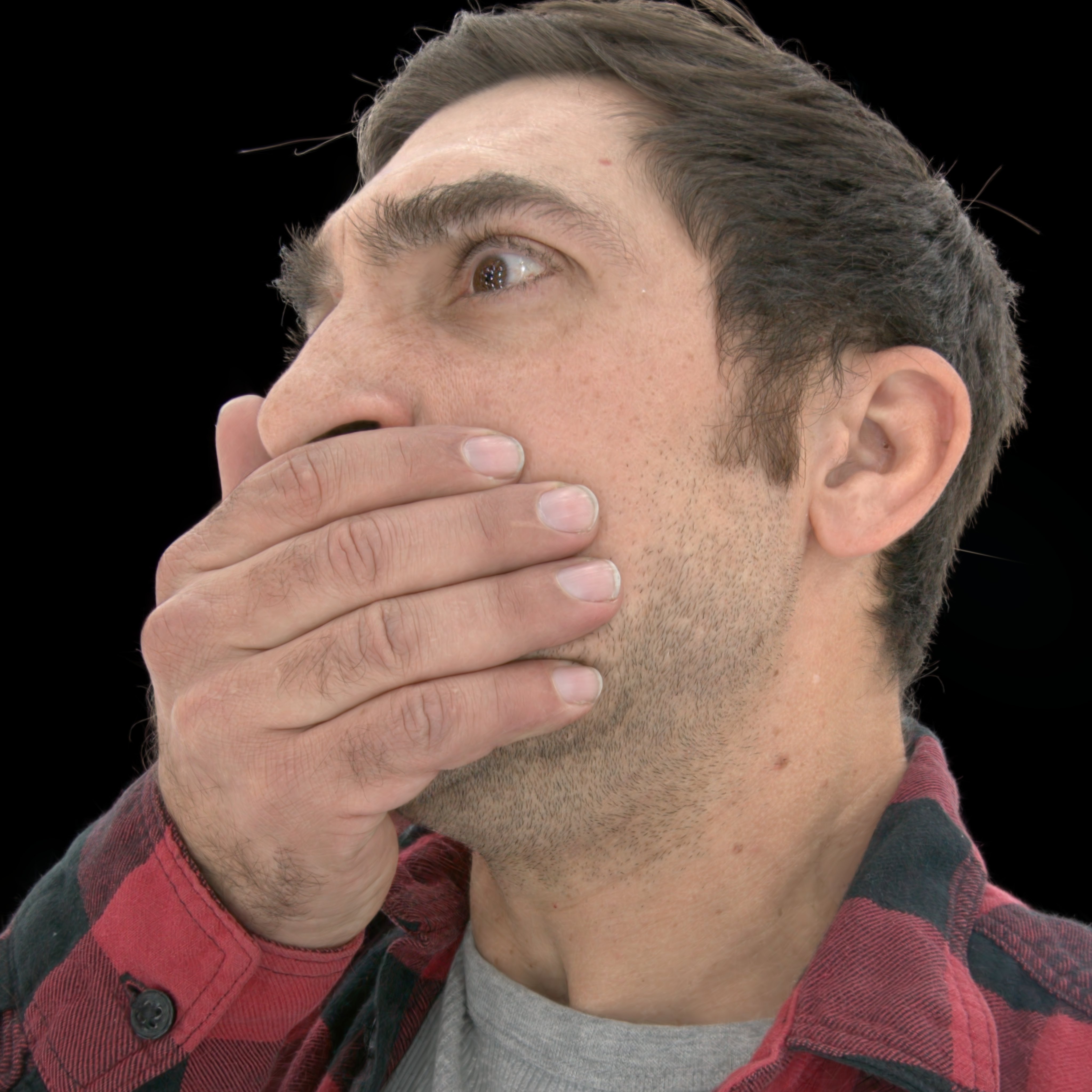} \\
\adjincludegraphics[trim= {0.52\width} {0.34\height} {0.38\width} {0.56\height}, clip, width=0.475\linewidth]{images/comparisons/seq10/HQ/cam0000_0012.jpg}
\hfill
\adjincludegraphics[trim= {0.63\width} {0.52\height} {0.27\width} {0.38\height}, clip, width=0.475\linewidth]{images/comparisons/seq10/HQ/cam0000_0012.jpg}
\end{minipage} &
\begin{minipage}{0.128\linewidth}
\centering
\adjincludegraphics[trim= {0.0\width} {0.0\height} {0.1\width} {0.0\height}, clip, width=1.0\linewidth]{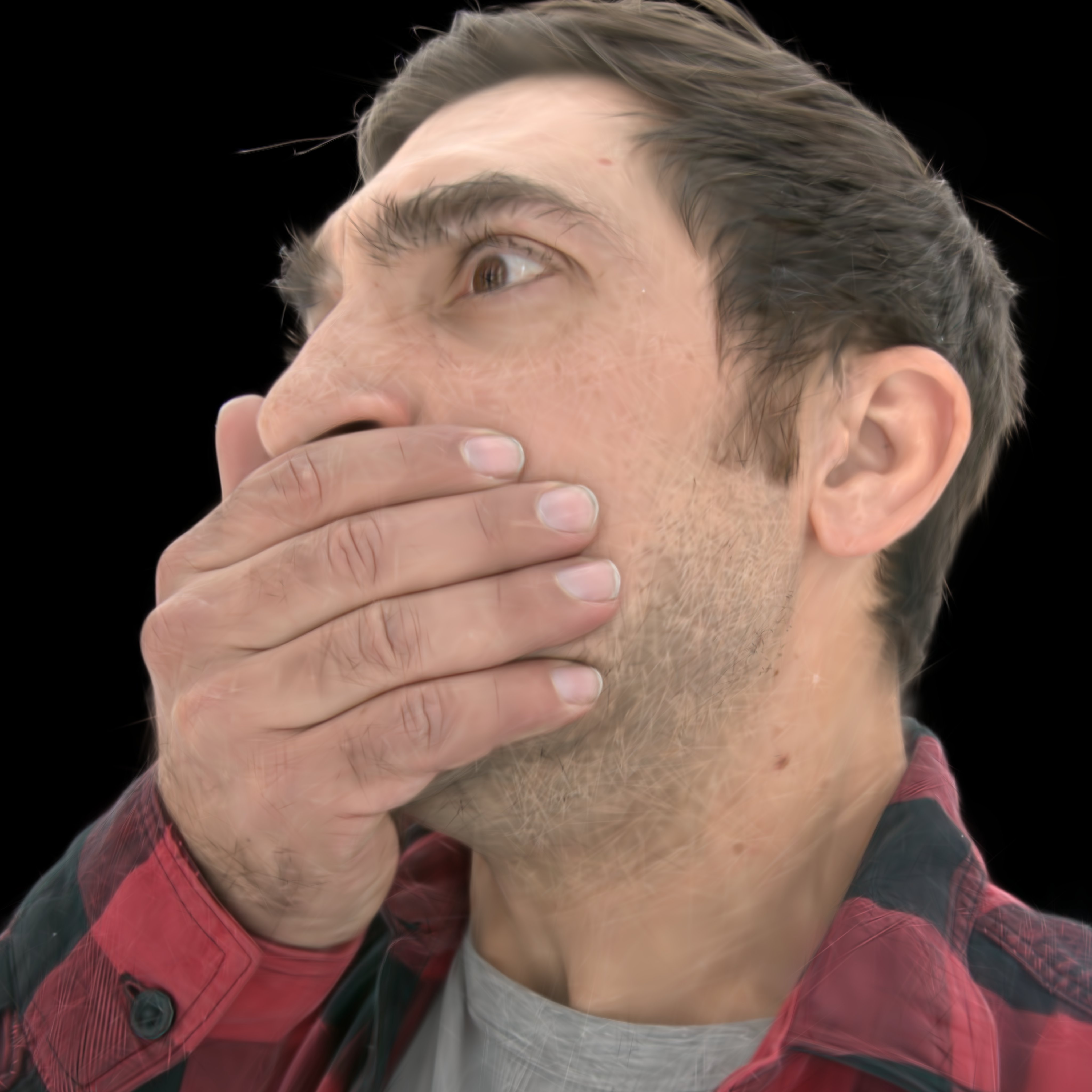} \\
\adjincludegraphics[trim= {0.52\width} {0.34\height} {0.38\width} {0.56\height}, clip, width=0.47\linewidth]{images/comparisons/seq10/LQ/cam0000_0012.jpg}
\hfill
\adjincludegraphics[trim= {0.63\width} {0.52\height} {0.27\width} {0.38\height}, clip, width=0.47\linewidth]{images/comparisons/seq10/LQ/cam0000_0012.jpg}
\end{minipage} &
\begin{minipage}{0.128\linewidth}
\centering
\adjincludegraphics[trim= {0.0\width} {0.0\height} {0.1\width} {0.0\height}, clip, width=1.0\linewidth]{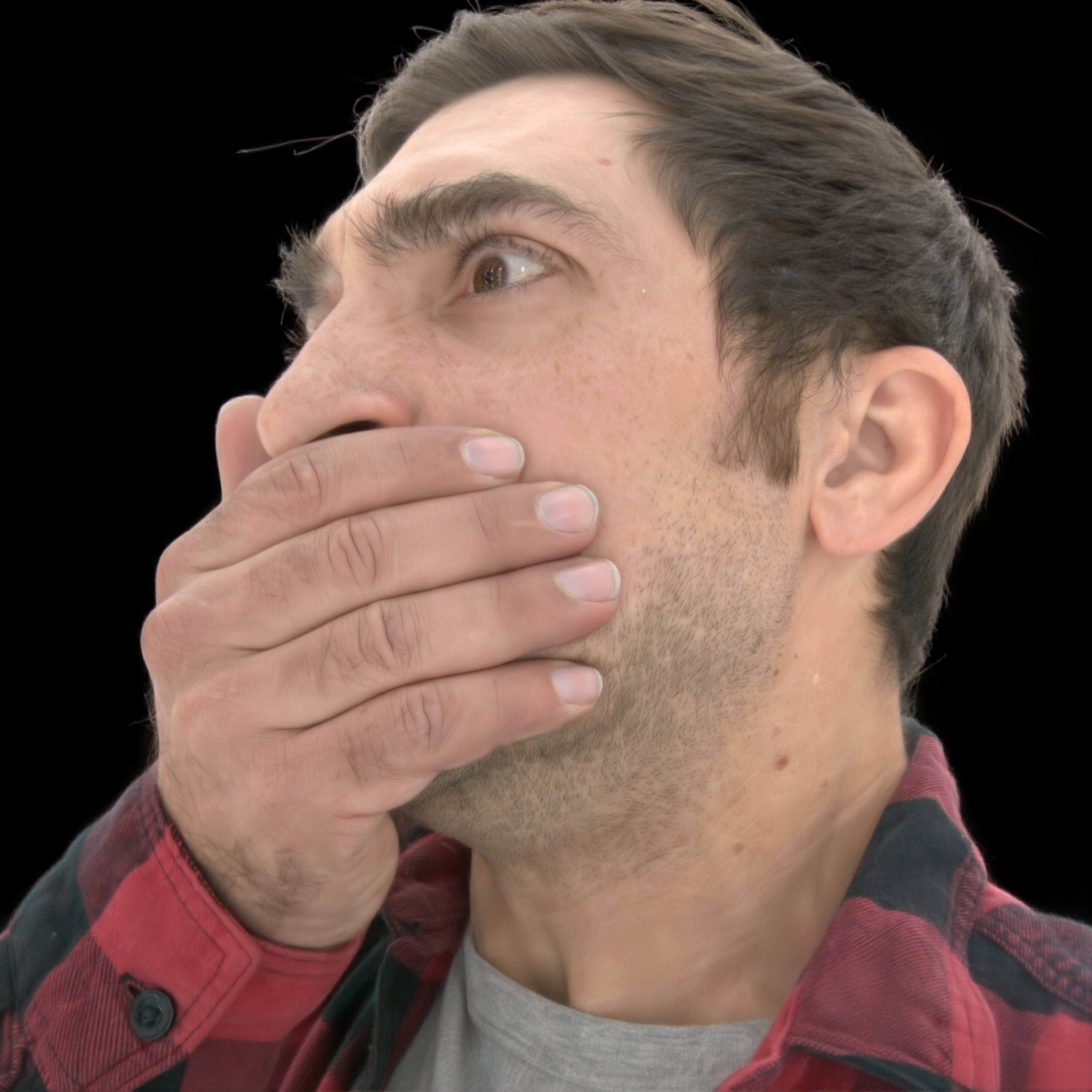} \\
\adjincludegraphics[trim= {0.52\width} {0.34\height} {0.38\width} {0.56\height}, clip, width=0.47\linewidth]{images/comparisons/seq10/Ours/cam0000_0012.jpg}
\hfill
\adjincludegraphics[trim= {0.63\width} {0.52\height} {0.27\width} {0.38\height}, clip, width=0.47\linewidth]{images/comparisons/seq10/Ours/cam0000_0012.jpg}
\end{minipage} &
\begin{minipage}{0.128\linewidth}
\centering
\adjincludegraphics[trim= {0.0\width} {0.0\height} {0.1\width} {0.0\height}, clip, width=1.0\linewidth]{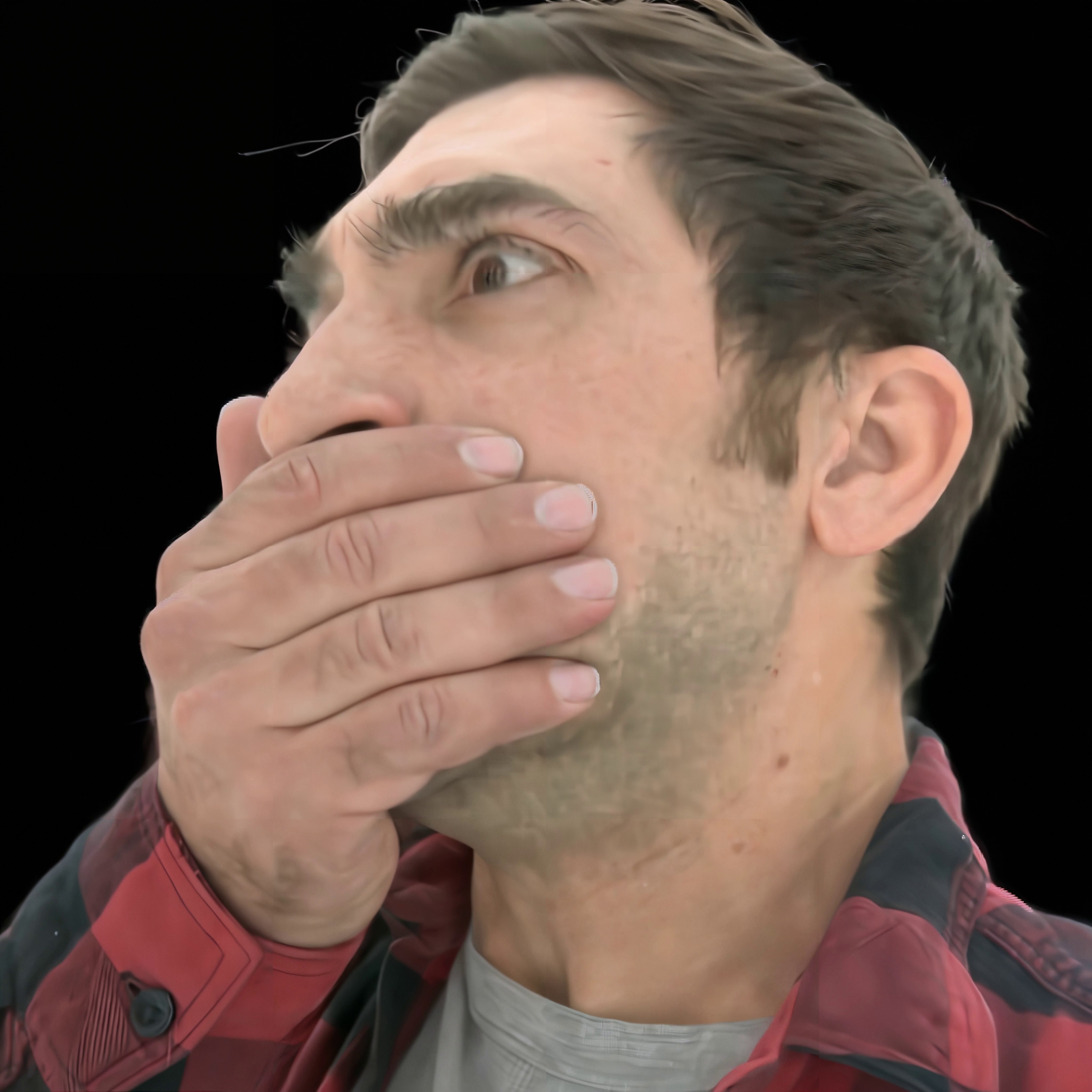} \\
\adjincludegraphics[trim= {0.52\width} {0.34\height} {0.38\width} {0.56\height}, clip, width=0.47\linewidth]{images/comparisons/seq10/UpscaleAVideo/cam0000_0012.jpg}
\hfill
\adjincludegraphics[trim= {0.63\width} {0.52\height} {0.27\width} {0.38\height}, clip, width=0.47\linewidth]{images/comparisons/seq10/UpscaleAVideo/cam0000_0012.jpg}
\end{minipage} &
\begin{minipage}{0.128\linewidth}
\centering
\adjincludegraphics[trim= {0.0\width} {0.0\height} {0.1\width} {0.0\height}, clip, width=1.0\linewidth]{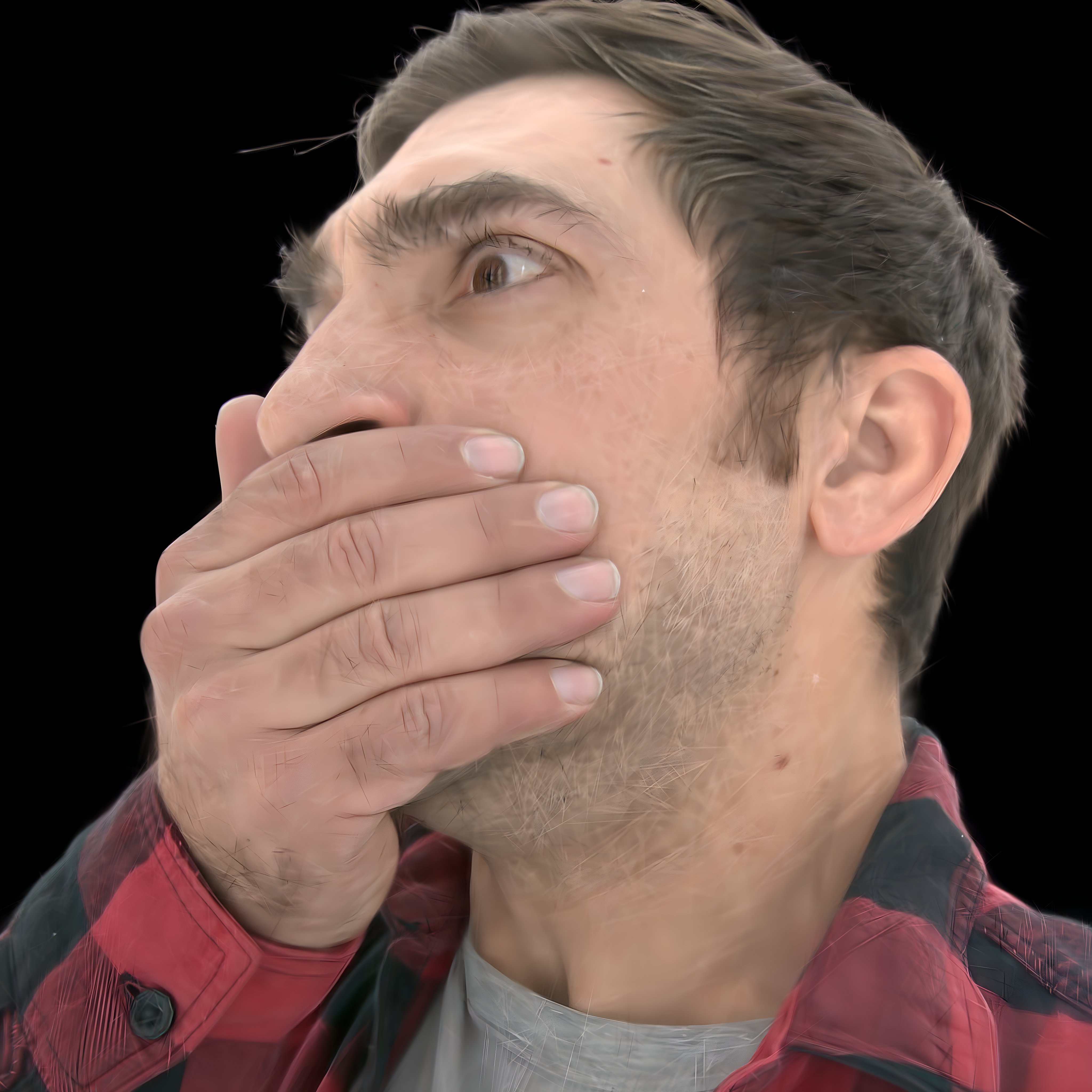} \\
\adjincludegraphics[trim= {0.52\width} {0.34\height} {0.38\width} {0.56\height}, clip, width=0.47\linewidth]{images/comparisons/seq10/KEEP/a/final_results/00000012.jpg}
\hfill
\adjincludegraphics[trim= {0.63\width} {0.52\height} {0.27\width} {0.38\height}, clip, width=0.47\linewidth]{images/comparisons/seq10/KEEP/a/final_results/00000012.jpg}
\end{minipage} &
\begin{minipage}{0.128\linewidth}
\centering
\adjincludegraphics[trim= {0.0\width} {0.0\height} {0.1\width} {0.0\height}, clip, width=1.0\linewidth]{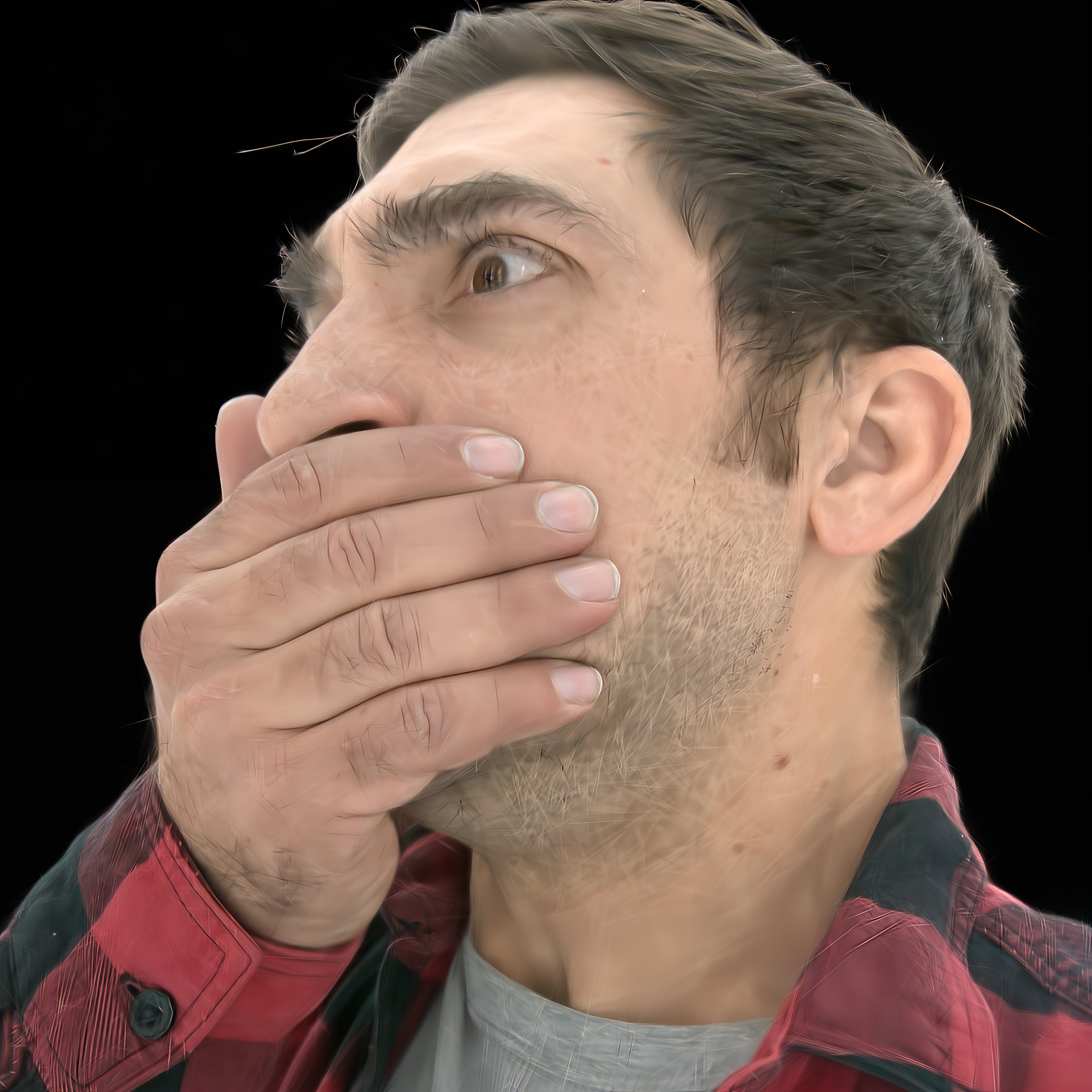} \\
\adjincludegraphics[trim= {0.52\width} {0.34\height} {0.38\width} {0.56\height}, clip, width=0.47\linewidth]{images/comparisons/seq10/ResShift/cam0000_0012.jpg}
\hfill
\adjincludegraphics[trim= {0.63\width} {0.52\height} {0.27\width} {0.38\height}, clip, width=0.47\linewidth]{images/comparisons/seq10/ResShift/cam0000_0012.jpg}
\end{minipage} &
\begin{minipage}{0.128\linewidth}
\centering
\adjincludegraphics[trim= {0.0\width} {0.0\height} {0.1\width} {0.0\height}, clip, width=1.0\linewidth]{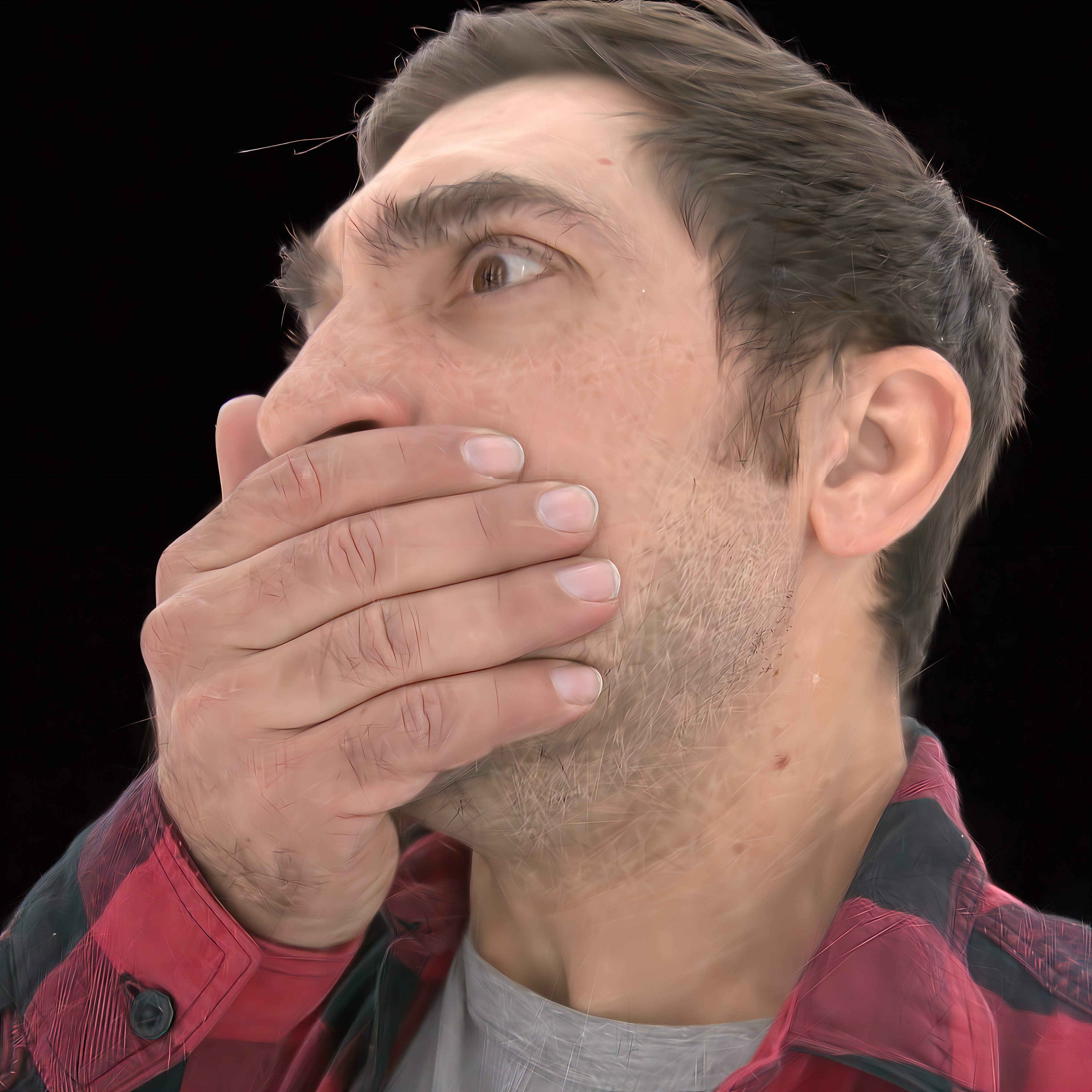} \\
\adjincludegraphics[trim= {0.52\width} {0.34\height} {0.38\width} {0.56\height}, clip, width=0.47\linewidth]{images/comparisons/seq10/SinSR/cam0000_0012.jpg}
\hfill
\adjincludegraphics[trim= {0.63\width} {0.52\height} {0.27\width} {0.38\height}, clip, width=0.47\linewidth]{images/comparisons/seq10/SinSR/cam0000_0012.jpg}
\end{minipage} \\


\begin{minipage}{0.128\linewidth}
\centering
\placeholderUnav \\
\hfill
\end{minipage} &
\begin{minipage}{0.128\linewidth}
\centering
\adjincludegraphics[trim= {0.20\width} {0.0\height} {0.0\width} {0.0\height}, clip, width=1.0\linewidth]{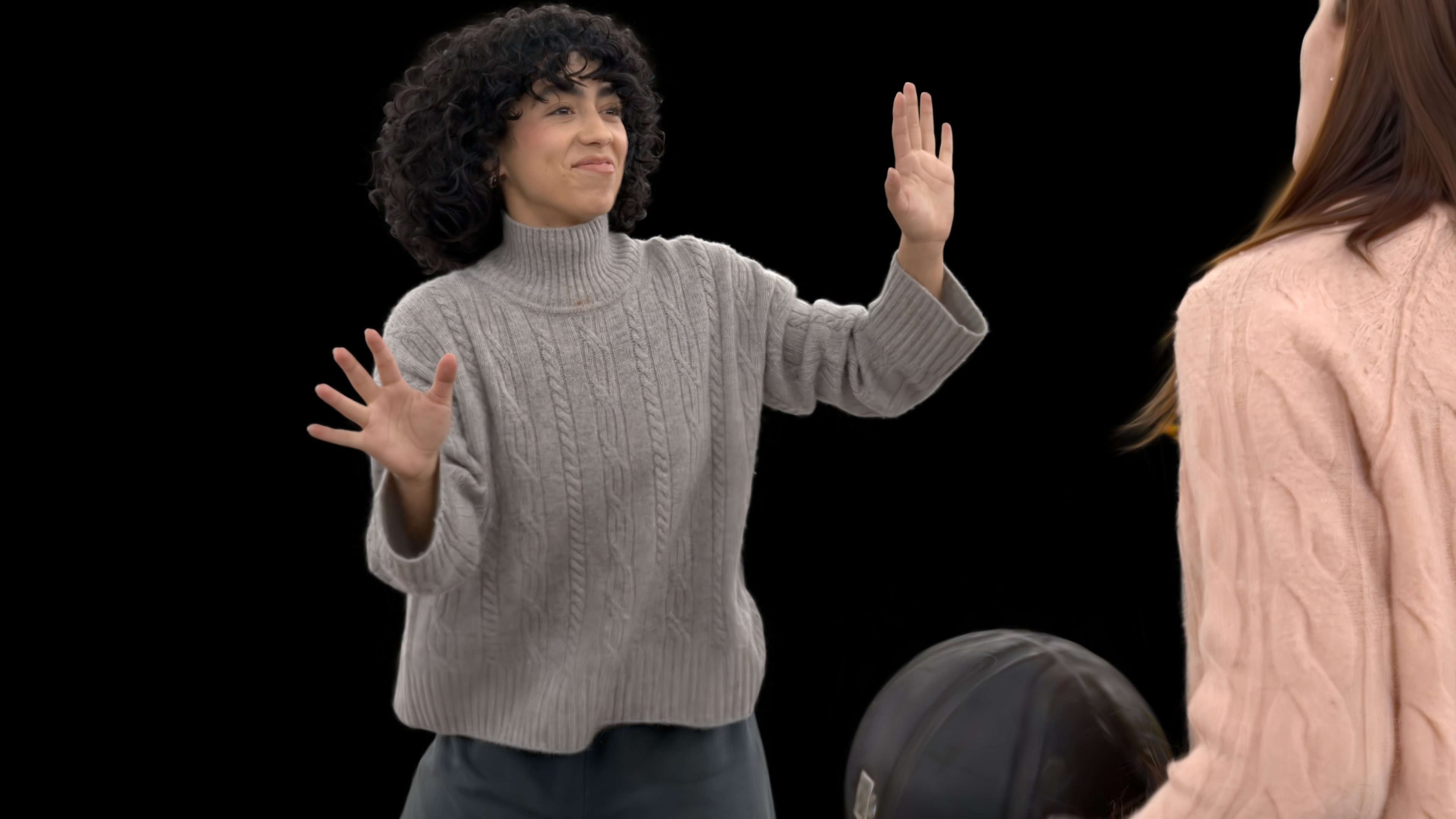} \\
\adjincludegraphics[trim= {0.37\width} {0.85\height} {0.60\width} {0.1\height}, clip, width=0.47\linewidth]{images/comparisons/seq06_wideangle_AV_JW/LQ/cam0000_0012.jpg}
\hfill
\adjincludegraphics[trim= {0.39\width} {0.77\height} {0.58\width} {0.18\height}, clip, width=0.47\linewidth]{images/comparisons/seq06_wideangle_AV_JW/LQ/cam0000_0012.jpg}
\end{minipage} &
\begin{minipage}{0.128\linewidth}
\centering
\adjincludegraphics[trim= {0.20\width} {0.0\height} {0.0\width} {0.0\height}, clip, width=1.0\linewidth]{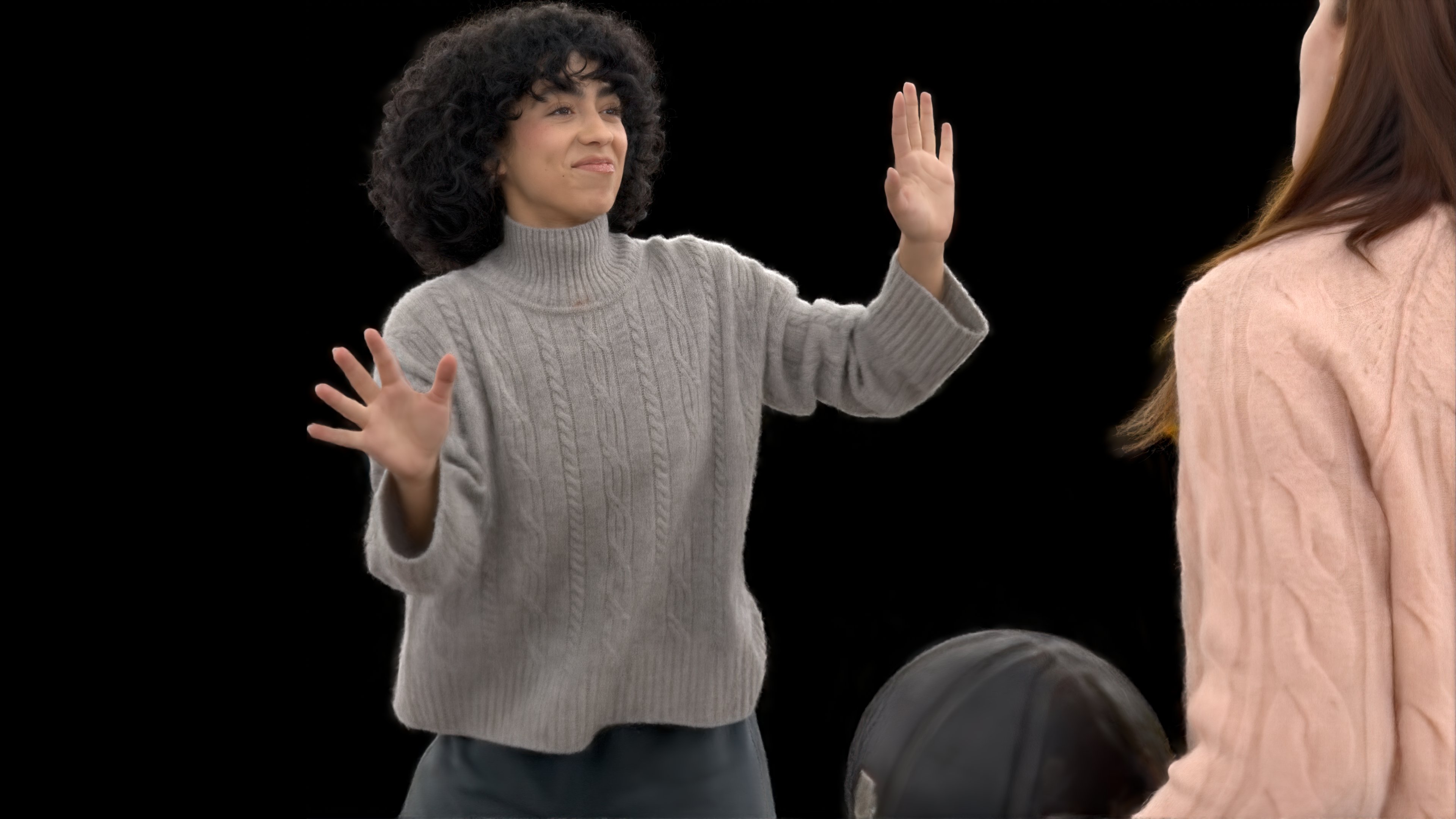} \\
\adjincludegraphics[trim= {0.37\width} {0.85\height} {0.60\width} {0.1\height}, clip, width=0.47\linewidth]{images/comparisons/seq06_wideangle_AV_JW/Ours/cam0000_0012.jpg}
\hfill
\adjincludegraphics[trim= {0.39\width} {0.77\height} {0.58\width} {0.18\height}, clip, width=0.47\linewidth]{images/comparisons/seq06_wideangle_AV_JW/Ours/cam0000_0012.jpg}
\end{minipage} &
\begin{minipage}{0.128\linewidth}
\centering
\adjincludegraphics[trim= {0.20\width} {0.0\height} {0.0\width} {0.0\height}, clip, width=1.0\linewidth]{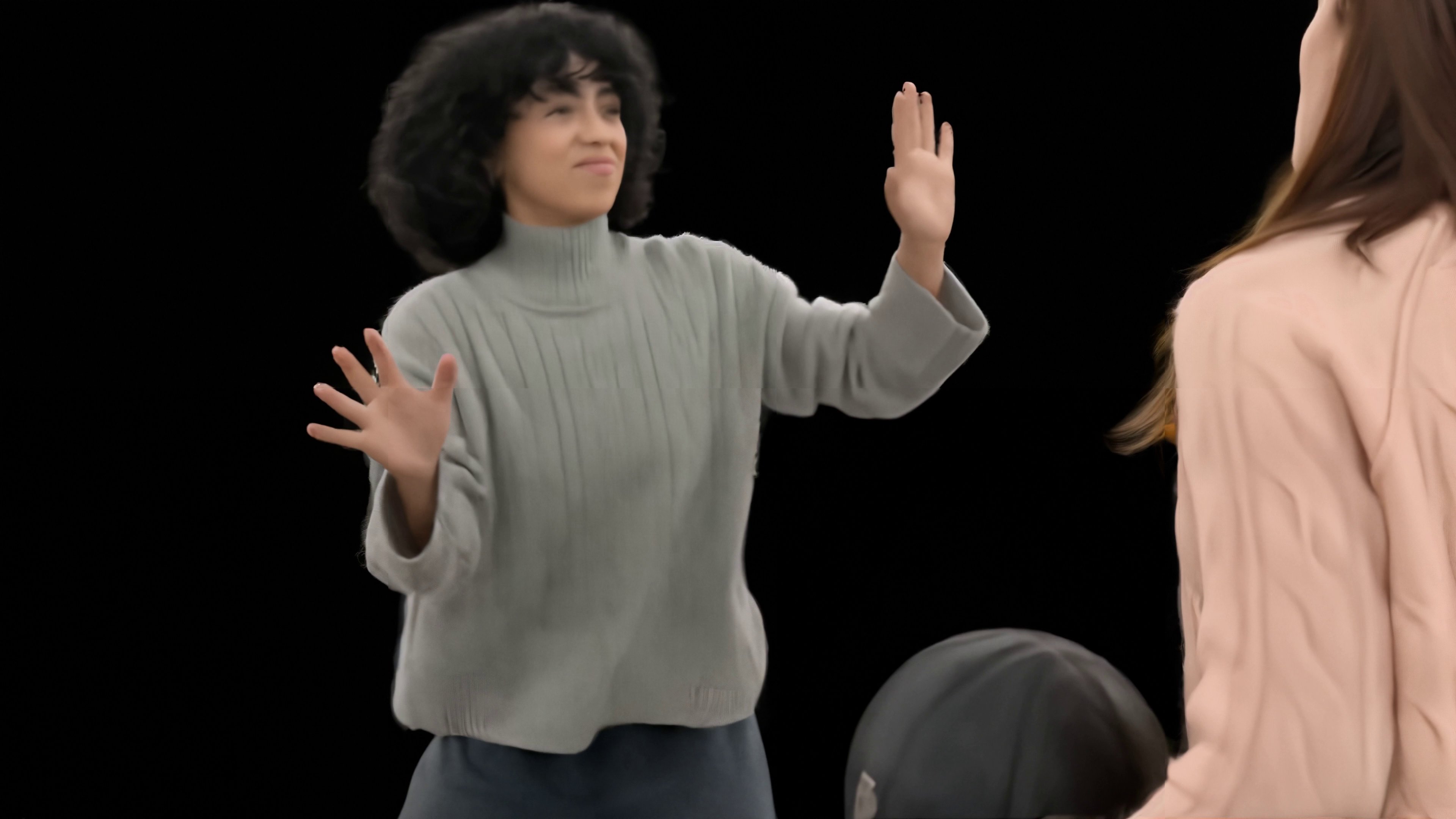} \\
\adjincludegraphics[trim= {0.37\width} {0.85\height} {0.60\width} {0.1\height}, clip, width=0.47\linewidth]{images/comparisons/seq06_wideangle_AV_JW/UpscaleAVideo/cam0000_0012.jpg}
\hfill
\adjincludegraphics[trim= {0.39\width} {0.77\height} {0.58\width} {0.18\height}, clip, width=0.47\linewidth]{images/comparisons/seq06_wideangle_AV_JW/UpscaleAVideo/cam0000_0012.jpg}
\end{minipage} &
\begin{minipage}{0.128\linewidth}
\centering
\adjincludegraphics[trim= {0.20\width} {0.0\height} {0.0\width} {0.0\height}, clip, width=1.0\linewidth]{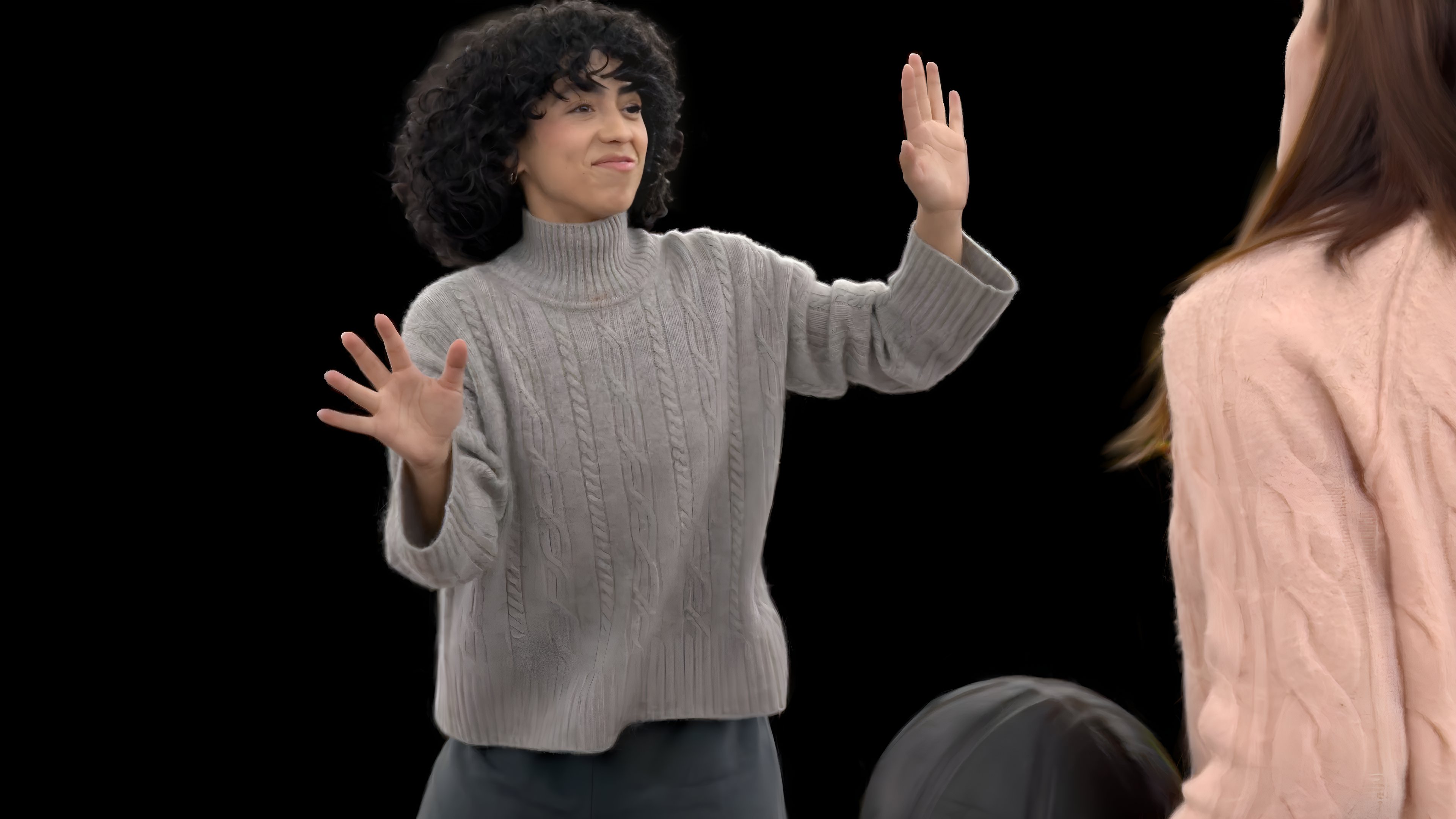} \\
\adjincludegraphics[trim= {0.385\width} {0.85\height} {0.585\width} {0.1\height}, clip, width=0.47\linewidth]{images/comparisons/seq06_wideangle_AV_JW/KEEP/a/final_results/00000012.jpg}
\hfill
\adjincludegraphics[trim= {0.405\width} {0.77\height} {0.565\width} {0.18\height}, clip, width=0.47\linewidth]{images/comparisons/seq06_wideangle_AV_JW/KEEP/a/final_results/00000012.jpg}
\end{minipage} &
\begin{minipage}{0.128\linewidth}
\centering
\adjincludegraphics[trim= {0.20\width} {0.0\height} {0.0\width} {0.0\height}, clip, width=1.0\linewidth]{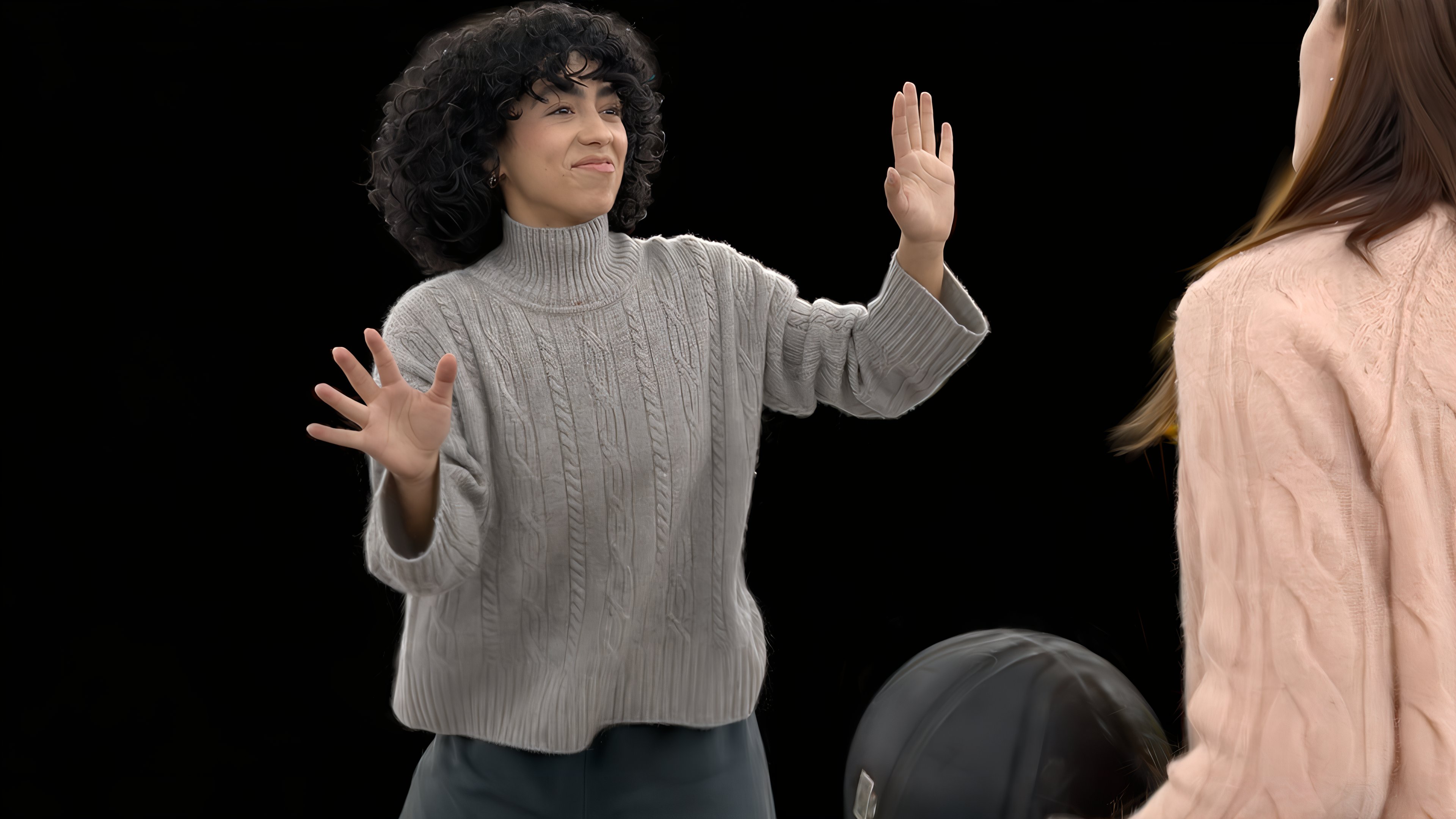} \\
\adjincludegraphics[trim= {0.37\width} {0.85\height} {0.60\width} {0.1\height}, clip, width=0.47\linewidth]{images/comparisons/seq06_wideangle_AV_JW/ResShift/cam0000_0012.jpg}
\hfill
\adjincludegraphics[trim= {0.39\width} {0.77\height} {0.58\width} {0.18\height}, clip, width=0.47\linewidth]{images/comparisons/seq06_wideangle_AV_JW/ResShift/cam0000_0012.jpg}
\end{minipage} &
\begin{minipage}{0.128\linewidth}
\centering
\adjincludegraphics[trim= {0.20\width} {0.0\height} {0.0\width} {0.0\height}, clip, width=1.0\linewidth]{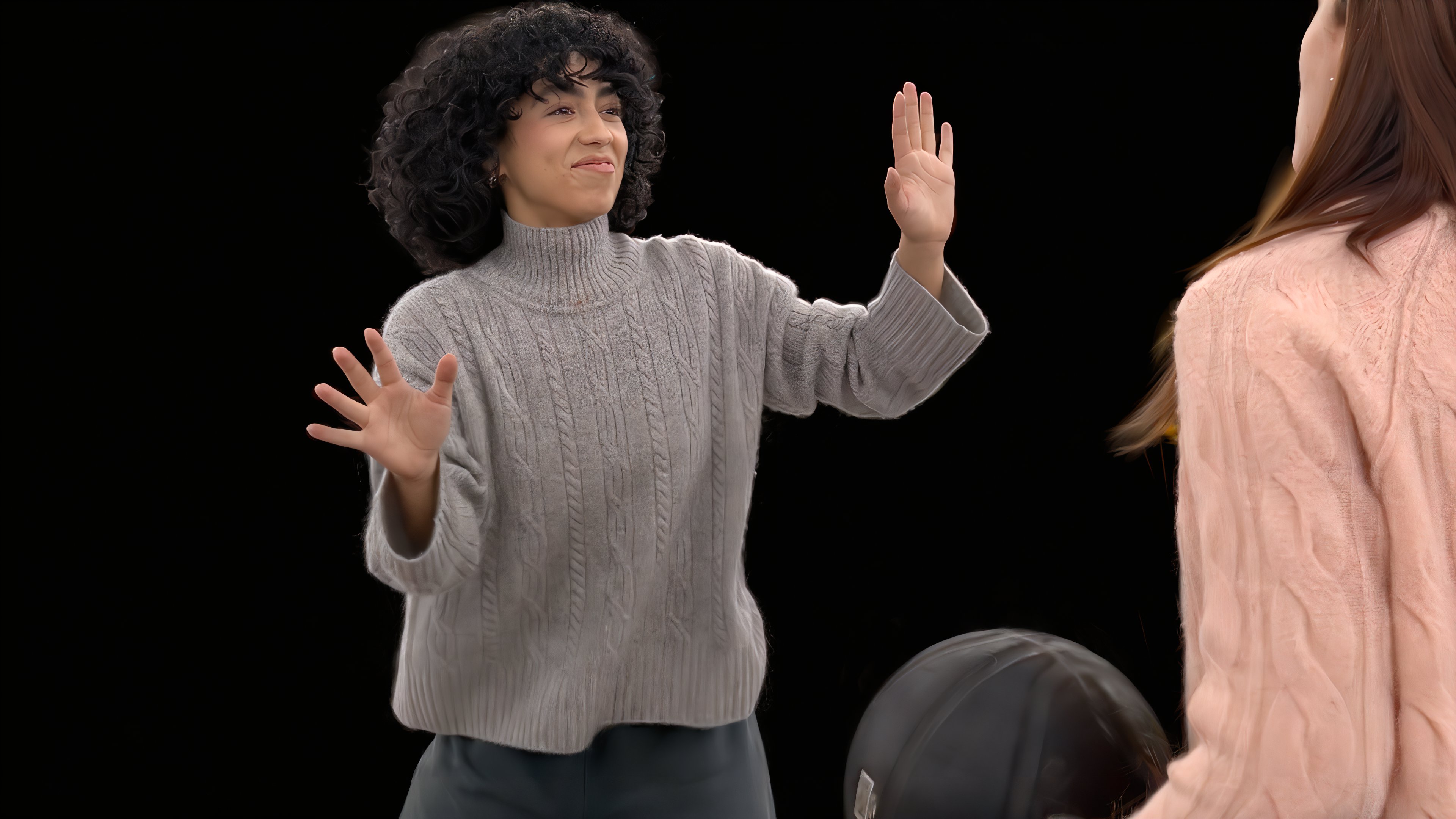} \\
\adjincludegraphics[trim= {0.37\width} {0.85\height} {0.60\width} {0.1\height}, clip, width=0.47\linewidth]{images/comparisons/seq06_wideangle_AV_JW/SinSR/cam0000_0012.jpg}
\hfill
\adjincludegraphics[trim= {0.39\width} {0.77\height} {0.58\width} {0.18\height}, clip, width=0.47\linewidth]{images/comparisons/seq06_wideangle_AV_JW/SinSR/cam0000_0012.jpg}
\end{minipage} \\


\begin{minipage}{0.128\linewidth}
\centering
\placeholderUnav \\
\hfill
\end{minipage} &
\begin{minipage}{0.128\linewidth}
\centering
\adjincludegraphics[trim= {0.07\width} {0.0\height} {0.13\width} {0.0\height}, clip, width=1.0\linewidth]{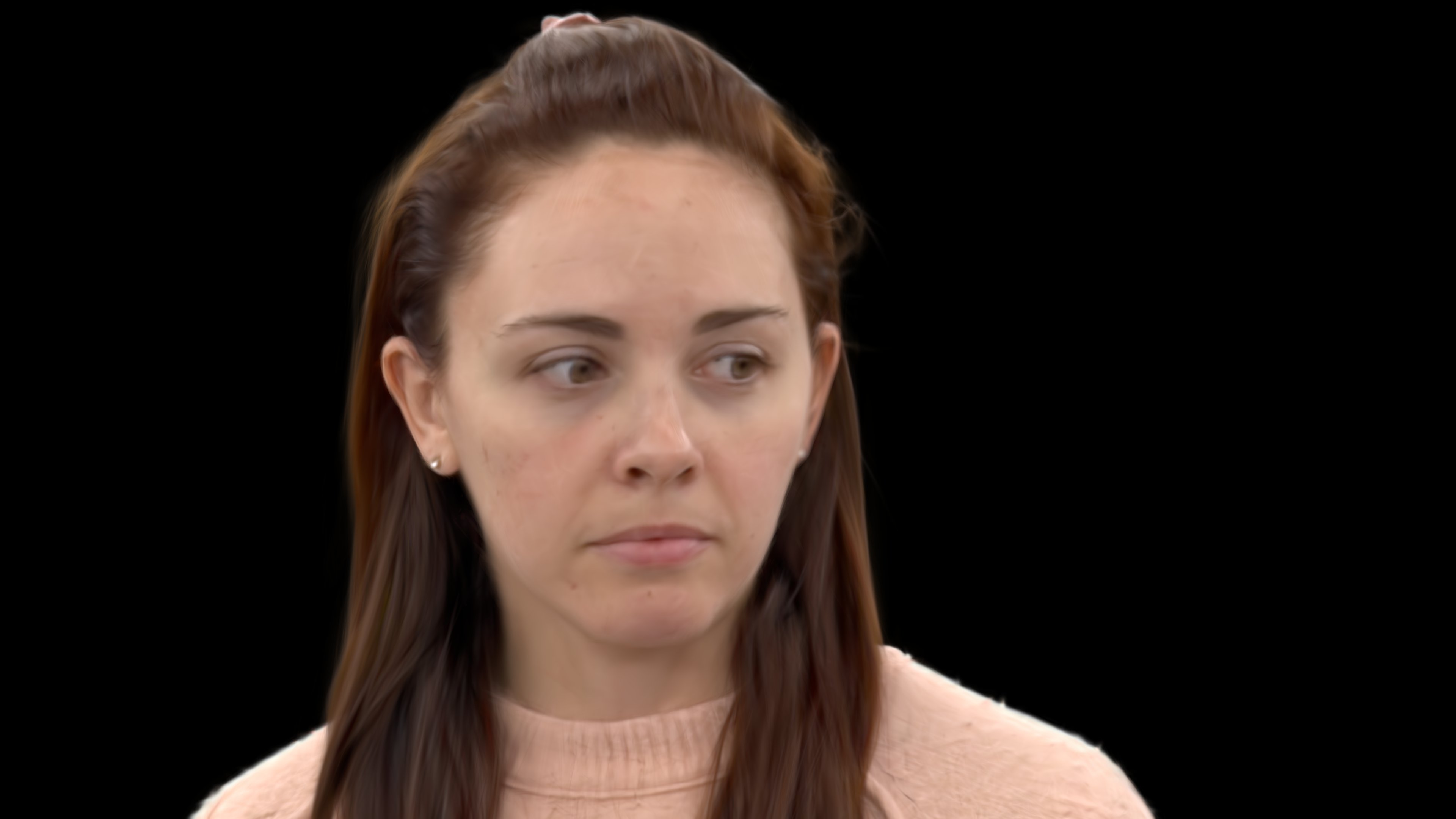} \\
\adjincludegraphics[trim= {0.475\width} {0.5\height} {0.475\width} {0.40\height}, clip, width=0.47\linewidth]{images/comparisons/seq02_closeup_TK/LQ/cam0000_0012.jpg}
\hfill
\adjincludegraphics[trim= {0.27\width} {0.4\height} {0.68\width} {0.5\height}, clip, width=0.47\linewidth]{images/comparisons/seq02_closeup_TK/LQ/cam0000_0012.jpg}
\end{minipage} &
\begin{minipage}{0.128\linewidth}
\centering
\adjincludegraphics[trim= {0.07\width} {0.0\height} {0.13\width} {0.0\height}, clip, width=1.0\linewidth]{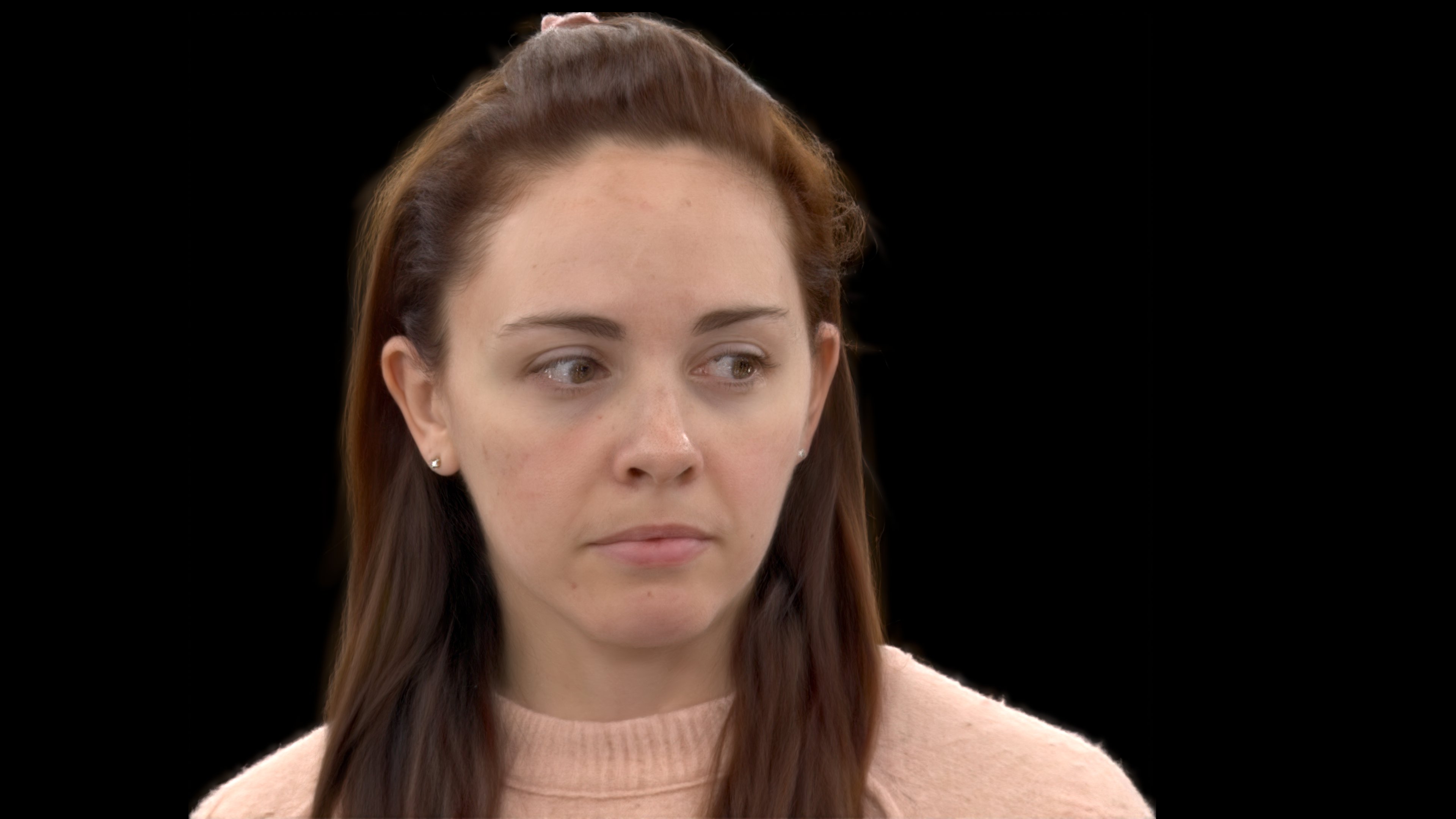} \\
\adjincludegraphics[trim= {0.475\width} {0.5\height} {0.475\width} {0.40\height}, clip, width=0.47\linewidth]{images/comparisons/seq02_closeup_TK/Ours/cam0000_0012.jpg}
\hfill
\adjincludegraphics[trim= {0.27\width} {0.4\height} {0.68\width} {0.5\height}, clip, width=0.47\linewidth]{images/comparisons/seq02_closeup_TK/Ours/cam0000_0012.jpg}
\end{minipage} &
\begin{minipage}{0.128\linewidth}
\centering
\adjincludegraphics[trim= {0.07\width} {0.0\height} {0.13\width} {0.0\height}, clip, width=1.0\linewidth]{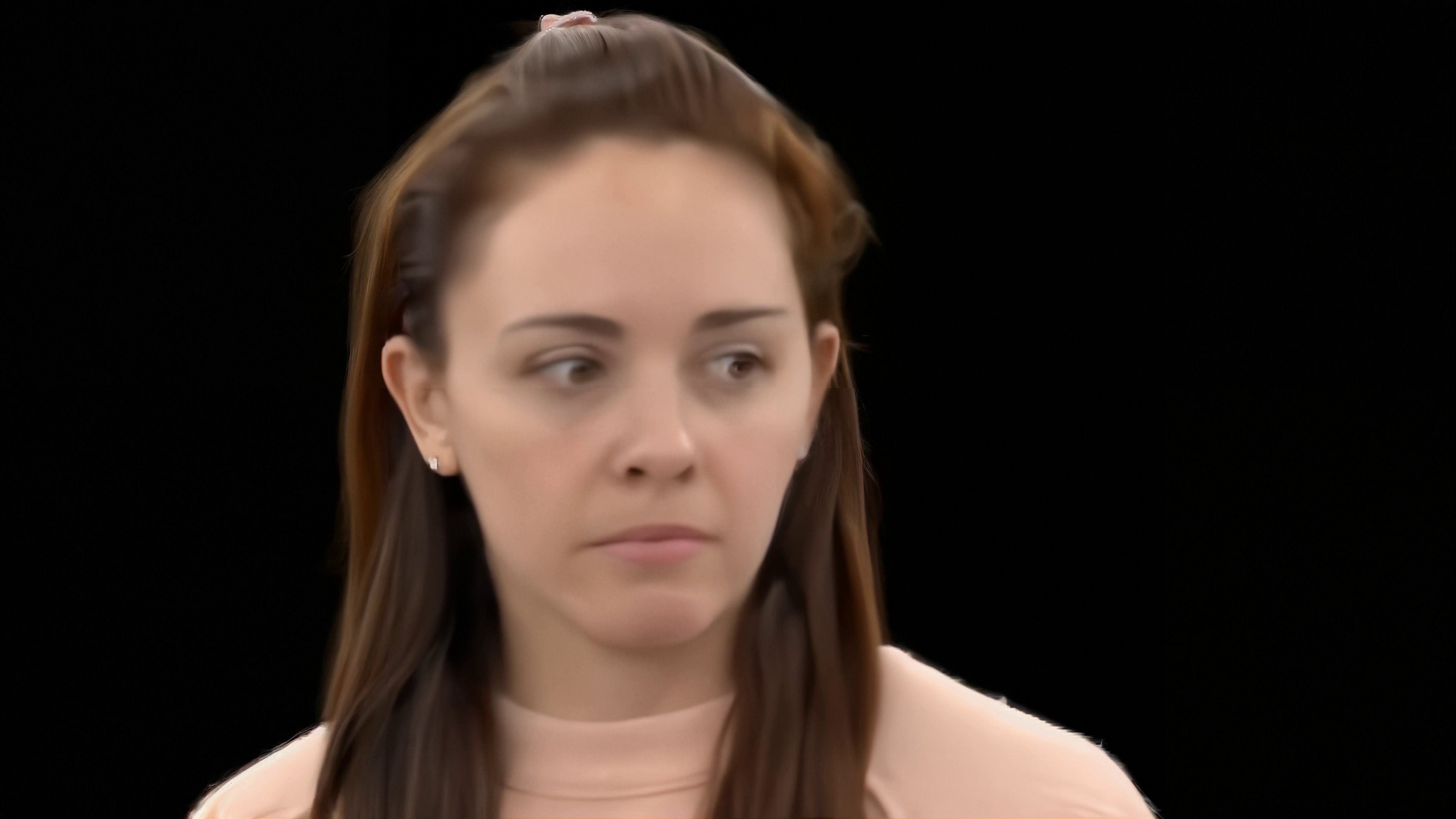} \\
\adjincludegraphics[trim= {0.475\width} {0.5\height} {0.475\width} {0.40\height}, clip, width=0.47\linewidth]{images/comparisons/seq02_closeup_TK/UpscaleAVideo/cam0000_0012.jpg}
\hfill
\adjincludegraphics[trim= {0.27\width} {0.4\height} {0.68\width} {0.5\height}, clip, width=0.47\linewidth]{images/comparisons/seq02_closeup_TK/UpscaleAVideo/cam0000_0012.jpg}
\end{minipage} &
\begin{minipage}{0.128\linewidth}
\centering
\adjincludegraphics[trim= {0.07\width} {0.0\height} {0.13\width} {0.0\height}, clip, width=1.0\linewidth]{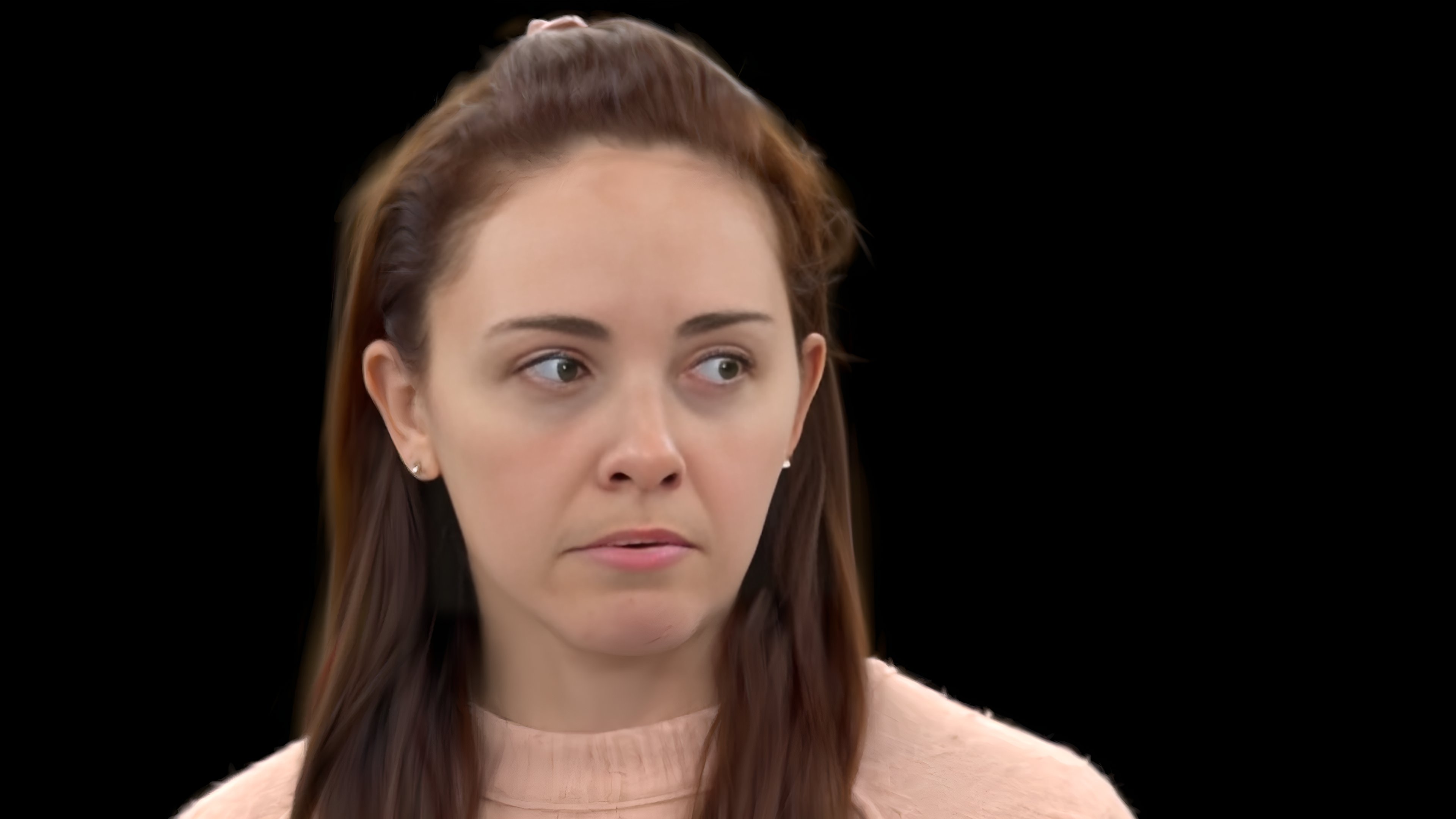} \\
\adjincludegraphics[trim= {0.475\width} {0.5\height} {0.475\width} {0.40\height}, clip, width=0.47\linewidth]{images/comparisons/seq02_closeup_TK/KEEP/a/final_results/00000012.jpg}
\hfill
\adjincludegraphics[trim= {0.27\width} {0.4\height} {0.68\width} {0.5\height}, clip, width=0.47\linewidth]{images/comparisons/seq02_closeup_TK/KEEP/a/final_results/00000012.jpg}
\end{minipage} &
\begin{minipage}{0.128\linewidth}
\centering
\adjincludegraphics[trim= {0.07\width} {0.0\height} {0.13\width} {0.0\height}, clip, width=1.0\linewidth]{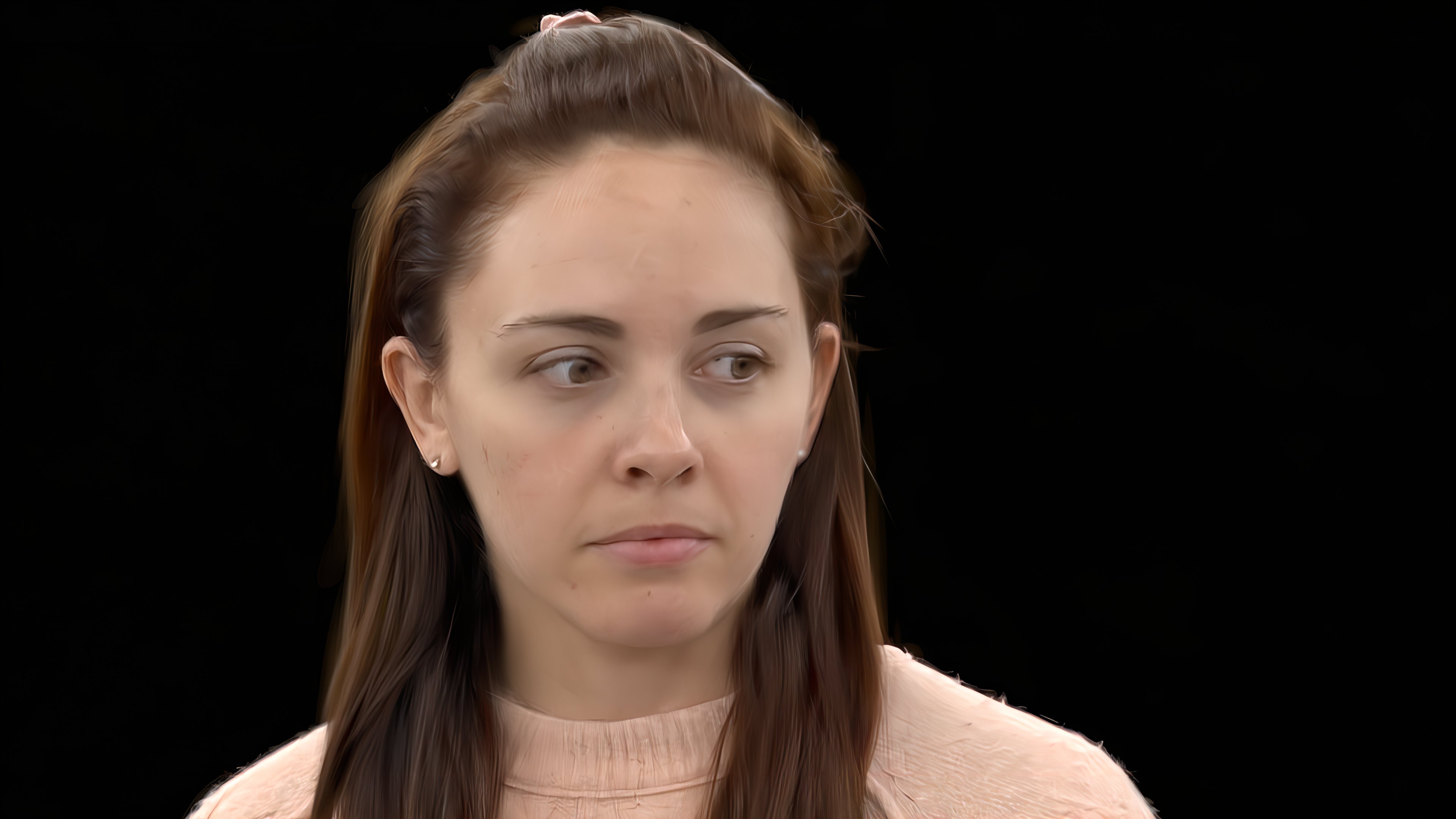} \\
\adjincludegraphics[trim= {0.475\width} {0.5\height} {0.475\width} {0.40\height}, clip, width=0.47\linewidth]{images/comparisons/seq02_closeup_TK/ResShift/cam0000_0012.jpg}
\hfill
\adjincludegraphics[trim= {0.27\width} {0.4\height} {0.68\width} {0.5\height}, clip, width=0.47\linewidth]{images/comparisons/seq02_closeup_TK/ResShift/cam0000_0012.jpg}
\end{minipage} &
\begin{minipage}{0.128\linewidth}
\centering
\adjincludegraphics[trim= {0.07\width} {0.0\height} {0.13\width} {0.0\height}, clip, width=1.0\linewidth]{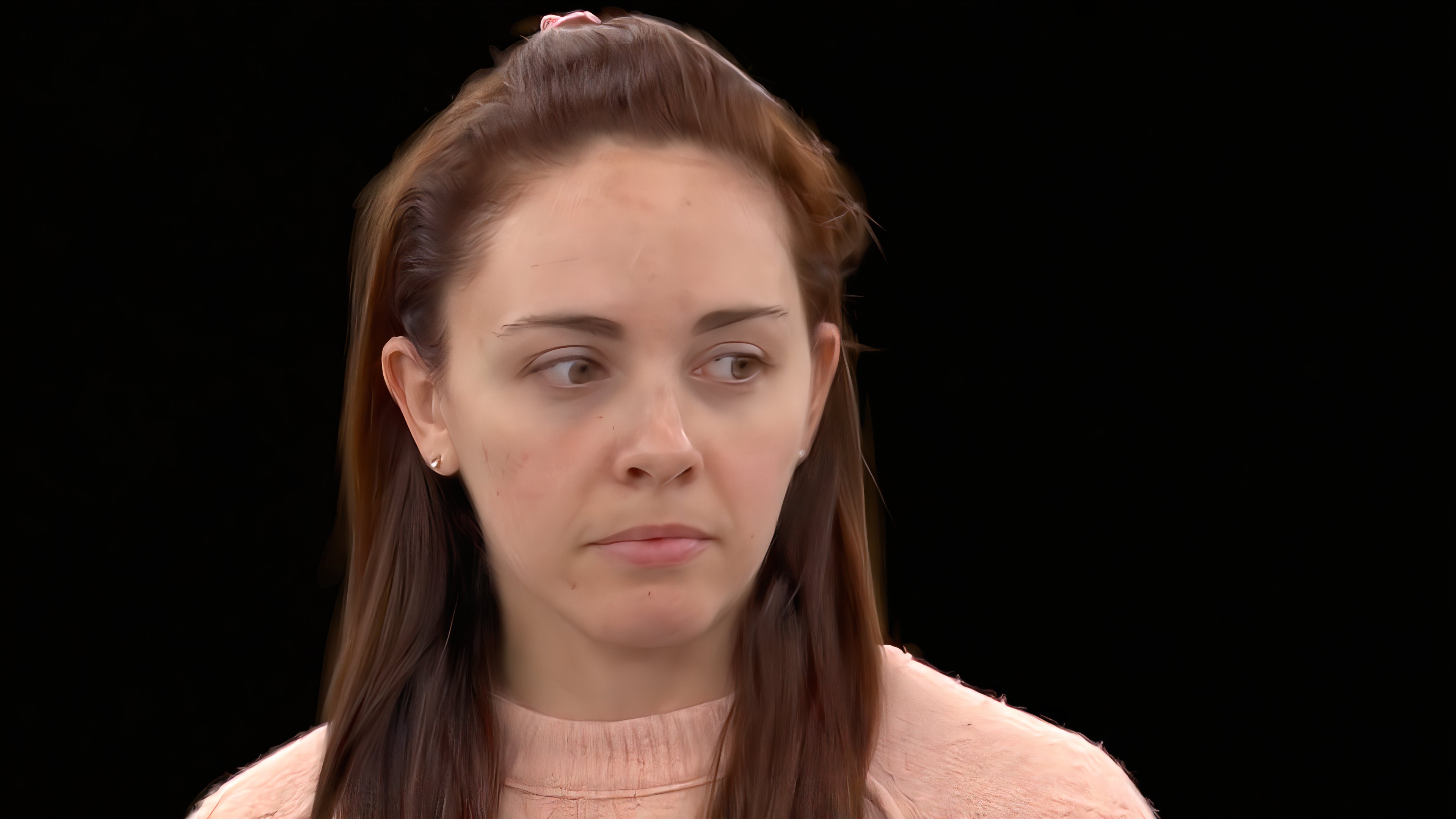} \\
\adjincludegraphics[trim= {0.475\width} {0.5\height} {0.475\width} {0.40\height}, clip, width=0.47\linewidth]{images/comparisons/seq02_closeup_TK/SinSR/cam0000_0012.jpg}
\hfill
\adjincludegraphics[trim= {0.27\width} {0.4\height} {0.68\width} {0.5\height}, clip, width=0.47\linewidth]{images/comparisons/seq02_closeup_TK/SinSR/cam0000_0012.jpg}
\end{minipage} \\

\end{tabular}

\caption{
Comparisons for our Detail Enhancement module. Top two images are from the validation set, while bottom two are from a test sequence.
From left to right: high resolution GS render, typically used as ground truth during training. A low res GS render used as input for the following. Our result. UpscaleAVideo \cite{zhou2024upscale}. Results from KEEP \cite{feng2024kalmaninspiredfeaturepropagationvideo}. Results using a ResShift\cite{yue2024resshift} model. And SinSR results \cite{wang2024sinsr}.}
\label{fig:ablations_sr}
\end{figure*}
\section{Experiments}

\subsection{Main Results}

\SA{We present the results of eight different performances from three groups of two or three actors.}
For each performance, we design virtual camera paths using a custom Maya \cite{maya} plugin.
We then render our reconstructions of the performances -- cropped to delete the background -- and enhance the renderings with our detail enhancement model.
In Fig.~\ref{fig:main_results} (best viewed full screen) we present qualitative image results. We show our reconstruction quality matches the capture resolution while providing better colors. \SA{The detail enhancement model effectively removes GS artifacts and adds significant missing detail both in the RGB and Alpha layers.
}
The results are then composited by an artist onto a background sequence using the same camera path. \SA{For backgrounds, we used 3DGS reconstructions or generated videos \cite{gen3}.}
Additional results are presented in the accompanying video.
\SA{We provide implementation details and estimates for the compute overhead of our method in the supplemental.}


\subsection{Baselines/Ablations and comparisons}
We conduct comparative and ablative analysis across different design choices in our method to evaluate the performance of the dynamic scene reconstruction and detail enhancement models. 

\begin{table}[h]
\centering
\caption{Ablation results of \textit{Poly4DGS} reconstruction. The best results are marked in \textbf{bold}. \SA{All our added components contribute to the final quality of the reconstruction. Using a per-frame GS instead of 4DGS leads to poorer reconstruction.}}

\renewcommand{\arraystretch}{0.8}
\begin{tabular}{l|c|c|c}
\toprule
Method & PSNR $\uparrow$ & SSIM $\uparrow$ & LPIPS $\downarrow$ \\
\midrule
Full & \textbf{27.29} & \textbf{.9752} & \textbf{.0160} \\
w/o black level & 24.11 & .9576 & .0200 \\
w/o expo. opt. & 25.50 & .9718 & .0161 \\
w/o dense pt. & 25.31 & .9576 & .0288 \\
per-frame GS &  \SA{26.11} & \SA{.9697} & \SA{.0204} \\
\FI{4D-Rotor \cite{4drotor}} &  \FI{27.08} & \FI{.9730} & \FI{.0171} \\
\bottomrule
\end{tabular}
\label{tab:ablation_recon}
\end{table}

\begin{table*}[t]
\centering
\caption{
\FI{Ablations and Comparison of the detail enhancement model using temporal and image quality metrics. MS, TF, MUSIQ and NIQE were copmputed on the \scenerig data in a No-Reference manner, while PSNR, SSIM, LPIPS, TPSNR, FID and FVD were computed on the test of the \facerig data. Best values are in \textbf{bold} and second best \underline{underlined}.}}

\small
\begin{tabular}{l|cccc|cccccc}
\hline
\textbf{Method} & \multicolumn{4}{c|}{\textbf{No-Ref metrics using \facerig data}} & \multicolumn{6}{c}{\textbf{Ref-based metrics using \scenerig data}} \\
\cline{2-5} \cline{6-11}
 & MS $\uparrow$ & TF $\uparrow$ & MUSIQ $\uparrow$ & NIQE $\downarrow$ & PSNR $\uparrow$ & SSIM $\uparrow$ & LPIPS $\downarrow$ & TPSNR $\uparrow$ & FID $\downarrow$ & FVD $\downarrow$ \\
\hline
Ours  & \textbf{95.34} & \textbf{94.29} & 51.79 & \textbf{5.28} & \underline{31.29} & \underline{0.8871} & 0.1812 & \underline{31.27} & \underline{102.77} & 213.52 \\
w/o OFW  & 94.72 & \underline{94.06} & \underline{52.24} & \underline{5.48} &  \textbf{31.43} & \textbf{0.8872} & \underline{0.1614} & \textbf{31.42} & 103.44 & \textbf{202.90} \\
w/o OFW, w/o LFS  & \underline{94.73} & 93.90 & \textbf{53.17} & 5.89 &  30.61 & 0.8604 & \textbf{0.1612} & 30.60 & \textbf{99.88} & \underline{209.85} \\
\hline
\hline
Ours  & \textbf{95.34} & \textbf{94.29} & 51.79 & \textbf{5.28} &  \underline{31.29} & \underline{0.8871} & \textbf{0.1812} & \underline{31.27} & \underline{102.77} & \textbf{213.52} \\
ResShift \citeyear{yue2024resshift} & 93.70 & 92.85 & 58.83 & 5.82 &  29.33 & 0.7771 & 0.3221 & 29.33 & 200.16 & 662.70 \\
ResShift (FT) \citeyear{yue2024resshift} &  93.36 & 92.34 & \underline{59.12} & 5.77 &  29.78 & 0.8517 & \underline{0.2146} & 29.78 & \textbf{87.93} & 405.16 \\
SinSR \citeyear{wang2024sinsr} & 93.89 & 93.08 & \textbf{62.59} & \underline{5.32} &  29.19 & 0.7070 & 0.3521 & 29.19 & 199.17 & 656.69 \\
\scalebox{.8}[1.0]{Upscale-A-Video} \citeyear{zhou2024upscale} & \underline{94.64} & \underline{93.66} & 34.87 & 7.87 &  30.30 & 0.8329 & 0.3620 & \underline{30.29} & 191.96 & 341.74 \\
KEEP \citeyear{feng2024kalmaninspiredfeaturepropagationvideo} & 94.09 & 93.20 & 45.33 & 6.16 &  30.08 & 0.8627 & 0.2996 & 30.07 & 176.90 & 350.91 \\
Input Poly4DGS Sequence & 94.18 & 93.25 & 37.22 & 6.89 & \textbf{31.72} & \textbf{0.9009} & 0.2474 & \textbf{31.72} & 150.51 & \underline{333.11} \\
\hline
\end{tabular}
\label{tab:comparison_baselines}
\end{table*}


\subsubsection{Dynamic Scene Reconstruction}

To evaluate \SA{our reconstruction pipeline, and main additions} we conduct an ablation study. We hold out 12 out of 140 cameras, covering both close-up and large-scale FoV.
Since our optimized GS model has a different exposure and black level with respect to the input images, we optimize the exposure and black level of the held-out views before comparing to ground truth.
We report the \textit{PSNR}, \textit{SSIM} and \textit{LPIPS} averaged over three sequences in Tab.~\ref{tab:ablation_recon} \SA{for different variants of our pipeline}. 
Without black-level optimization (w/o black level) and exposure optimization (w/o expo. opt.) the view-inconsistent effects such as lens glare and exposure variations get baked into the splats, leading to low contrast and poor colors, also shown in Fig.~\ref{fig:ablations_recon}.
Initializing from dense point cloud (w/o dense pt.) helps to reconstruct fine details.
\FI{Last we compare with using per-frame 3DGS \cite{kerbl3Dgaussians} reconstruction and 4D-Rotor \cite{4drotor} both modified to benefit from dense point cloud initialization and black level and exposure optimization.
We observe that using the \textit{Poly4DGS} formulation achieves better performance}.
This is because 4DGS allows Gaussian primitives to be shared across frames \FI{with a smoother interpolation}, leading to a higher per-frame gaussian count.
Most importantly, as shown in the supplementary video, our 4DGS produces more temporally stable results compared to per-frame 3DGS.

\subsubsection{Detail enhancement}

\SA{
We compare our detail enhancement model to existing SoTA super-resolution methods, which can be categorized in image-based \cite{wang2024sinsr,yue2024resshift} and video-based categories \cite{zhou2024upscale,feng2024kalmaninspiredfeaturepropagationvideo}. We also compare our approach to a version of ResShift \cite{yue2024resshift} finetuned (FT) on 512-resolution crops of our dataset, which was trained from scratch for approximately 520K steps, nearing convergence.
Additionally, we provide 2 ablated versions of our method: 1.) removing the Optical Flow Warping mechanism (w/o OFW) 2.) removing the Optical Flow, and the Low Frequency Swapping (LFS) (w/o OFW, w/o LFS), which corresponds to an image model conditioned on the low quality inputs.

We show quantitative results in Tab.~\ref{tab:comparison_baselines}.
\FI{We quantitatively evaluate our method using both No-Reference metrics and Reference based Metrics.
The no-reference scenario allows us to evaluate our model in its natural usage setting using renderings of the \scenerig reconstruction as input.
We used all our render paths for evaluation. We first provide temporal stability metrics, namely motion smoothness (MS) and temporal flickering (TF) from the widely adopted VBench \cite{vbench} benchmark. We also provide two no-reference image quality metrics, MUSIQ \cite{vbench} and NIQE \cite{6353522}.}

\FI{Referenced based metrics are computed on \facerig test data, left out of the training, and allows us} to evaluate the PSNR, SSIM, LPIPS \cite{zhang2018lpips}, TPSNR \cite{10.1111:cgf.13919}, FID, and FVD \cite{conf/iclr/UnterthinerSKMM19} metrics.
PSNR, SSIM, LPIPS, and temporal PSNR (which is PSNR computed
on the temporal finite differences) were calculated per frame and averaged over all test frames and sequences.
For both FID and FVD we extract features from patches and compute the distance over all patches.

From the ablations (Tab.~\ref{tab:comparison_baselines}, first three lines) we can see that our method does better at temporal consistency on the MS and TF metrics at the cost of a slight decrease in image quality compared to removing the Optical Flow Warping, this is due to a slight image smoothing.
Swapping the Low Frequency without Optical Flow Warping strongly helps temporal consistency, though the Warping further improves it.

Our method greatly outperforms all other methods and baselines in terms of reference-based image quality and is competitive with Upscale-A-Video in temporal stability, though this method produces much less details results as can be seen in our supplemental video.
\FI{Our method is also the best in the NIQE metric. ResShift and SinSR beat us on MUSIQ, which we attribute to the overly smooth and sharp nature of SinSR and ResShift; they exhibit less details and are less realistic as seen in our supplemental video. We are significantly better than the input Poly4DGS on these metrics though.}
\FI{The input Poly4DGS without detail enhancement beats our result in terms of PSNR and SSIM, but this is expected given it is mostly a “blurred” version of the GT target, while our method generates details, such as hair strand, that might not be aligned with the ground truth.
Our method has a better LPIPS metric, which captures the visual similarity to the ground truth.}

We present qualitative comparisons in our supplemental video and Fig.\ref{fig:ablations_sr}. We can see our method is temporally stable and produces higher quality, detailed images compared to other methods.
}

\section{Conclusion, Limitations, and Future Work}
We have presented a novel 4D performance capture pipeline that bridges the gap between scalable dynamic capture and production-quality rendering \SA{using a detail enhancement model}.
\SA{We demonstrate added fine details thanks to our enhancement model, allowing us render frames that match production requirements.
Nonetheless, our method has some limitations.}
It is hardware and resource-intensive, making it not universally accessible, but still appropriate for use in professional production.
\SA{While our temporal stability mechanisms greatly help, some flicker may remain on silhouette.
\SA{We did not explore relighting in this work, leaving it to future research (e.g.  \cite{DBLP:conf/siggrapha/HeCTMPXRYBYD24}), which will allow for better integration into various backgrounds.}
Eye reflections showing the sparse lighting of the \facerig are transferred with detail enhancement, suggesting that future work should provide a way to synthesize realistic eye reflections for a given environment.}
\FI{A specialized model could be used \cite{EyeNeRF} and adding continuous lighting patterns from LED panels could mitigate the sparsity of the reflections.}

\clearpage
\bibliographystyle{ACM-Reference-Format}
\bibliography{bibliography}

\end{document}